\documentclass[12pt]{article}
     \usepackage{amsmath}
  \usepackage{graphicx}
  \newlength{\abstractwidth}
  \newcommand{\be}{\begin{equation}}
  \newcommand{\bea}{\begin{eqnarray}}
  \newcommand{\eea}{\end{eqnarray}}
  \newcommand{\beq}{\begin{equation}}
  \newcommand{\ee}{\end{equation}}
  \newcommand{\eeq}{\end{equation}}

  \newcommand{\half}{{1\over 2}}
  
\def\la{\label}

\def\32{{3 \over 2 } }

  \def\ba{\begin{eqnarray}}
  \def\ea{\end{eqnarray}}

  \def\lb{\label}
 \def\simleq{\; \raise0.3ex\hbox{$<$\kern-0.75em
      \raise-1.1ex\hbox{$\sim$}}\; }
 \def\simgeq{\; \raise0.3ex\hbox{$>$\kern-0.75em
      \raise-1.1ex\hbox{$\sim$}}\; }

\def\XM{q}
\def\YM{p}

\def\WM{\YM}
\def\ZM{{\YM'}}

\def\Malpha{\gamma}
\def\Mbeta{\delta}

\def\nref#1{(\ref{#1})}

\begin{document}

\begin{titlepage}
  \bigskip

  \bigskip\bigskip

  \bigskip

\begin{center}
{\Large \bf {Cosmological Collider Physics
 }}
 \bigskip
{\Large \bf { }}
    \bigskip
\bigskip
\end{center}

  \begin{center}

 \bf { Nima Arkani-Hamed and   Juan Maldacena }
  \bigskip \rm
\bigskip

   Institute for Advanced Study,  Princeton, NJ 08540, USA  \\
\rm

\bigskip
\bigskip

  \end{center}

 \bigskip\bigskip
  \begin{abstract}

  We study the imprint
of new particles on the primordial cosmological fluctuations.
New particles with masses comparable to the Hubble scale produce a
distinctive signature on the non-gaussianities.
This feature arises in the squeezed limit of the correlation functions of primordial fluctuations.
It consists of particular power law, or oscillatory, behavior that contains information about the masses
of new particles. There is an  angular dependence that gives information about the spin.
We  also have a relative phase that crucially depends on the
quantum mechanical nature of the fluctuations and can be viewed as arising from
the interference between two processes.
While some of these features were noted before in the context of specific inflationary scenarios, here we give a general description
emphasizing the role of symmetries in determining the final result.

 \medskip
  \noindent
  \end{abstract}
\bigskip \bigskip \bigskip

  \end{titlepage}



\section{Introduction}

Inflation seems to be the highest energy observable natural process.
 The Hubble scale, $H$,  during inflation could be as high as
$10^{14}$ GeV. This  is much larger than any energy  we can expect to achieve with particle
accelerators in the foreseeable future.  It is therefore interesting to understand  how to extract information about
the laws of physics at those scales.
We can view inflation as a particle accelerator. The physics during inflation is very weakly coupled, as compared to
usual particle physics. On the other hand going between the physics during inflation and astronomical observations
we can do today is
somewhat involved. It is   similar to the process of going between the
underlying high energy process and the  detector signals  in a hadron collider, except that in cosmology
the final state involves galaxies instead of hadrons.
In a hadron collider we look at jets or patterns of energy deposition on the detector. In cosmology we similarly look
for patterns in the distribution of galaxies or in the cosmic microwave background. These patterns appear to arise
from a single  fluctuation mode,  the curvature or adiabatic mode conventionally called $\zeta$.
We are going to take for granted the very delicate process of going from present day astrophysical observations and the primordial shape of  $\zeta$.
In order words, we are going to imagine
that astrophysicists tell us what $\zeta(\vec x) $ is at the end of inflation, before it reenters the horizon.
 What can we learn from this?. How do we extract the physics during inflation from this?

The discipline of collider physics involves going from the direct collider observables   to the underlying lagrangian
of the theory. One of the simplest questions one can ask is how to recognize the presence of new particles.
In colliders the answer to this question is simple, one collects groups of particles (pairs, for example) and one plots the
invariant mass. If  a bump is seen in this distribution, one says that there is a new particle. One can also look at the
angular distribution of the particles and read off the spin of the new particle.

In this paper we answer  a similar  question for the ``cosmological collider''.
 Namely, we ask: { \it  how do we recognize the
presence of new particles during inflation ? } The simplest version of this
 question is whether we had any new light or massless
 fields during inflation. These are models which have more than one field and can have rather striking signatures,
 such as isocurvature fluctuations or violations of the inflationary consistency condition in the squeezed limit
 of the three point function. There is an extensive literature on this case, see \cite{Wands:2007bd,Langlois:2012sz}
for reviews.
 We will concentrate on the case where we have particles whose mass is comparable to the Hubble scale.
 Such particles can be produced by quantum fluctuations during inflation and they can then decay to ordinary
 inflatons. This process leaves a statistical imprint on the spectrum of primordial fluctuations. More precisely, this
 process gives rise to non-gaussianities, see \cite{Chen:2010xka} for a review.
A particular model of this kind,  ``quasi-single-field inflation'',
 was studied in \cite{Chen:2009zp,Baumann:2011nk,Assassi:2012zq} and some of
the features we discuss below were described in \cite{Noumi:2012vr}.

  Of course, even self interactions of the inflaton, or interactions of the
 inflaton with gravity,  can give rise to non-gaussianities.
 We want to know:

 {\it Which specific non-gaussianities are a signature of new particles during inflation},

 \noindent
as opposed to signatures that arise due to inflaton self interactions.

 Now, if we think about very massive particles, $m\gg H$, then we can integrate them out and they produce new
 terms in the effective lagrangian for the light fields. Since we do not know the original lagrangian, it is clear that we
 are not going to discover them in this way. For masses of order $H$, $m \sim H$, the situation is different,
 because we can produce the particles giving rise to non-local effects which cannot be mimicked by changing the
 interaction lagrangian of the inflaton. The fact that this is an interesting question was addressed in
\cite{Chen:2009zp,CurvatonModel,Baumann:2011nk,Assassi:2012zq,Noumi:2012vr,Suyama:2010uj}.

    \begin{figure}[h]
\begin{center}
\includegraphics[scale=.5]{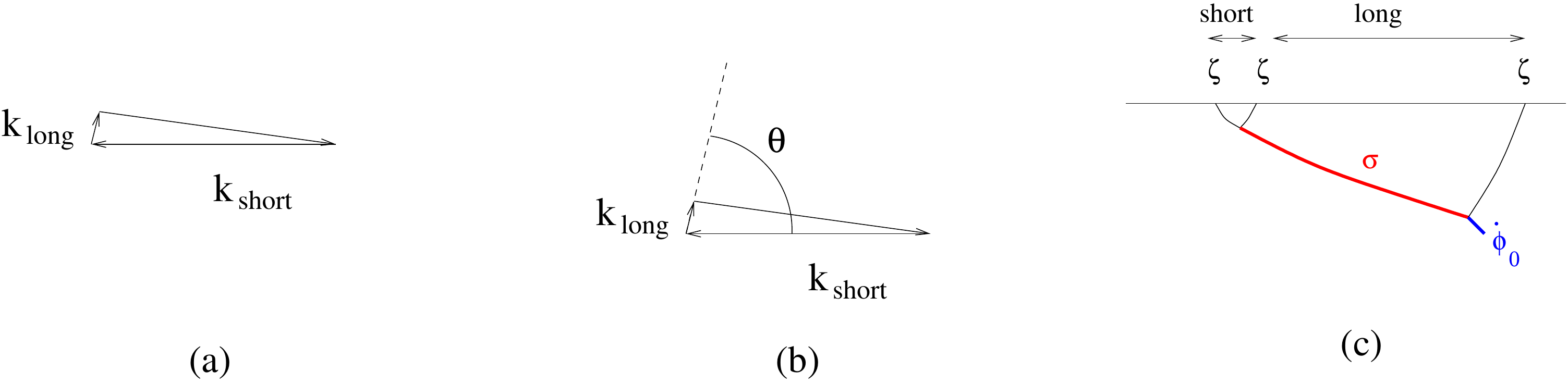}
\caption{ (a) The three momenta of the three point function form a closed triangle. In the squeezed limit,
 one
of the sides, $k_{\rm long}$, is much smaller than the other two, $k_{\rm short}$. (b) In this limit we can also define
the angle $\gamma$ between the short and long mode momentum vectors. (c) In position space we have
two insertions at a short distance from each other, associated to $k_{\rm short}$ and one at a longer distance, $k_{\rm long}$.
We are interested in considering the effects of massive fields, $\sigma$,
 that can decay into pairs of inflatons. In an inflationary background we can replace an inflaton fluctuation by
 the classical $\dot \phi_0$ background, so that we get a contribution to the three point function.     }
\label{SqueezedIntro}
\end{center}
\end{figure}

 The large masses, this effect is suppressed by $e^{-\pi m/H}$, which is why it isn't
  captured by the effective field theory which is an expansion in powers of
 $(H/m)$.

 The simplest non-gaussian observable is the three point function $\langle \zeta_{\vec k_1} \zeta_{\vec k_2}
 \zeta_{\vec k_3} \rangle$ in Fourier space. Translation invariance implies that the momenta form a closed triangle.
 The near scale invariance of the fluctuations implies that this is a function of the shape of the triangle. In particular
 we can form the ratio between the smallest side and the largest side $k_{\rm long}/k_{\rm short}$. See figure
 \ref{SqueezedIntro}(a).
 As a function of this ratio the three point function can display interesting power law behavior for small values of this ratio.
 This is called the ``squeezed'' limit. We find
 \be \la{3ptExin}
 { \langle \zeta \zeta \zeta \rangle \over \langle  \zeta \zeta  \rangle_{\rm short} \langle \zeta \zeta \rangle_{\rm long} }
 \sim \epsilon \sum_i w_i
 \left(  { k_{\rm long} \over k_{\rm short} } \right)^{\Delta_i }  ~,~~~~~~~~{\rm for} ~~~~
 { k_{\rm long} \over k_{\rm short}  } \ll 1
 \ee
 where $\Delta_i$ can be real or complex, $w_i$ are some
coefficients and $\epsilon$ is a slow roll parameter (see \cite{Chen:2009zp}). This form of the correlator is a
 consequence of the slightly broken conformal symmetry of the late time wavefunction of the inflationary universe.

  Such power law behavior is characteristic of conformal field theories, where  the $\Delta_i$ that occur are  the
  (anomalous) dimensions of operators in the theory. Here we have the same situation because the de-sitter isometries
  act on late time expectation values in the same way as the conformal group in one less dimension.
  In inflation,   these
  approximate symmetries govern the behavior in the squeezed limit.

     The actual values of
 $\Delta_i$ that occur in \nref{3ptExin} give us the desired information about the spectrum of new particles.
 Namely, a single new particle of mass $m > { 3 H \over 2}$ and spin $s$ gives rise to a pair of terms with  \cite{Noumi:2012vr}
 \bea \la{Delpmin}
 \Delta_\pm &=& { 3 \over 2} \pm i \mu ~,~~~~~~~~ \mu = \sqrt{ { m^2 \over H^2 } - {  9 \over  4 } }
 \\
 { \langle \zeta \zeta \zeta \rangle \over \langle  \zeta \zeta  \rangle_{\rm short} \langle \zeta \zeta \rangle_{\rm long} }
  &\sim & \epsilon e^{-\pi \mu} |c(\mu)| \left[ e^{ i \delta(\mu)}
 \left(  { k_{\rm long} \over k_{\rm short} } \right)^{{ 3 \over 2} + i \mu  }
 + e^{ - i \delta(\mu)}
 \left(  { k_{\rm long} \over k_{\rm short} } \right)^{{ 3 \over 2} - i \mu  } \right]  P_s(\cos \theta) ~~~~~~~~ \la{OscilFa}
 \eea

  A single particle gives rise to two dimensions, $\Delta_\pm$,
   which are complex conjugates of each other, which is good, so that
  we can get a real answer for the correlator\footnote{ The correlator is real for the following reason. $\zeta(\vec x)$ is real
  in position space. Therefore, in momentum space it obeys $\zeta_{\vec k}^* = \zeta_{-\vec k}$. If we complex conjugate
  a correlator, we get the same answer as reversing the sign of all momenta. For a two or three point function, we can
  relate the correlator with reversed mometa to the original one by a  two dimensional rotation, so that they should have
  the same value.}. An interesting fact is that there is an interesting calculable phase in the coefficients
  $w_\pm = |w_\pm|e^{ \pm i \delta} $ that appear in \nref{3ptExin}. This phase, $\delta(\mu)$,
   depends only on the mass of the
  particle. The oscillatory behavior as a function of the ratio of scales is a   quantum effect.
   It arises from the quantum  interference between two processes. One process is the ordinary gaussian quantum evolution
   of the inflaton. The other is the creation of a pair of massive particles that subsequently decay to inflaton modes.
   We know that we can assign a characteristic time during inflation to a given momentum $k$. It is the time when this
   mode crossed the horizon during inflation. The mode $k_{\rm long}$ crossed the horizon earlier in than the mode
   $k_{\rm short}$. The time measured in number of e-folds between these two events is
   $ N =  \log ( k_{\rm short}/k_{\rm long} )$. What we see in \nref{OscilFa}  is the behavior
   of the amplitude of the wavefunction of the massive particle as a function of time, or $N$. The wavefunction oscillates
   as we expect for a massive particle. In addition the $3/2$ in \nref{Delpmin} gives rise to the expected dilution
   factor due to the expansion of the universe. Namely the square of the wavefunction should go like
 $e^{- 3 N} \sim 1/$Volume.
   The fact that we have effects involving the amplitude of the wavefunction of the massive particle as opposed to its square
   is due to the fact that this is a quantum interference effect.  Another related fact is the dependence of the overall
   factor in \nref{OscilFa} on the mass. Keeping the coupling between the inflaton and the fields fixed, this factor
  goes as $e^{ - \pi \mu}$. This should be compared to the thermal factor going like $e^{ - 2 \pi \mu}$. This last factor
  is proportional to the {\it probability} for creating a pair of massive particles. The fact that we get $e^{ - \pi \mu}$ is
  again due to the fact that we are seeing an interference effect,
sensitive to the amplitude and containing
phase information.

   Needless to say, the observation of non-gaussianities with such oscillations and with the precisely predicted phase
   would be a striking signature of the quantum mechanical nature of cosmological fluctuations. We can view this as a kind
   of ``cosmological double slit experiment''.  The precise prospects for actually measuring this will not be evaluated
   precisely in this paper. However, let us make some comments.    For natural  values of the couplings
   this is harder to measure than the standard inflationary three point function \cite{Maldacena:2002vr}, which does
   not appear measurable in the near future. However, it is possible to have larger couplings, so that it would be
   interesting to look for such  templates in the data. Of course, detection of non-gaussianity is limited, in an irreducible
   fashion, by the cosmic variance noise (the fact that we observe only one universe). The squeezed region we are
   studying has a smaller signal to noise ratio than
 that of a more  equilateral triangle. Therefore it is likely that one
   would first find evidence for non-gaussianity and then, by a more sensitive measurement, find the detailed behavior
   in the squeezed limit. Our reason for emphasizing the squeezed limit is not that this is the easiest to measure, the
   reason is that this limit contains very direct information about the spectrum of particles during inflation.
   This is somewhat similar to the fact that at the LHC we concentrate on events with large transverse momentum. This is
   small subset of the events, but is the subset that can display the existence of new particles a direct way.
   Note that the
   standard non-gaussianity for the single field model \cite{Maldacena:2002vr} can be understood as arising
   from the effects of graviton exchange, as it was explicitly shown in \cite{Kundu:2014gxa}. This also shows that
   the effects we are after are expected to be smaller than the ones of the standard non-gaussianity, at least if the
   couplings are all suppresed by the Planck scale. In other words, we can consider couplings to the inflaton of the form
   $ \lambda \int (\nabla \phi)^2 \sigma$, where $\sigma$ is the massive field.
   We generically expect such couplings to be present with $\lambda \propto 1/M_{pl}$. But of course, the couplings
   could be larger. Special models with possibly larger values of non-gaussianity due to fields of masses of order
   Hubble were discussed in \cite{Chen:2009zp,Baumann:2011nk,Assassi:2012zq,Noumi:2012vr}
 under the name of ``quasi-single-field inflation''. In those papers the relevance
   of the squeezed limit for reading off the presence of new particles was emphasized.

   If the particle is light, with $m <  3 H/2$, we still get two powers as in \nref{Delpmin} \nref{OscilFa}
   except that now $\Delta_\pm$ are
   both real and $\Delta_-$, being smaller, dominates the squeezed limit. Here also the relative coefficient between
   the two terms is completely fixed by the value of the mass.

   It is also possible to study the spin of the particle. If the massive particle has spin, $s$,
 then we get a factor of
   a Legendre polynomial, $P_s(\cos \theta)$, in front of \nref{OscilFa}, where $\cos \gamma $ is the angle between
   the small momentum $\vec k_{\rm long}$ and the large one $\vec k_{\rm short}$.

\begin{figure}[h]
\begin{center}
\includegraphics[scale=.4]{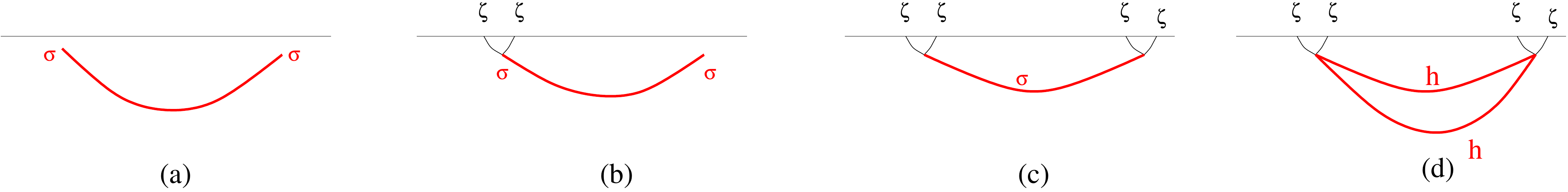}
\caption{ (a)  Two point function of massive fields (b) Three point functions of two massless fields and a massive field.
(c) Four point function of massless fields with an exchange of a massive field. We can view this diagram as the
amplitude to create a pair of massive fields which then decay into inflatons. (d) Contributions from Higgs field exchange, or
the exchange of any other gauge non-invariant field.    }
\label{TwoThreeFour}
\end{center}
\end{figure}

   So far we have discussed the effects of single particles which can directly couple to the inflaton via $\int (\nabla \phi)^2 \sigma$.
     There are situations where such a coupling is not allowed by a gauge symmetry. For example we can have a
     coupling of the form $ \lambda \int (\nabla \phi)^2 h^\dagger h $ where $h$ is the Higgs field. If the electroweak
     symmetry is unbroken, we can only couple to a pair of Higgs fields, not  to a single Higgs field. See figure
     \ref{TwoThreeFour}(d).
       This situation also leads to an expansion of
     the form \nref{3ptExin}, except that now the leading dimensions that appear reflect the presence of a pair of particles and
     have the form
     \be
      \Delta = 3 + 2 i \mu ~,~~~~~~~\Delta =3 - 2 i \mu ~,~~~~{\rm or} ~~~\Delta = 3
      \ee
      where $\mu$ is given in terms of the mass of the Higgs field by \nref{Delpmin}.
      This gives rise to more rapidly decaying terms in the squeezed limit, with a dilution factor of $3$ instead of $3/2$.
      In a similar way one can consider couplings to the Yang Mills fields or pairs of fermions.

      A prevailing attitude in the study of inflationary models is the idea that we have a single light field. However, we
      know that now we do not have a single light field. Where were all the fields we have now during inflation?. They could
      be very massive and unobservable, or some of them could be massless (the gauge fields or the fermions) and
      have observable consequences in the squeezed limit. In this case, the natural values of the coupling are smaller,
      $ { c \over M_{pl}^2} \int (\nabla \phi)^2 h^\dagger h $, but they could also be bigger!.

More generally, the powers appearing in \nref{3ptExin} seem to correspond to all the quasinormal modes
in the static patch of de-Sitter space. Therefore, {\it in principle}, the squeezed limit contains the full information
about the spectrum of the theory.  Of course, in practice we are limited by cosmic variance.

      This paper is organized as follows. We start with a discussion of the two point functions of massive fields.
      See figure \ref{TwoThreeFour}(a). We then discuss three point functions of two massless fields and a massive field,
      see figure \ref{TwoThreeFour}(b).
      We then discuss the effects of a massive field on a massless  field in de-Sitter, see figure \ref{TwoThreeFour}(c).
       We study
      the de-Sitter case first because the result will be largely determined by the de-Sitter symmetries. These
      de-Sitter symmetries act like conformal symmetries on the late time three dimensional slices of the universe.
      In de-Sitter, the main effect of the massive field appears in the four point function of the massless  field.
      In this case, the interesting limit is analogous to an operator product expansion were we consider the four point
      correlator in a kinematic configuration where, in position space, the four points form two pairs which are
      far from each other, see figure \ref{TwoThreeFour}(c). In this limit the answer will be related to products of three
      point functions.   It turns out that the formulas are simplest when we consider
      a conformally coupled field, $m^2 = 2 H^2$, instead of a massless field.
      Therefore we analyze this case first, and then the
      massless case is simply obtained from the conformally coupled case.   We also discuss general aspects of the
      analytic structure of correlators in momentum space. It is interesting to understand where the correlator is smooth
      and where it is singular. As a toy problem we also first briefly discuss equal time flat space correlators.

      We then go to the inflationary case, which is related to the case we  discussed above by replacing
      one of the external massless fields by the classical inflaton solution $\dot \phi_0(t)$, see figure \ref{SqueezedIntro}(c).

      We treat both the case of an intermediate massive scalar field and also a massive field with spin.
       The discussion
      of the spinning case is a bit more technical and the reader might want to skip
 it in a first
reading.
     There are two physical motivations for including fields with spin. There may be extra dimensions with size not far from the Hubble scale during inflation, in which case we can expect the inflation to couple to graviton Kaluza-Klein modes. It is
     also conceivable to have a very ``stringy'' model of
     inflation where the string scale is comparable to the Hubble scale, with a tiny string coupling. We do not have an explicit model of this type, but
     that could be due simply to a ``looking under the lamppost''' effect, namely an explicit construction of such a
     model lies somewhat outside the range of our technical tools (for some attempts in this direction see
\cite{Maloney:2002rr}). Of
     course,  if we were to detect the contribution of a spin 4 state in the non-gaussianity, it would be a strong indication
     of string theory during inflation, since we suspect  that a structure like that of string theory follows when we have weakly
     interacting particles with spin $s>2 $.

 \section{Generalities}

In the first part of this paper we consider quantum fields propagating on a fixed
de-Sitter space. In other words, we do not have dynamical gravity.
We write the metric   as
\be \lb{DSmetric}
ds^2 = { - d\eta^2 + d \vec x^{ \, 2} \over \eta^2 }
\ee
and we  set the Hubble scale to one, $H=1$. We will restore it later.

De Sitter has a large group of isometries. Some are manifest in \nref{DSmetric}. The two that are not so manifest correspond to
the Killing vectors
\be
{\rm D}: ~ -\eta \partial_\eta - x^i \partial_{x^i} ~,~~~~~~~~ \vec b .  \vec  K : ~~ - 2 (\vec b . \vec x ) \eta \partial_{\eta} -2 (\vec b . \vec x )
 \vec x .  \partial_{\vec x }  - (\eta^2 - |\vec x|^2 )  \vec b . \partial_{\vec x }
\ee
If we have a quantum field theory in de Sitter,  in a de Sitter invariant vacuum, then the correlation functions are invariant under the action of these
isometries.

We will be often interested in the late time limit of correlators where some points approach the asymptotic future at $\eta =0$ keeping $\vec x $ fixed.
In many circumstances one can argue that the late time dependence of the operator is a sum of  simple exponentials of proper time of the form
$ e^{ - \Delta t}$ or $\eta^\Delta$. This is related to the quasi normal mode behavior in the static patch.
In other words, inside a correlation function, a four dimensional  operator can be expanded as
\be
\phi(\eta , \vec x ) \sim \sum_{ \{ \Delta \} } \eta^\Delta O_\Delta(\vec x )
\ee
where we sum over various values of $\Delta$. Here ${\cal O}(\vec x)$ behave as three dimensional operators. If $\phi$ is an elementary free field we sum over two values of $\Delta$, as we will discuss later.

  Under these circumstances, we find that the
above Killing vectors lead to an action of the form
\be \lb{Kgen}
{\rm D}: ~ - \Delta  - x^i \partial_{x^i} ~,~~~~~~~~ \vec b . \vec K  : ~~  -2 \Delta (\vec b . \vec x )    -2 (\vec b . \vec x )
 \vec x .  \partial_{\vec x } +  |\vec x|^2    \vec b . \partial_{\vec x }
\ee
on the operators $O_\Delta(\vec x)$. Here
  we have  used that $\eta \to 0$. This is precisely the form of the generator of the conformal group in three dimensions for an operator of dimension
$\Delta$. In conclusion, if the late time expectation values of values of operators in de-Sitter contain simple exponentials of proper time, then the coefficients of
these exponentials behave as primary operators of a conformal field theory with dimension $\Delta$. This is a consequence of de Sitter dynamics and we
are not using any hypothetical duality here\footnote{Of course, these properties are necessary for a dS/CFT duality  \cite{Witten:2001kn,Strominger:2001pn} and were discussed previously in that context.
 In AdS we have a similar property,
except that we are computing a special transition amplitude where, for elementary fields,  the boundary conditions imply that one of the operators vanishes. }.
Here $O_\Delta(\vec x)$ are standard operators acting on the Hilbert space of the quantum field theory that lives  on de Sitter space.

\begin{figure}[h]
\begin{center}
\includegraphics[scale=.3]{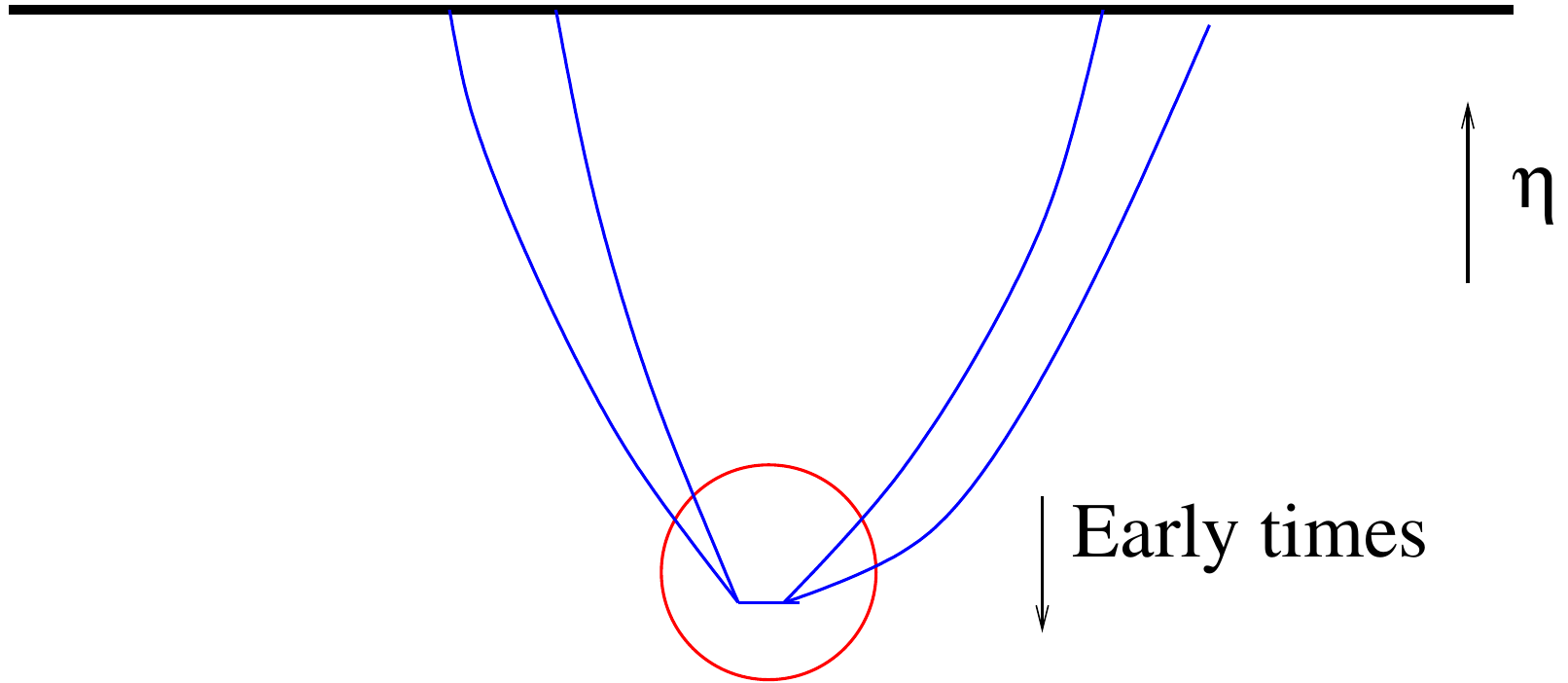}
\caption{  The singularity at $k_t\to 0$ arises when the whole diagram occurs at very early times. This translates into very short distances where we can make
the flat space and high energy approximation, obtaining \nref{FlatLi}.      }
\label{FlatSpaceLimit}
\end{center}
\end{figure}

The fact that de-Sitter gives a scale invariant spectrum of fluctuations is well known.
 It is also true that the spectrum is conformal invariant. The extra symmetries are
  the special conformal generators $K_j$ defined in \nref{Kgen}.
Of course the manifest translation symmetries $x^i \to x^i + \epsilon^i$ imply that it is  convenient to go to Fourier space. Then the
consequence of the symmetry is momentum conservation $ \sum \vec k_i =0$. Note that these are three dimensional vectors. Any correlator in fourier space has the form
\be
\langle O_{\vec k_1} \cdots O_{\vec k_n } \rangle = ( 2 \pi)^3 \delta^3 ( \sum_i \vec k_i ) \langle O_{\vec k_1} \cdots O_{\vec k_n } \rangle'
\ee
where the prime indicates that we have extracted the delta function prefactor associated to momentum conservation.
When we compute scattering amplitudes in flat space we have a similar delta function except that we also have an energy conservation delta function.
De Sitter correlators  have an interesting analytic structure which we will discuss in this paper. One feature that we would like to
point out here is that perturbative correlators have a singularity when $ k_t \to 0$, where $k_t =| \vec k_1| + |\vec k_2| + \cdots |\vec k_n|$. Of course this cannot happen
for non zero real values of the $\vec k_i$. However it can happen if we analytically continue the $\vec k_i$. The coefficient of this singularity is related to the
high energy limit of the flat space scattering amplitude
\be \la{FlatLi}
\langle O_{\vec k_1} \cdots O_{\vec k_n } \rangle' \sim { { \cal A}' \over k_t^{\rm power}  }  + c.c. ~,~~~~~~{\rm for}~~~k_t \to 0
\ee
Here ${\cal A}'$ is the high energy limit of the amplitude, with all four energy momentum conservation $\delta $ functions stripped off. The amplitude arises when we evaluate the all $+$
contribution of the in-in (or Keldysh) contour and the c.c. part comes from the all $-$ part of the contour.
The singularity at $k_t\sim 0$
 comes from a region where all interactions of the perturbative diagram are happening at early times, as in figure \ref{FlatSpaceLimit}.
(See also \cite{Raju:2012zr}).

When we perform a cosmological measurement we are limited by cosmic variance. This can be alleviated by doing the same measurement in different
patches and averaging over the patches.
 When we compute the correlators in Fourier space we are typically performing  this spatial average already. This is one reason that it is
convenient to think in terms of Fourier space. If we had the exact conformal symmetry, it is sometimes easier to go to Mellin space (see e.g. \cite{Fitzpatrick:2011ia}),
 and it would be interesting to
explore it in the de-Sitter context. Here we will stick with fourier space because   we will also use it  when we go away from the exact conformal symmetry in
the context of inflation.
It is convenient to know how the generators of dilatations and conformal transformations act in Fourier space.
Doing a simple Fourier transformation of the action of \nref{Kgen} we find
\bea
D: && ~ -( \Delta -3)  + \vec k  . \vec \partial_{k}  ~,~~~~~~~~  \la{Dexpr}
\\
\vec b . \vec  K : &&~~
- ( \Delta -3) 2 \vec b . \vec \partial_k -  [  ( \vec b . \vec k)( \vec \partial_k . \vec \partial_k ) - 2( \vec k . \vec \partial_k )(\vec b . \vec \partial_k )]    \la{SpecConfP}
\eea
These are the generators acting on each scalar operator\footnote{See \nref{SpecConfSpin} for the case with spin.}.
 The object that annihilates the amplitude $\langle O \cdots O \rangle'$ is the sum of each generator acting on each operator.
For the dilatation we have that   $ -3 + \sum D_i $ annihillates the stripped correlator, where
$D_i$ is given by \nref{Dexpr} with $\vec k\to \vec k_i$ and $\Delta \to \Delta_i$ where $i=1,\cdots,n$ runs
over the $n$ operators in the correlator.
 For special conformal transformations  $\sum_i  \vec b . \vec K_i$
annihilates the stripped correlator.

 \section{Two point functions in de Sitter}

The two point function for a scalar field in de-Sitter space depends only on the proper distance
between the two points. It can be obtained by imposing the equation of motion on the time
ordered expectation value
\be \lb{KGeqn}
[  \nabla_x^2 - m^2  ]  \langle \phi(x) \phi(x') \rangle = { i \over \sqrt{-g(x)} } \delta^4(x-x')
\ee
In addition we need to impose that the function is regular at the antipodal point.
This completely determines the solution
\bea
\langle  \phi(x) \phi(x') \rangle &= &  { \Gamma({ 3 \over 2 }  + i \mu) \Gamma( { 3 \over 2 } - i \mu) \over ( 4 \pi)^2  } ~_2F_1({ 3 \over 2 }  + i \mu ,
   { 3 \over 2 } - i \mu , 2 ; 1-{ 1 \over w}  )  \lb{TwoPointPro}
   \\
   {w }  &= &   4  \left[ {  \eta \eta'  \over - ( \eta - \eta')^2 + |\vec x - \vec x'|^2 + i \epsilon } \right]   \lb{Defw}
  \\
  \mu &\equiv&   \sqrt{ {m^2 \over H^2} - { 9 \over 4 } }
   \eea
   The antipodal point corresponds to $w=1$ and we see that the function is regular there. The coincident point singularity is
   at $w = \infty$ and the prefactor in \nref{TwoPointPro} ensures  that \nref{KGeqn} is obeyed.
   For heavy enough fields $\mu$  is  real. However, these formulas are also valid for purely imaginary $\mu$, which is the case for
   light fields.
    We will be often interested
   in the limit when we keep the comoving distance $|\vec x - \vec x'| $ fixed and we send $\eta , ~\eta' \to 0$.
   In this limit, It is easy to see from \nref{TwoPointPro} that  the correlator behaves as
   \be \lb{LongPosTwo}
    \langle  \phi(\eta , \vec x)  \phi( \eta', \vec 0)  \rangle \sim    { 1 \over4 \pi^{5/2}   }  \left[
      \left(  { \eta \eta' \over |\vec x |^2 } \right) ^{ { 3 \over 2} + i \mu }  { \Gamma( -   i \mu) \Gamma({3 \over 2} + i \mu ) }
      + \left({ \eta \eta' \over |\vec x |^2 }\right) ^{ { 3 \over 2} -i \mu }  { \Gamma(    i \mu) \Gamma({3 \over 2} - i \mu ) } \right] + \cdots
    \ee
    For real $\mu$ these are the leading terms. We see that we get  two simple exponentials of proper time $e^{ -(  \32  \mp i \mu ) t } $, with $\eta = - e^{ -t}$.
    As we discussed above, each coefficient of the exponential
    is expected to behave as correlators of  operators with the corresponding scaling dimensions. This is indeed the case in \nref{LongPosTwo}.
    Note that each field gives rise to two operators with dimensions
    \be \la{DelMu}
      \Delta = \32 + i \mu ~,~~~~~~~~{\rm and} ~~~~~  \tilde \Delta = 3-\Delta =
   \32 - i \mu
   \ee
   These operators simply correspond to the  annihilation and creation operators with respect to the out vacuum, respectively. The factor $\eta^{\Delta}
   \sim e^{ - ( \32 + i \mu ) t }$ is the expected time dependence multiplying the out vacuum annihilation operator.
   We can contrast \nref{LongPosTwo} to the corresponding long spatial distance expectation value in flat Minkowski space for a massive field. In that
   case the expectation value decreases exponentially as $e^{ - m d }$, while in de-Sitter we have the oscillatory behavior which is due to the fact that the
   expansion of the universe created pairs of particles.

   For imaginary $\mu$, $\Delta $ and $\tilde \Delta $ are  both real but one of these will give the dominant contribution at late times.
   For the special case of $\mu =0$, we see that both
   terms in \nref{LongPosTwo}  are similar, but  divergent. Taking the limit  carefully we find a behavior of the form
   $ w^{ \32 } +  w^{\32 } \log w $ (see \nref{Defw}).
    This can be understood as a logarithmic operator \cite{Gurarie:1993xq}. Namely, we have a pair of operators such that the
   dilatation operator has the Jordan form $ D = \left( \begin{array}{cc}  { 3 \over 2}   & 1 \\ 0  & { 3 \over 2 }  \end{array} \right) $.
   The dilatation operator is not unitary but  this is allowed, since the three dimensional theory is not
arising as the analytic continuation of a Lorentzian 2+1 dimensional field theory.
    Of course, for $\mu >0$, the two eigenvalues are complex.
    The $(\eta \eta')^{3\over 2} $ factor reflects the dilution of these massive particles due to the expansion of the universe.  If we square the wavefunction we will find that the probability of finding the particle goes as $1/a^3$, where
   $a \sim 1/\eta$ is the expansion factor of the universe.
    More generally, both the leading term, as well as the subleading terms in the expansions are related to
   the quasi normal modes for the field $\phi$ in the static patch, see e.g.  \cite{Du:2004jt,Jafferis:2013qia}.
   The imaginary part represents the oscillation of the wavefunction of a massive particle
   $\Psi \sim e^{ - i m t } $ and the real parts represent the dilution factor. The quasi normal modes are $\32 + i \mu + n$. The $n$ comes from the
   expansion of the wavefunction in power series around $r=0$ in the static patch, which is in agreement with the dilatation eigenvalues of derivatives of the
   primary operator $O_\Delta(x) $.

   Another important property of \nref{LongPosTwo} is the fact that for large $\mu$ it is suppressed as $e^{ - \pi \mu }$. If we square this we get
   $e^{ - 2 \pi \mu }$ which is the Boltzman factor of a massive particle. In addition, \nref{LongPosTwo} contains a relative phase between the two terms
   which varies with $\mu$ between $e^{ i \pi } $ for small $\mu$  and $e^{i  \pi /2}$ for large $\mu$.  This detailed structure is present for the standard
   Euclidean vacuum.
   It is also interesting to consider the correlator in the limit that the two points are separated by a large timelike distance, with $|\vec x - \vec x'|=0$ and
   $|\eta| \ll |\eta'|$. We can
   get into this regime by replacing $w \sim { \eta \eta' \over |x|^2} \to { \eta \over \eta'} e^{ - i \pi }$   in \nref{LongPosTwo}, see \nref{Defw}.
   This extra $e^{ i \pi}$ gives factors of
   $e^{ \pm \pi \mu }$, which imply that the first term in \nref{LongPosTwo} is now not suppressed exponentially for large $\mu$, while the second is suppressed
   by $e^{ - 2\pi \mu}$. The first term is the one expected   where we create a particle at $\eta' $ and destroy it at $\eta > \eta'$. In fact,  setting $\eta = - e^{ -t}$
   we see that the first term goes as $e^{ - i \mu (t-t')} $.


   It is also interesting to look at the two point function in fourier space. One way to obtain it is to Fourier transform \nref{TwoPointPro}.
   However, we rederive it by first expanding each Fourier mode of  the field in terms of creation and annihilation operators
   \bea \lb{FieldExp}
   \phi_{\vec k}(\eta ) &=& f_k(\eta) a^\dagger_{\vec k} + \bar f_{k}(\eta ) a_{ - \vec k}
   \\ \notag
   f_{k}(\eta) &=& (-\eta)^{ 3 \over 2 } h_{i \mu} ( k \eta )  ~,~~~~~~~ h_{i\mu} ( k \eta) = { \sqrt{\pi} \over 2 } e^{   \pi \mu/2 }  H^{(2)}_{i \mu} ( - k \eta ) = h_{- i \mu}(k \eta)
   \\ \notag
    \bar f_{k}(\eta) &=& (-\eta)^{ 3 \over 2 } \bar h_{  i \mu} ( k \eta )  ~,~~~~~~~
    \bar h_{  i\mu} ( k \eta) = { \sqrt{\pi} \over 2 }  e^{- \pi \mu/2 }  H^{(1)}_{i \mu} ( - k \eta )   = \bar h_{- i \mu}
    \\ \notag
    &&
    \bar h_{- i \mu}( k \eta) = -
     h_{i\mu}( e^{ - i \pi }  k \eta   ) = ( h_{i \mu}( k \eta ) )^*
   \eea
   where we have also summarized some of the properties of the functions.
   These creation and annihilation operators are the ones appropriate for the early time vacuum, at $\eta \to -\infty$. They should not be confused with the
   out vacuum creation and annihilation operators we discussed earlier. The ones in \nref{FieldExp} are defined so that the mode functions $f_k$ have a simple
   oscillatory behavior at early times, $\eta \to -\infty$.
   These functions are classical solutions of the equations of motion for each fourier mode.
   Note that all these equations,  \nref{FieldExp} ,  are true   for both  real and imaginary $\mu$, here
$^*$ denotes complex conjugation.
   As usual,  the normalization is chosen so that we get the right commutation relations. These can be most
   easily fixed by looking at the early time limit, $\eta \to -\infty$.
   The vacuum selection condition is given by choosing the solution of the Bessel equation with the right oscillatory behavior for early times, or very negative
   $\eta$. From the small argument expansion of these functions we can compute the small $\eta$, $\eta'$ behavior of the fourier transform of the correlator
   \bea
   \langle \phi_{\vec k} (\eta)\phi_{\vec k'} (\eta')\rangle & = & ( 2 \pi)^3 \delta^3(\vec k + \vec k') \left[ \bar f_{k}(\eta) f_{k}(\eta') \theta(\eta-\eta') +
     f_{k}(\eta) \bar f_{k}(\eta') \theta(\eta'-\eta) \right] ~~~~ \lb{ProFour}
   \\
  \langle \phi_{\vec k} (\eta)\phi_{-\vec k} (\eta')\rangle'
  &= &
   { (\eta \eta')^{3\over 2} \over 4 \pi}   \left[  \Gamma(- i \mu)^2    \left( {  k^2 \eta \eta' \over 4 } \right)^{   i \mu } +
    \Gamma( i \mu)^2  \left( {  k^2 \eta \eta' \over 4 } \right)^{   - i \mu }
\right] + {\rm local }~~~   \lb{LongMomTwo}
   \eea
   Recall that the prime in $\langle \cdots \rangle'$
    means that we drop the factor of $(2 \pi)^3 \delta^3(\sum_i \vec k_i )$. By  ``local'' we mean terms that are analytic in $k$ which do not
   give rise to long distance correlations in position space.
   Notice that the correlator \nref{ProFour} depends on the sign of $\eta - \eta'$. On the other hand, the long distance behavior \nref{LongMomTwo}, \nref{LongPosTwo}
   does not depend on this sign. This is because operators commute outside the lightcone.
   The terms denoted by ``local'' in \nref{LongMomTwo} do depend on the
   sign of $\eta - \eta'$.
     Of course we can directly go from \nref{LongPosTwo} to \nref{LongMomTwo} by using the formula
   \be \la{FTPower}
    \int d^3 x e^{ i \vec k . \vec x } |x|^{ - 2 a } =   8 \pi^{3/2} 2^{ - 2 a } k^{ 2 a -3}   { \Gamma(\32 -a) \over \Gamma(a) }
   \ee

   In order to perform perturbative evaluations of expectation values it is useful to use the in-in or Keldysh formalism. In this formalism one performs a time
    ordered
   integral from the initial time to the time of interest and then one goes back to the initial time in an anti-time ordered fashion. Each of the paths corresponds to
   the perturbative evaluation of the bra and the ket respectively.
   Perturbativative computations reduce to the standard prescription but with the path ordering along this
   special contour. For this purpose we need to introduce the correlators with points along different parts of the contour
   \bea
   \langle \phi_k (\eta) \phi_k(\eta')   \rangle'_{++} &=&  \bar f_{k}(\eta) f_{k}(\eta') \theta(\eta-\eta') +
     f_{k}(\eta) \bar f_{k}(\eta') \theta(\eta'-\eta)
     \cr
     \langle \phi_k (\eta) \phi_k(\eta')   \rangle'_{+-} &=&
     f_{k}(\eta) \bar f_{k}(\eta')
     \cr
       \langle \phi_k (\eta) \phi_k(\eta')   \rangle'_{-+} &=&   \bar f_{k}(\eta) f_{k}(\eta')
       \cr
     \langle \phi_k (\eta) \phi_k(\eta')   \rangle'_{--} &=&  \bar f_{k}(\eta) f_{k}(\eta') \theta(\eta'-\eta) +
     f_{k}(\eta) \bar f_{k}(\eta') \theta(\eta-\eta') \la{KeldyshC}
   \eea
   Here the $+$ corresponds to the time ordered part of the contour and the $-$ to the anti-time ordered part.

   \subsection{Two point functions of fields with spin}

\begin{figure}[h]
\begin{center}
\includegraphics[scale=.6]{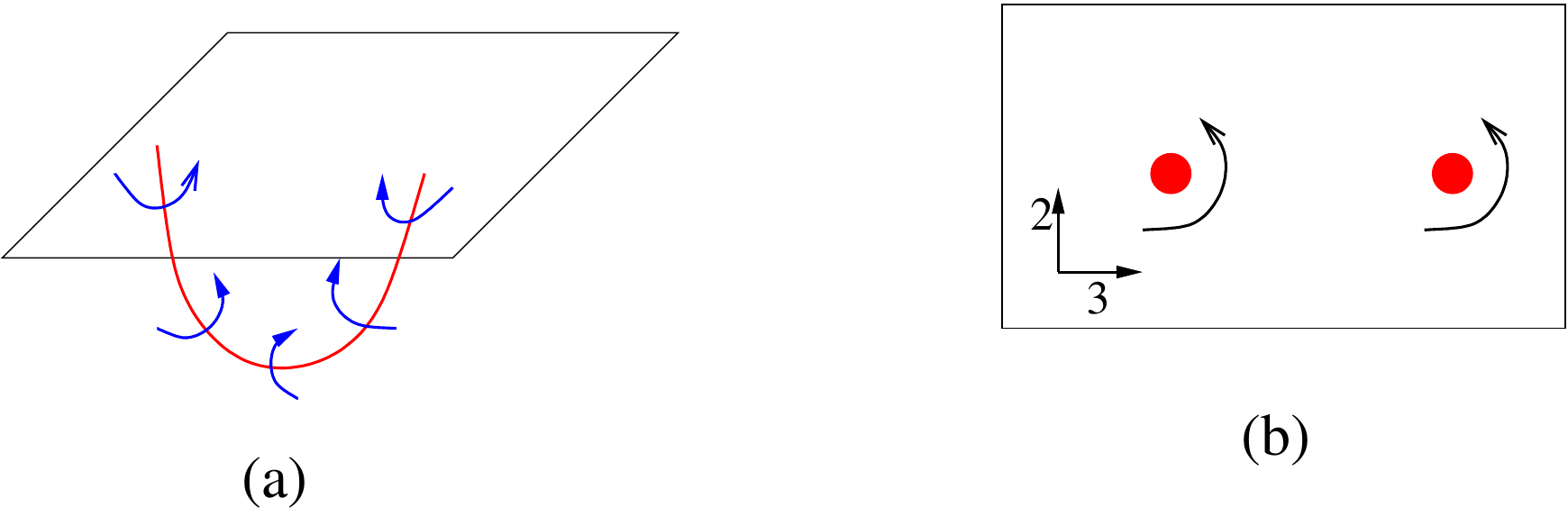}
\caption{ (a) We see a pair of created particles that approach a given comoving points at late time in de-Sitter. There is a subgroup
of the conformal group that preserves this two points. These include transformations which act as rotations around each of the
two points as indicated.
(b) If two points are
 separated along the direction $3$, then the spins of the physical particles  in directions, say $J_{23}$ are the same.
  So the two particles would be spinning as shown. The spins in
 the direction $J_{12}$ are opposite.  }
\label{SpinCorrelation}
\end{center}
\end{figure}

   For the case with spin we will not write down the full propagator in de Sitter. Instead we will focus only on the late time behavior which is governed by
   conformal symmetry. For an operator with spin it is convenient to contract its indices with a three dimensional null vector $\epsilon $, $\epsilon^2 =0$.
   This takes care of the fact that we want to look at traceless tensors. It is then simple to write the general two point function, in position space, for a
   general spin $s$ \cite{Polyakov:1974gs}
   \be  \lb{TwoSpinPos}
   \langle \epsilon^s . O(\vec x)  \tilde \epsilon^s . O(\vec 0) \rangle = { \left[  (\vec \epsilon . \vec { \tilde   \epsilon} )  - 2
    ( \vec \epsilon . \hat  x) ( \vec { \tilde \epsilon } . \hat x ) \right]^s
   \over |x|^{ 2 \Delta   } } ~,~~~~~~{\rm with} ~~ \hat x \equiv  { \vec x \over |\vec x| }
   \ee
   where $\epsilon^s . O \equiv \epsilon^{i_1} \cdots \epsilon^{i_s} O_{i_1 \cdots i_s} $.   Let us make a few remarks about the funny looking numerator in this
   expression. This contains the information about the spin correlation between two massive particles. There is a subgroup of the conformal group that preserves
   the two points where the massive particles are sitting. This subgroup includes an SO(3) subgroup. This is easy to see if we view the boundary as $S^3$ and we put
   the two points at the north and south pole. But it is still an $SO(3)$ subroup for any choice of two points. These $SO(3)$ elements are not ordinary rotations in space but combinations of various elements of the conformal group. Around each of the two points these rotations act
   like the ordinary rotation group.
    However, the rotation on one of the points is conjugated by reflection along the $\hat x$ direction relative to the other point.
   This is related to the term with a factor of 2 in the numerator of \nref{TwoSpinPos}.  Let us say that $\hat x $ lies along the $\hat 3$ direction.
   This implies that if one particle has a given eigenvalue  under the generator $J_{13}$, then the other particle has the {\it same} eigenvalue under $J_{13}$ where
   $J_{ij}$ are the usual rotation generators. If one particle has some eigenvalue under $J_{12}$ then the other particle has the {\it opposite} eigenvalue.
   See figure \ref{SpinCorrelation}.

  We now transform \nref{TwoSpinPos}   into Fourier space.
  We then have a preferred vector which is $\vec k$ and we can decompose the operator in terms of the spin around the
   $\vec k $ axis.
   More precisely, we can say that $\vec k$ lies along the $\hat 3 $ axis, $\vec k = ( 0,0,k)$ and choose the polarization vectors as
   \be \la{EpEptil}
   \vec \epsilon = ( \cos \psi , \sin \psi , i) ~,~~~~~  \vec { \tilde \epsilon } = ( \cos \psi' ,  \sin \psi' , - i )
   \ee
   Note that we only care about $\epsilon$ up to an overall scale, therefore we can generically  take it to be of this form.
   Then we find that the two point function in Fourier space is proportional to (see appendix \ref{TwoPtApp})
   \bea
   \langle  \epsilon^s . O_{\vec k}      \tilde \epsilon^s . O_{-\vec k} \rangle'  &\propto &k^{ 2 \Delta -3} I_2(\vec \epsilon, \vec {\tilde \epsilon} , \hat k )
   ~,~~~~~~~~~{\rm with} ~~~ \hat k \equiv { \vec k \over |\vec k | }
   \cr
    I_2(\vec \epsilon, \vec {\tilde \epsilon} , \hat k )&\equiv &  \sum_{m=-s}^s
    e^{ im(\psi -\psi')}  \left( {    (2 s)!\over (s-m)! (s+m)!}  \right)  I_2(\Delta,m) \la{ItwoEpDef}
    \\
 I_2(\Delta,m)  &\equiv &
      { \Gamma(\Delta   - 1  +m ) \over \Gamma(2-\Delta +m)  } { \Gamma(2 -\Delta    +s) \over \Gamma(\Delta  -1+s)}
    \la{Itwodef}
   \\
I_2(\vec \epsilon, \vec {\tilde \epsilon} , \hat k )&=&       e^{ - i s (\psi-\psi')} ~_2F_1( \Delta -s-1 , - 2 s , 2-s -\Delta ;- e^{ i  (\psi-\psi')}  )  \lb{HelAmpl}
\eea
   Here $m$ indicates the  angular momentum of the mode around the $\vec k$ direction.
   We sometimes call this the ``helicity'' of the mode.
   The factor in parenthesis in \nref{ItwoEpDef} is an uninteresting normalization factor,
   see appendix \nref{TwoPtApp}. The $I_2(\Delta,m)$ factor is a phase for $\Delta = \32 + i \mu$, with $\mu$ real.
    Note that the hypergeometric function in \nref{HelAmpl} is  a polynomial.

   This formula contains an interesting lesson. First, let us recall the formula that gives the dimension in terms of the mass for a spin $s$ field \cite{Garidi:2003ys}
   \be
   \Delta_\pm = \32 \pm \sqrt{  \left( s- \half \right)^2 - { m^2 \over H^2} }  = \32 \pm i \mu    \lb{MassFor}
   \ee
   where the last equality defines $\mu$.
  When  $m=0$ we have a gauge symmetry in the bulk. In this case, $\Delta_+ = 1 + s$ which is the dimension of a conserved
   current. Let us now consider the massive case, but with a small enough mass so that
    $\Delta_\pm  $ are real. The leading late time behavior of the expectation value of the field is given in terms of the component with dimension
    $\Delta_-$,  which is associated to the
     more slowly decaying function.
     An interesting feature of the coefficients in \nref{HelAmpl} is that they can change sign.  First, let us see this in a concrete example.  Consider the $s=2$ case
     \be
   \langle \epsilon^2 . O_{\vec k} \tilde \epsilon^2 . O_{-\vec k} \rangle' \propto   k^{ 2 \Delta -3}  \left[
     e^{ - 2 i \chi} + { 4 (3-\Delta) \over \Delta} e^{ - i \chi } + { 6 (3-\Delta)(2-\Delta ) \over (\Delta -1) \Delta }  +  { 4 (3-\Delta) \over \Delta} e^{   i \chi }  +
     e^{ 2 i \chi }  \right]  \lb{SpinTwoEx}
\ee
with $\chi = \psi - \psi'$. Note that for any integer spin $I_2$ in \nref{Itwodef}  is a ratio of simple polynomials of $\Delta$.  Let us start with $\Delta = { 3 \over 2}$
where all terms are positive.
    As $\Delta$ decreases we see that the middle term changes sign at $\Delta =1$. The fact that it diverges at $\Delta=1$ is related to the
   fact that  the kinetic term for this mode becomes zero. This is the phenomenon of partial masslessness discussed in \cite{Deser:1983mm,Deser:2001us}.
   For $\Delta_- < 1$ we have a negative sign. This negative sign is a problem because the term corresponds to the expectation value of a field and its complex
   conjugate\footnote{Now that for $\psi'=\psi$, $\vec {\tilde \epsilon }= \vec \epsilon^*$ in \nref{EpEptil}.},
   which should be positive. Note that here we are using that for real $\Delta$ we can focus on the $\Delta_-$ contribution because it is the largest at late
   times, forgetting about the contribution from $\Delta_+$. We should emphasize that the relative sign between the various terms is determined by the
   special conformal symmetry (de Sitter isometries) and cannot be changed.
   In the language of the wavefunction of the universe, we can say that the wavefunction is not normalizable.
In other words, as positive sign in the two point function indicates that the mode has a standard gaussian
wavefunction $\Psi \sim e^{ - \phi^2}$ while a negative sign implies $\Psi \sim e^{ + \phi^2} $.
    Note that for the particular case of a massless spin two, we get $m=0$ and $\Delta_- =0$. In that case we see that all components with helicity less than two
    diverge. However, in this case these modes are pure gauge and this divergence is not an obvious problem. So the $m=0$ case is fine.
    The  conclusion is that very massive spin two fields in dS make sense. However,   masses within the window
    \be
    {\rm BAD: } ~~~~ 0 < { m^2 \over H^2} < 2 \lb{Forbidden}
     \ee
     are not allowed \cite{Higuchi:1986py}. It remains to be seen if
     one can make sense of the case ${m^2/H^2} = 2 $ (with $\Delta_- = 1$). If it made sense, it would be in the context
     of a theory with an extra gauge symmetry, since there is a null state. Recent explorations in this direction can be found
    in \cite{PartiallyMassless}.  In this hypothetical  theory,  the graviton mass would be related to the cosmological constant.

     There is one particular example where it is very clear that the range \nref{Forbidden}   should not be allowed. Consider a Kaluza Klein
     reduction with a big internal space. If the internal space has a size which is larger than the Hubble radius, then we expect to have massive Kaluza Klein gravitons
     which  lie in the range \nref{Forbidden}. However, if the internal manifold is larger that the Hubble radius, we expect that also the internal  dimensions should be inflating.

     For general spins the forbidden range is \cite{Deser:2001us}
     \be \la{BADSpin}
     {\rm BAD:} ~~~   0 < { m^2 \over H^2} < s (s-1)
     \ee
      This can again be found by starting with the case that $\Delta_-$ is real and equal to 3/2.
        For this value all the coefficients in \nref{HelAmpl} have the same sign.  As the mass
     continues to decrease, $\Delta_-$ will also decrease, and when $\Delta_- $ crosses one, then one of the terms changes sign.
     Namely, the term with
     $m=0$ associated  to the longitudinal polarized field
      \be
      I_2(\Delta_- , m=0) = { \Gamma(\Delta_-   - 1    ) \over \Gamma(2-\Delta_-  )  } { \Gamma(2 -\Delta_-    +s) \over \Gamma(\Delta_-  -1+s)}
       \ee
       When $\Delta_-$ crosses one this becomes negative. So, the region $\Delta_-<1 $ is not allowed,
        which from \nref{MassFor} leads to \nref{BADSpin}.
      As expected,  for $s=1$ any mass is allowed.

One might be puzzled by the following question. For large mass we do not expect any problem. However,
   when $\Delta = \32 + i \mu $, with $\mu$ real,  the coefficients are complex. If we we had a problem with negative coefficients, why don't we have  a problem
   with complex coefficients?!. In the range where $\Delta_-$ was real, we could argue that  the coefficient of  the dominant power of $\eta$ at late times is a
   hermitian operator whose square should have positive expectation value. On the other hand, when we have the oscillating factors $\eta^{ \32 \pm i \mu}$
   we cannot argue that the prefactors of each power are  hermitian operators.  For that reason we cannot apply the same argument. The hermitian conjugate of
   the prefactor of $(-\eta)^{ \32 + i \mu}$ is the prefactor of $(-\eta)^{\32 - i \mu}$. But conformal symmetry does not allow a long distance two point
   function between  operators with different scaling dimensions. Such a two point function would be purely local and it is not governed by \nref{HelAmpl}.
   \nref{HelAmpl} governs the two point function of  two operators with the same dimension, namely the piece behaving like
   $ \eta^{ \32 + i \mu} \eta'^{\32 + i \mu }$ for example. The analysis in \cite{Deser:2001us} implies that there is no problem in this case, see \cite{WittenMassSpin}
   for a more algebraic discussion.

     Recently the Vasiliev version gravity \cite{Vasiliev:1999ba,Vasiliev:1988sa} in de-Sitter has been explored, see e.g. \cite{Anninos:2011ui}.
     In that case,
 the higher spin fields are exactly massless and correspond to
     generalized gauge fields. The existence of a forbidden range \nref{BADSpin} implies that  we cannot continuously break the higher spin symmetry so that we give a
     small mass for the higher spin gauge fields. This is possible in AdS and it corresponds to dual field theories that are weakly coupled. This is saying that
     the dual to inflation cannot be a weakly coupled large $N$ theory (as in \cite{McFadden:2010na}).
 If a dual exists, it should be sufficiently strongly coupled so that the anomalous dimensions
     for the higher spin currents are large enough to pull the bulk higher spin  fields outside the forbidden range \nref{BADSpin}.
\footnote{  In the dS/CFT example for Vasiliev theory \cite{Anninos:2011ui},
the boundary operators defined through the wavefunction of the universe can sometimes have anomalous
dimensions of order
$1/N$.
 However, even in that case the bulk higher spin fields are still exactly massless. The argument
leading to \nref{BADSpin} used expectation values, not the wavefunction.
Also the argument as stated, makes sense only to leading order
in $G_N$ expansion since we ignored the fluctuations of the background metric.
 At higher orders we need to be more careful about the proper gauge invariant
definition of the observable, etc. }

   Finally, note that in the limit $\Delta \to \infty$ we get  $I_2(\Delta, m) = (-1)^{s -m}$ and
   \be
   I_2(\vec \epsilon, \vec { \tilde \epsilon } , \hat k ) = (\vec \epsilon . \vec {\tilde \epsilon } - 2 \vec \epsilon . \hat k \vec {\tilde \epsilon } . \hat k )^s
   \ee
   which has  the same form as the position space numerator of \nref{TwoSpinPos}, with $\hat x \to \hat k $,
   and the discussion around figure  \ref{SpinCorrelation} applies.

  \section{Three point functions in de Sitter }

In position space, the general form of the correlation function of two scalar operators and a spin $s$ operator is given by \cite{Polyakov:1974gs}
\be
\langle O_1(\vec x_1) O_2(\vec x_2) \epsilon^s . O_3(\vec x_3 ) \rangle ={
\left( |x_{23}|^2 \vec \epsilon . \vec x_{13}   - |x_{13}|^2 \vec \epsilon . \vec x_{23}   \right)^s
 \over |x_{13}|^{\Delta_3 +s + \Delta_1 -\Delta_2 }
|x_{23}|^{\Delta_3 +s + \Delta_2 -\Delta_1 } |x_{12}|^{  \Delta_1 + \Delta_2-\Delta +s  }}    \lb{ThreePos}
\ee
Note that, when $s$ is odd, this function is odd under the exchange $ 1 \leftrightarrow 2$. Therefore, if the operators are identical only even spins can arise.

In the rest of this section we will simply explore these three point functions in Fourier space.
For our eventual application to cosmology we will be mostly interested in the case $\Delta_1 = \Delta_2 = 3$. However, we will also consider
the case $\Delta_1 = \Delta_2 = 2$ which corresponds to a conformally coupled field in the bulk, or
a scalar field with mass $m^2 = 2 H^2$.
We will consider this case because it is simpler than the
general case but still non-trivial.  Furthermore, the fourier transform for the case $\Delta =3$ will follow simply from the one with $\Delta =2$. There have been numerous previous
 discussions of conformal correlators in Fourier space, for a sample see
 \cite{Polyakov:1974gs,Kehagias:2012pd,Bzowski:2012ih,Anninos:2014lwa}

Given that \nref{ThreePos} is so simple one can wonder why we complicate our lives
to compute its Fourier transform. For the cosmological application it is customary to
express the correlators in Fourier space, specially since the inflationary universe is also breaking
the scale invariance.  Even from purely theoretical reasons, the work done computing the Fourier
transforms pays off when we need to consider conformal blocks, since they essentially  multiply in Fourier
space \cite{Polyakov:1974gs}.

\subsection{Two conformally coupled fields and a general scalar}
\la{TwoConfOneScalar}

Here we set $\Delta_1 = \Delta_2 =2$ and consider a scalar with general $\Delta_3 =\Delta$. In this case the correlator in Fourier space is just a function of
$k_i \equiv |\vec k_i |$ which are the lengths of the triangle formed by the three momentum vectors. It is a closed triangle due to momentum conservation.
Scaling implies that it can be a function of only two variables
\be \la{XYdef}
 \XM  = { k_1 - k_2 \over k_3 } ~,~~~~~~ \YM =  {k_1 + k_2 \over k_3}
 \ee
 with
 \be \lb{Exprsf}
 \langle \varphi(\vec k_1)  \varphi(\vec k_2)\sigma_{ \Delta   } ( \vec k_3) \rangle' = k_3^{\Delta -2} G(\XM , \YM )
 \ee

\begin{figure}[h]
\begin{center}
\includegraphics[scale=.5]{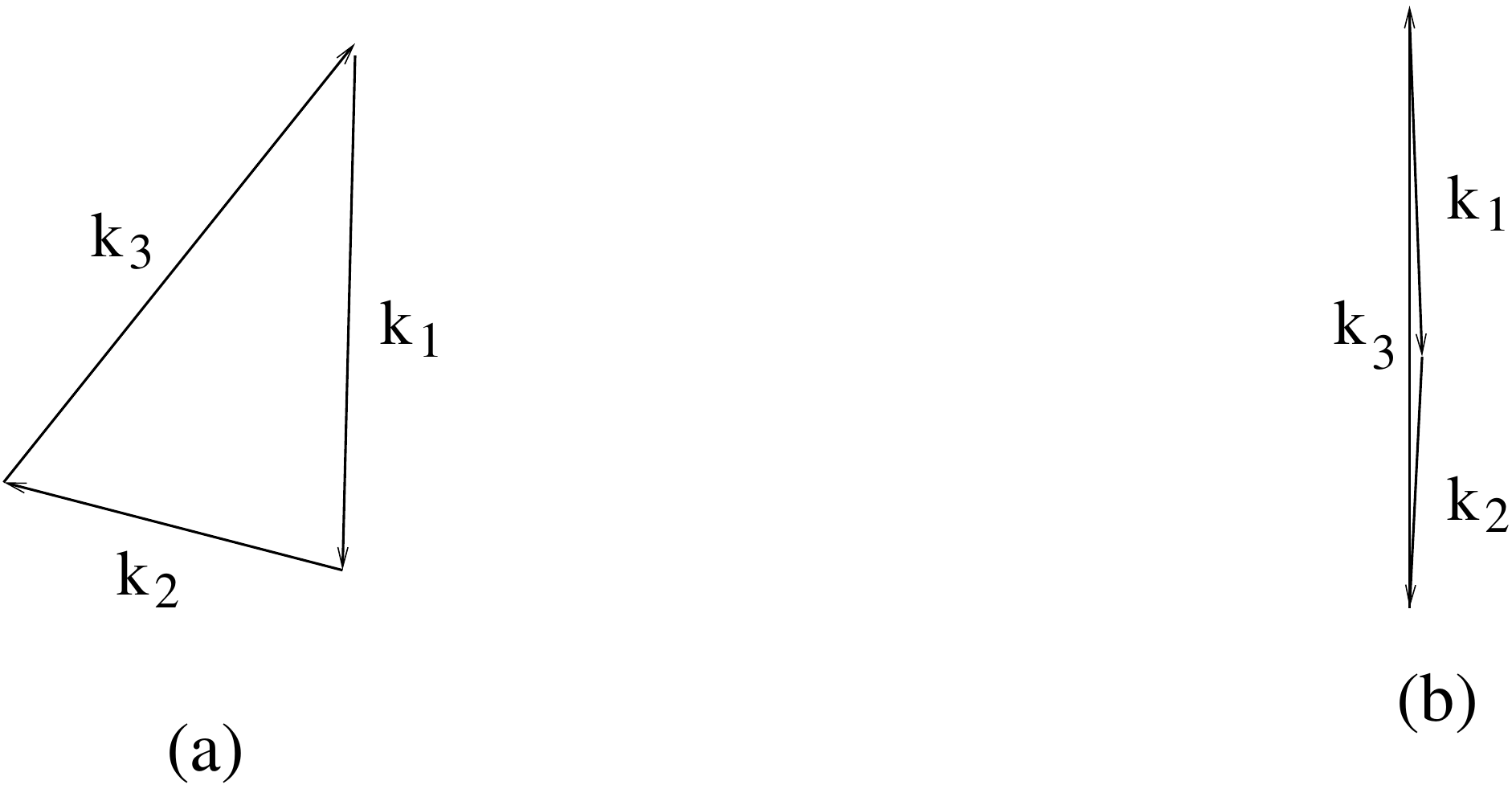}
\caption{  (a) Generic configuration of momenta for the three point function. (b) Nearly ``collapsed triangle'' configuration with $k_3 \sim k_1 + k_2 $.     }
\label{CollapsedTriangle}
\end{center}
\end{figure}

 Instead of doing the Fourier transform we will obtain the answer by imposing the conformal invariance conditions.
When the correlator depends only on the magnitude of the $\vec k_i$,  the special conformal generator \nref{SpecConfP}
simplifies and becomes
\be \lb{SpeConfSca}
 \vec b . \vec K = ( \vec b . \vec k) \left[ -2(\Delta -2) { 1 \over k} \partial_k + \partial^2_k \right]
 \ee
 where $k   \equiv |\vec k|$.
 The invariance condition becomes
 \be \lb{SpCfEq}
  ( \vec b . \vec K_1+ \vec b . \vec K_2 + \vec b . \vec K_3 ) k_3^{\Delta -2} G(\XM , \YM ) =0
  \ee
  where $\vec b . \vec K_i$ denotes \nref{SpeConfSca} with $\vec k \to \vec k_i$, which is the momentum of each of the operators.
    For each case we use \nref{SpeConfSca} with the
  corresponding scaling dimension. If we first set $\vec b$ so that it is orthogonal to $\vec k_3$, and we use the momentum conservation
  condition, we find
  \be \la{xyeqn}
    (\partial_{k_1}^2 - \partial_{k_2}^2 )  G(\XM , \YM )=0 ~~~\to ~~~~ \partial_\XM \partial_\YM  G(\XM , \YM ) =0
    \ee
   Here we have used that $\Delta_1 =\Delta_2 =2$. This implies that either $G$ is independent of $\XM$ or $\YM $.
   Since the equation is linear, we can first find  the solution independent of $\XM$. Once we find it, we can find the second solution by  the
   analytic continuation
    $k_2 \to - k_2$.
   Now, we assume $G$ is a function of only $\YM$, insert it in \nref{SpCfEq}, and find the equation
   \be  \lb{CoCouSc}
    ( \YM^2 -1) \partial^2_\YM G(\YM) + 2 \YM \partial_\YM G(\YM)   - [\Delta(\Delta -3) +2 ] G(\YM) = 0
    \ee
     This equation has two solutions. We are interested in the one that is regular at $\YM =1$.
     $\YM =1$ is a special kinematic configuration, it corresponds to a ``collapsed'' triangle,
      where $k_3 = k_1 + k_2$, see figure \ref{CollapsedTriangle}(b). Of course, for general triangles
     we have $k_3 < k_1 + k_2 $.  The absence of a singularity for the collapse triangle is related to
the adiabatic vacuum condition\footnote{It was shown explicitly in  \cite{Chen:2006nt,Holman:2007na,LopezNacir:2011kk,Flauger:2013hra,Aravind:2013lra} that departures from the adiabatic vacuum lead to
singularities for the collapsed triangle configuration in figure \ref{CollapsedTriangle}(b).}.
     Then the unique solution is proportional to
     \be \lb{HypThPt}
     G(\YM ) \propto    ~_2F_1(2-\Delta, \Delta -1, 1 ; { 1- \YM  \over 2 } ) =  ~_2F_1(\half - i \mu, \half + i \mu, 1 ; { 1-\YM  \over 2 } )
     \ee
     One interesting property of $G(y)$ is that it is invariant under $\mu \to -\mu$. This is not the case with the whole correlator due to the explicit factor of
     $k_3^{\Delta-2}$ in \nref{Exprsf}.
Of course, in momentum space, in order to convert a correlator with an operator of
dimension $\Delta $ to one with
     an operator of dimension   $\tilde \Delta = 3 - \Delta$ all
     we need to do is to multiply it by a suitable factor of $k$. More explicitly
     \be \la{Conversion}
     \langle O_1(\vec k_1) O_2(\vec k_2) O_{3 \, \tilde \Delta}( \vec k_3) \rangle' =
      \langle O_1(\vec k_1) O_2(\vec k_2) O_{3 \, \Delta}( \vec k_3) \rangle' k_3^{ 3 - 2 \Delta}
     ~,~~~~~~\tilde \Delta = 3 - \Delta
     \ee

     We  now discuss the same result but from the bulk point of view. By a slight abuse in notation let us also denote by $\varphi$ and $\sigma$ the bulk
     fields. Let us further postulate an interaction term in the Lagrangian of the form $\int  \lambda  \varphi^2 \sigma $.
     The conformally coupled scalar has a de-Sitter propagator which is simply obtained from the flat space one
     \be \lb{ConfProp}
     \langle \varphi_{\vec k}(\eta) \varphi_{-\vec k}(\eta') \rangle' =  \eta \eta'  { e^{ -i k |\eta -\eta'| } \over 2 k }
     \ee
     Of course, this can also be obtained from  \nref{ProFour} setting $\mu = i/2$.

    We   now compute $\langle \varphi_{\vec k_1}(\eta_0) \varphi_{\vec k_2}(\eta_0) \sigma_{\vec k_3}(\eta_0 ) \rangle'$ for a late time $\eta_0$.
    This computation can be done using the
    in-in formalism and it involves two integrals over the interaction lagrangian
    \bea \lb{ThPtPlusIn}
    \langle 3pt \rangle' |_+ &=&
    { \eta_0^2  \over4  k_1 k_2 } \left[ \bar f_{k_3}(\eta_0) I_+ + f_{k_3}(\eta_0) I_- \right]
\\
I_+ &=& i  \lambda \int_{-\infty}^{\eta_0} { d \eta \over \eta^2 } e^{ i (k_1 + k_2) \eta} (-\eta)^{3/2} h_{i \mu}(k_3 \eta)
\la{Int3G}
    \\
& \propto &      i  \lambda  { \eta_0^2  \over k_1 k_2 } \left[ { 1 \over \sqrt{k_3} } \bar f_{k_3}(\eta_0)  G(\YM) \right]
\la{OneToTwoc}
    \eea
    We see that the integral depends only on $k_1 + k_2$ due to the simple form of the propagators for the conformally coupled scalars
    \nref{ConfProp}.
    The prefactor that depends on $f_{k_3}(\eta_0)$ contains both powers of $\eta_0$ and gives rise to the two operators with dimensions $\Delta_\pm =
    \32 \pm i \mu$
    \be  \la{modefunex}
    { 1\over \sqrt{k_3}}  \bar f_{k_3}(\eta_0) \propto  A (-\eta_0)^{ \32 + i \mu } k_3^{-\half +  i \mu } + B (-\eta_0)^{\32 - i \mu} k_3^{- \half -i \mu}
    \ee
    where $A$ and $B$ are some constants.
     In addition,  in \nref{modefunex} we see the powers of   $k_3^{\Delta_\pm -2}$ appearing
in \nref{Exprsf}.
 The factor $\eta_0^2 $ in \nref{OneToTwoc} comes from two factors of $\eta_0$ one for each dimension one operator $\varphi$. The factors of ${ 1 \over k_1 k_2}$
 in \nref{OneToTwoc}  are
 converting the three point function involving dimension two operators in square brackets in \nref{OneToTwoc}, to one appropriate to dimension one operators.
 Notice that $1 = 3-2$, so that we can use \nref{Conversion}. In \nref{ThPtPlusIn} we have kept the leading $\eta_0$ dependence. A more complete
 expression can be obtained by keeping subleading terms from the propagator \nref{ConfProp}, which lead  to an  additional factor of $e^{ - i(k_1 + k_2)\eta_0}$.
 Expanding this exponential gives additional expected contributions from the dimension two part of $\varphi$.
    The integral \nref{Int3G} can be evaluated as the hypergeometric function in
   \nref{HypThPt}. It is interesting to note   the early time behavior of $h_{i\mu}(k_3 \eta) \propto(-\eta)^{-\half}   e^{ i k_3 \eta }$. Then we see that the whole integral
   behaves as $ \int { d \eta \over \eta} e^{ i ( k_1 + k_2 + k_3) \eta } $ for early times.
    As usual we make the integral convergent by rotating the contour in to the suitable part of the
   complex $\eta$-plane,
    $ \eta \sim  - e^{ - i \epsilon}$(positive). An important point is that nothing special happens when $k_3 = k_1 + k_2$, therefore the
   integral is regular there. This is related to the adiabatic vacuum condition:  there are no particles at short distances. This correlator would be singular at
   this point in   an $\alpha$-vacuum  \cite{Mottola:1984ar,Allen:1985ux} for the field $\sigma$, since we would have both of $e^{ \pm i k_3 \eta}$ in the propagator.
   This also explains why we do not need to consider   solutions of
    the conformal invariance equation \nref{xyeqn} that depend  only on $\XM $. They have
     singularities at $\XM =1$ and or $\XM =-1$.    These correspond to collapsed triangles with $k_1 - k_2 \mp k_3 =0$, where
    the integral representation is perfectly regular.
     In any case, we also see that if we analytically continue $k_3$ to negative values, then we can indeed find a
   singularity when $k_t \equiv k_1 + k_2 + k_3 \to 0 $. This comes from the large $|\eta| $ region of the integral and it represents the on shell three point
    scattering amplitude  at high energies. The exponential has some $\eta$ dependent prefactors that imply that the actual singularity in that region goes as
    $\log k_t $. Indeed, we can check that the singularity of $G(\YM )$ at $\YM =-1$ is logarithmic.

    The full expectation value involves also the minus contour which is the complex conjugate of \nref{Int3G}.

    \subsection{Three point functions involving two massless scalars and a massive scalar }

    For our inflationary application we will be particularly interested in considering massless scalars, as an approximation to
    the inflaton. Let us denote the massless scalar by $\xi$.
    It is natural to consider a coupling of the form
    \be
    \lb{BIder}
    \int  \sigma (\nabla \xi)^2
     \ee
     in the Lagrangian. This is consistent with the shift symmetry, $\xi \to \xi +$constant, which can be an approximate symmetry during inflation.

    The mode functions \nref{FieldExp} for the massless scalar take a simple form
    \be \la{OneWf}
     f_{k}(\eta) = { 1 \over \sqrt{ 2 k^3} } ( 1 - i k \eta ) e^{ i k \eta } ~,~~~~~~~~ \bar f_k(\eta) = { 1 \over \sqrt{ 2 k^3} } ( 1 + i k \eta ) e^{ - i k \eta }
     \ee
     Using these mode functions,  the integrals producing  the three point function are
     \bea
       \int (\nabla \xi)^2 \sigma & = & \int { d \eta \over \eta^2 } [ - ( \partial_\eta \xi)^2 + ( \partial \xi)^2 ] \sigma  \lb{DerCoup}
 \\
&  =&  \int { d\eta \over \eta^2 }  [ - k_1^2 k_2^2 \eta^2 - \vec k_1 . \vec k_2 ( 1 - i k_1 \eta) ( 1 - i k_2 \eta) ] e^{ i (k_1 + k_2) \eta} \sigma(\eta , k_3)
 \cr
 &  = &O_{12} \int { d\eta \over \eta^2 } e^{ i k_{12} \eta } \sigma(\eta , k_3)    ~,~~~~~~~~~~~~
 \cr
 k_{12} & \equiv  & k_1 + k_2  \la{Defk12}
 \\
\cr
O_{12} &\equiv &  - \half k_1 k_2 (k_3^2 -k_{12}^2 ) \partial_{k_{12}}^2 +\half ( { k_1^2 + k_2^2 - k_3^2 } ) ( 1 - k_{12} \partial_{k_{12} } )
\la{OpZetak}
\\
& =&    { k_3^{2 }  \over 8}   \left[ ( \XM^2 -\YM^2) (1-\YM^2) \partial_\YM^2 + 2 ( \YM^2 +\XM^2 -2) ( 1 - \YM \partial_\YM) \right]
     \lb{OpZeta}
  \eea
where we used the momentum conservation condition.
  The main point to note is that the results for the massless scalar are obtained by
     applying a simple differential
  operator, $O_{12}$, to the result for a conformal scalar. It is also important that if we had the integral with the other mode functions, the complex conjugates of
  \nref{OneWf}, then we would have obtained exactly the same operator \nref{OpZeta}. This means that we can pull the operator $O_{12}$ out of the
 in-in  contour sum.
      Notice that now the result also depends on $\XM $ but in a simple quadratic fashion, the complicated
  dependence is on $\YM $ \footnote{When the dimensions are $\Delta_1 = \Delta_2 = 2 + n$ we get a polynomial of degree $n$ in
  $\XM^2$.}.  The $\XM$ dependence of the Fourier transform for general scaling dimensions is more complicated (see the appendix of \cite{Bzowski:2012ih} for an explicit expression).
   Fortunately we will not need it for the inflationary
  discussion.

    It is also possible to derive this result from the analysis of the conformal invariance condition on the three point function.
    See appendix \ref{ConfInvDel3}.

    The solution of the conformal invariance condition with the right physical conditions is unique, up to an overall constant. On the other hand
    we could have considered another type of bulk interaction such as
    \be
    \lb{BIcub}
     \int \xi^2 \sigma
    \ee
    instead of \nref{BIder}. In this case, an exercise similar to what we did in    \nref{DerCoup} would also give us the answer in terms of
    an operator acting on the result for the conformally coupled scalar. However, due to the presence of terms that go like
    like $1/\eta^4$ in the integrand, now the operator also involves integrating the result for the conformally coupled scalar.
    Nevertheless, using the equation \nref{CoCouSc} we can check that the result is also proportional to \nref{OpZeta}. This can be seen at the level
    of the lagrangian since \nref{BIcub} and \nref{BIder} are the same on shell.
     More manifestly,  we can start from \nref{BIder}, do some integrations by parts and use the equations of motion for the  massless and
      massive  scalars   to obtain
     \be \la{TwoVertexF}
     \int \sigma (\nabla \xi)^2 = - \int \nabla \sigma \xi \nabla \xi = - \half  \int \nabla \sigma  \nabla ( \xi^2) = \half \int (\nabla^2 \sigma) \xi^2
     = { m^2 \over 2  } \int \sigma \xi^2
     \ee
     where $m$ is the mass of $\sigma$.

     \subsection{ Two  scalars and a massive higher spin state  }

     We  now turn our attention to the case of two scalars with the same mass and a massive state with spin.
     The position space answer is very simple and was written in \nref{ThreePos}. We are interested in the fourier transform
     of \nref{ThreePos}.

     We will focus first on the case  when the momentum of the operator with spin is much smaller than the other two.
     One naive answer is to set the momentum to zero and write an expression in terms of the remaining momenta
       \be \la{Sqan}
      \left. \langle O_1(\vec k_1) O_2(\vec k_2) \epsilon^s .O_3(\vec k_3) \rangle'\right|_{\vec k_3 \to 0 }
        \propto  |k_1|^{ \Delta_1 + \Delta_2 + \Delta-6 }   ( \vec \epsilon . \hat k_1)^s
      \ee
      This would be correct if the Fourier transform was analytic at $k_3=0$. Indeed \nref{Sqan} captures the analytic part at $k_3=0$. However,
      the Fourier transform contains also non-analytic terms in $k_3$. These arise from the integration region where the operator with
      spin is much   away than the other two, $|x_{31}|, ~|x_{32}| \gg |x_{12}|$.
     Without any loss of generality we can set $\vec x_2=0$. Then we can approximate \nref{ThreePos} as
     \be
    \left.  \langle O_1(\vec x_1) O_2(\vec 0 ) \epsilon^s . O_3(\vec x_3 ) \rangle \right|_{|x_3|\gg |x_1| }   \sim
     { \left[ x_3^2 (\vec \epsilon . \vec x_1 ) - 2 (\vec \epsilon . \vec x_3) (\vec x_1 . \vec x_3)\right]^s \over |x_3|^{ 2 \Delta + 2 s} |x_1|^{\Delta_1 + \Delta_2 -\Delta+ s}
     }
     \ee
     We now perform first the fourier transform with respect of $\vec x_1$. This can be expressed as
     \bea
     &&\int d^3 x_1  e^{ i \vec k_1 . \vec x_1 }  ( \vec V. \vec x_1)^s { 1 \over |x_1|^{\Delta_1 + \Delta_2 -\Delta+ s} }
     \propto ( \vec V . \vec k_1 )^s |\vec k_1 |^{ \Delta_1 + \Delta_2 -\Delta - s - 3 }
     \\
     && \vec V \equiv x_3^2 \vec \epsilon - 2 (\vec \epsilon . \vec x_3) \vec x_3 ~,~~~~~~~~~{\rm where}~~~~\vec V . \vec V =0
     \eea
     An important property that we used here is that the vector $\vec V$ is null. This implies that all the $s$ factors of $\vec k_1$ in $(\vec V . \vec k_1)^s$
      are combining into a spin
     $s$ tensor.
     We can now perform the final integral
     \be
     |\vec k_1|^{ \Delta_1 + \Delta_2 -\Delta - s - 3 }   \int d^3 x_3 {
     ( \vec V . \vec k_1)^s e^{ i \vec k_3 . \vec x_3}  \over    |x_3|^{ 2 \Delta + 2 s}   }
     \ee
     This integral has precisely the same form as the integral we would need to do to perform the Fourier transform of the
      two point function \nref{TwoSpinPos}, except that $\vec {\tilde  \epsilon}$ needs to be replaced by $\vec  k_1  $.
      We can then say that the three point function is given by
      \be \la{Sqnonan}
      \left. \langle O_1(\vec k_1) O_2(\vec k_2) \epsilon^s . O_3(\vec k_3) \rangle'\right|_{\vec k_3 \to 0 }  \propto
       |k_1|^{ \Delta_1 + \Delta_2 -\Delta   - 3 }   |k_3|^{ 2 \Delta -3}  I_2( \epsilon , \hat  k_1 , \hat k_3 )
      \ee
      where $I_2(\vec \epsilon, \vec {\tilde \epsilon}, \hat k_3) $ is the fourier
      transform of the two point function
       \nref{Itwodef}\footnote{
We might worry that in \nref{Itwodef}
      $\vec {\bar \epsilon}$ was null, whereas here it is being replaced by the non-null vector $\hat k_1$.
      We can take care of that by projecting onto the spin $s$ component,  $ (\hat k_1)^s|_{{\rm spin} ~s} $.}.
      Recall that $\hat k_i = { \vec k_i \over |\vec k_i | } $.

       The full small $\vec k_3$ behavior is the sum of \nref{Sqan} and \nref{Sqnonan}
       \be \la{Sqannonan}
       \left. \langle O_1(\vec k_1) O_2(\vec k_2) \epsilon^s . O_3(\vec k_3) \rangle'\right|_{\vec k_3 \to 0 }  \propto   |k_1|^{ \Delta_1 + \Delta_2 -\Delta   - 3 }   |k_3|^{ 2 \Delta -3}  I_2( \epsilon , \hat  k_1 , \hat k_3 )  +  |k_1|^{ \Delta_1 + \Delta_2 + \Delta-6 }   ( \vec \epsilon . \hat k_1)^s
       \ee
       where we have not calculated the coefficient in front of each term.

      We will also be interested in some of these correlation functions
      functions away from the squeezed limit. Though the general form seems somewhat complicated, we can
      focus on the component with maximal angular momentum  around the $\vec k_3$ direction.  In other words, we can write the fourier transform of the
      three point function as
      \bea
      \langle O_1(\vec k_1) O_2(\vec k_2) \epsilon^s .O_3(\vec k_3) \rangle' &=&
       [ \vec \epsilon . (\vec k_1 - \vec k_2) ]^s |k_3|^{\Delta + \Delta_1 + \Delta_2 -s -3} a_s( \XM , \YM) + \cdots
      \eea
      where the dots represent terms with lower angular momentum.
      These neglected terms have the form $[\vec \epsilon .(\vec k_1-\vec k_2)  ]^{s - i} ( \vec \epsilon . \vec k_3)^i $ with $i>0$.
       $\XM,~\YM$ are defined as in \nref{XYdef}.

      When the two scalars have dimensions $\Delta_1 =\Delta_2 =2$, we find that $a_s$ is a function of only $\YM$ which obeys an  equation
      similar to \nref{CoCouSc}. See appendix \ref{ThreeMomSpin}. More precisely, defining $a_s(\XM,\YM) = G_s(\YM)$ we get
      \be \lb{CoCouSpin}
       ( \YM^2 -1) \partial^2_\YM G_s(\YM) + 2(1+s) \YM \partial_\YM G_s(\YM)   - [ \Delta(\Delta -3) +2 -s(s+1) ] G_s(\YM) = 0
      \ee
     This can be derived using the special conformal invariance condition.

     In the case of massless scalars,  when $\Delta_1 = \Delta_2 = 3$, one can also show (see appendix \nref{DelThreeSpAp}) that the maximal angular momentum
      component can be written as
     \bea  \la{MaxHel3}
        a_s(\XM,\YM) &= &U_s(\YM) + { 1 \over 4 } ( \YM^2 -\XM^2) G_s(\YM) ~,~~~~~~~~\partial_\YM U_s(\YM) = - \YM G_s(\YM)
    \cr
    U_s(\YM) &=& { - 2 s \YM^2 G_s + (\YM^2 -1) ( G_s - \YM \partial_\YM G_s )  \over \Delta(\Delta-3) - s (s-3) }
     \eea
     where $G_s(\YM)$ also obeys the equation \nref{CoCouSpin}.

\section{Four point functions}

\subsection{Flat space equal time four point functions}

In order to get some intuition for   four point functions, let us first  compute some  flat space
four point functions. We consider four operator insertions at time
$\eta=0$ and we fourier transform in
the spatial variables.
Let us consider first a massless $\tilde \lambda \varphi^4$ theory in flat space.
The four point function is
\be \lb{FourFlat}
\langle \varphi_{\vec k_1}  \cdots \varphi_{\vec k_4}  \rangle  \propto  (2 \pi)^3 \delta^3 ( \sum_{i=1}^4 \vec k_i )
 { 1  \over k_1 k_2 k_3 k_4 } {\tilde  \lambda \over k_t} ~,~~~~~~k_t = k_1 + k_2 + k_3 + k_4
\ee
 Notice that  we do not have
 an energy conservation delta function since we are doing an observation at a particular time.
 We see that we get a singularity when $k_t \to 0$.  We can access this singularity by analytically continuing
 the momenta.
 The coefficient of this singularity is proportional to
  the flat space four point scattering amplitude, ${\cal A}'\sim \tilde \lambda $.

 \begin{figure}[h]
\begin{center}
\includegraphics[scale=.7]{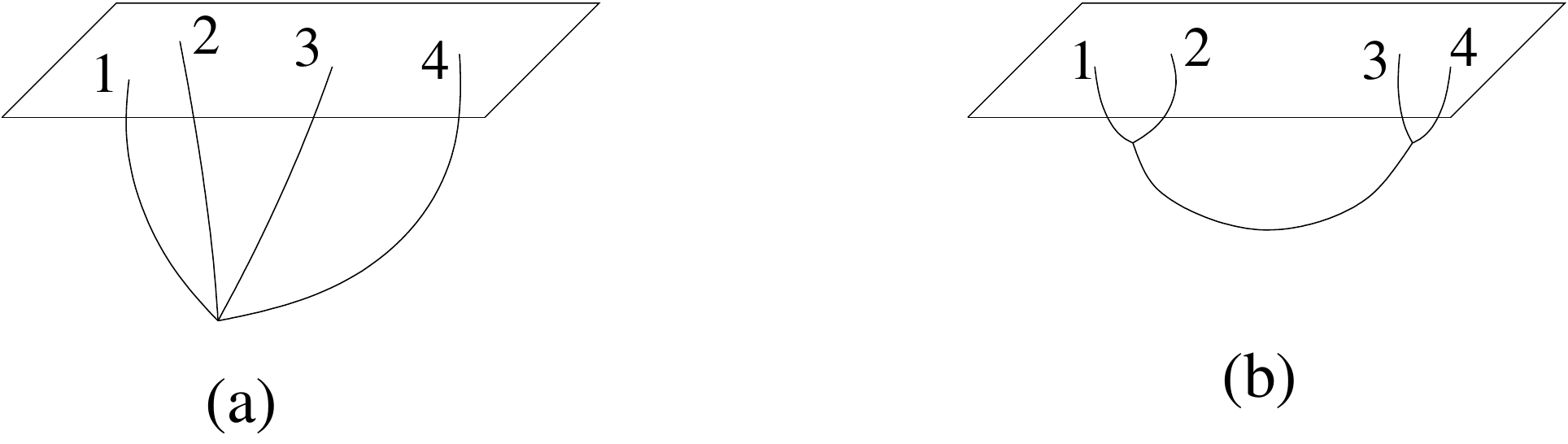}
\caption{ Diagrams for the computation of the four point function. (a) Diagram for the $\varphi^4$ theory. (b) One of the diagrams
for the $\varphi^3$ theory. There are two other diagrams that arise by permuting the external labels.}
\label{FourDiagrams}
\end{center}
\end{figure}

 We now consider a $\lambda \varphi^3$ theory. In this case, the contribution of the diagram in figure \ref{FourDiagrams}(b), is
 \bea
 \langle  \varphi_{\vec k_1}  \cdots \varphi_{\vec k_4}  \rangle' &=&
   { \lambda^2 \over 4 k_1 k_2 k_3 k_4 }\left[ \sum_{\pm \pm} I^{\rm flat}_{\pm \pm}  (k_{12},k_{34},k_I)  \right] \la{FlatTri}
   \\
  I^{\rm flat}_{++}   & =&  I^{\rm flat}_{--} = 2 T_1 + T_2 ~,~~~~~~ I^{\rm flat}_{+-} = I^{\rm flat}_{-+} = T_2
  \cr
 T_1&=&   { 1   \over 2 k_t   (k_{12} + k_I) (k_{34} + k_I) }  ~,~~~~~~~~~~T_2 =    { 1 \over 2 k_I  (k_{12} + k_I) (k_{34} + k_I) }
 \la{T1T2}
 \eea
   where
   \be
   \la{defVect}
    \vec k_I \equiv \vec k_1 + \vec k_2~,~~~~~~k_{12} \equiv k_1 + k_2 ~,~~~~~~k_{34} \equiv k_1 + k_4 ~,~~~~k_i =|\vec k_i|
    \ee
    where the $\pm$ subindices indicate the sum over all pieces of the in-in contour.
   Here we see a few features. First, we also see a singularity when $k_t \to 0$. Its coefficient is again the flat space amplitude
   $ \lambda^2/[ (k_{12})^2 - k_I^2 ]$, where we have used that $k_t \to 0$ to write $k_{34} = - k_{12}$.
   Notice that \nref{T1T2} is singular when $k_{12} + k_I \to 0$. This corresponds to setting one of the three point vertices on shell,
   so that energy is conserved at the vertex. Of course, this cannot be done for real values of the 3-momenta, but we can reach this
   point by an analytic continuation from $k_I \to - k_I$, for example.

   The interaction points in figure \ref{FourDiagrams}  should be integrated on both branches of the in-in contour, both branches for
   $\eta < 0$. Alternatively, they   can be integrated over all of spacetime.
   For this flat space computation both methods give the same answer, since
 the vacuum expectation value coincides with the vacuum to vacuum transition amplitude in the presence of the operator insertions.

 \subsection{ Four point functions in de Sitter}

   Let us first consider the case of a conformally coupled scalar. If we have a bulk $\tilde \lambda \varphi^4$ interaction, then the
   result is essentially the same as \nref{FourFlat}, since the $\varphi^4$ interaction preserves the conformal
     symmetry. The result is
     \be \lb{FourDSff}
\langle \varphi_{\vec k_1}(\eta_0)  \cdots \varphi_{\vec k_4}(\eta_0)  \rangle  \propto  (2 \pi)^3 \delta^3 ( \sum_{i=1}^4 \vec k_i )
 { \eta_0^4  \over k_1 k_2 k_3 k_4 } { \tilde \lambda \over k_t}
\ee
     We only needed to
     multiply by   one factor of $\eta_0$ per external field, where $\eta_0$ is the time at which we compute the correlator in
     de Sitter space.\footnote{ The expression \nref{FourDSff} gives the leading late time correlators. By putting the operators at
     different times, $\eta_0^i$, $i=1,\, 2,\, 3,\, 4$,  then one could extract the a subleading term in
     the $\eta_0^i$ expansion which contains the part of $\varphi$ with dimension two.   }

     As a next step we can consider a $\lambda \varphi^3$ interaction. Let us first consider the three point function
      \bea \la{ThreePole}
       \langle  \varphi_{\vec k_1}(\eta_0)   \varphi_{\vec k_2}(\eta_0) \varphi_{\vec k_3}(\eta_0)\rangle'
        &=& { \lambda \eta_0^3 \over 8 k_1 k_2 k_3 } \left[ i \int_{-\infty}^{\eta_0}{ d\eta \over \eta }
        e^{ i k_t \eta } - i \int_{-\infty}^{\eta_0}  {d\eta \over \eta }  e^{ - i k_t \eta} \right] =
        \cr
        &=&  { \lambda \eta_0^3 \over 8 k_1 k_2 k_3 } \left[ i  \log( i k_t  ) - i \log (- i k_t  ) \right]
   =  { \lambda \eta_0^3 \over 8  k_1 k_2 k_3 } (-\pi) ~~~~~~~
     \eea
      The $i$s in the logarithms comes from being careful about the $i \epsilon$ prescription for the convergence of the
      integral for early times.
       Note also that \nref{ThreePole} does not have a singularity at $k_t=0$.
 In this case we have a cancellation between the two terms in \nref{FlatLi}, there is a cancellation
between the $+$ and $-$ part of the in-in contour.  Indeed,  the first
       first term in the bracket in \nref{ThreePole}, coming from the $+$ contour,
displays the expected  singularity.
      There was no such cancellation in \nref{FourDSff}.

     Let us now consider the four point function. This  involves    integrals of the form
     \be
     I_{+ +} =-  \int_{-\infty}^{\eta_0} { d\eta \over \eta} \int_{-\infty}^{\eta_0}
      { d\eta \over \eta'} e^{ i k_{12} \eta } e^{ i k_{34} \eta' } { e^{ - i k_I |\eta - \eta'|}  \over 2 k_I}
      \ee
      where the $++$ index indicates the type of in-in contour contribution.
     For small $\eta_0$,  these differ from the corresponding flat space integrals only by the factors of $1/\eta$ and $1/\eta'$.
      These factors  can be introduced by starting with  the flat space result
      \nref{FlatTri} and integrating it with respect to $k_{12}$ and $k_{34}$
      \bea
      && \langle  \varphi_{\vec k_1}  \cdots \varphi_{\vec k_4}  \rangle' _{dS} = { \eta_0^4 \lambda^2
      \over  4  k_1 k_2 k_3 k_4 }  \int_{k_{12}}^\infty dk_{12}'
      \int_{k_{34}}^\infty \sum_{\pm \pm } - (\pm 1)(\pm 1)  I_{\pm \pm }( k'_{12},k'_{34},k_I)
      \\
      & &=  { \eta_0^4 \lambda^2
      \over  4  k_1 k_2 k_3 k_4 }   \left[ 2 {\cal T}_1 + {\cal T}_2 \right] \la{FourPoly}
      \\
      {\cal T}_1 &\equiv & { 1\over 2 k_I } \left[
   {\rm Li}_2( { k_{12}- k_I \over k_t} ) + {\rm Li}_2 ( { k_{34} - k_I \over k_t} )  +
 \log { (k_{12} + k_I) \over k_t } \log({ k_{34} + k_I \over  k_t  } )  - { \pi^2 \over 6 }   \right] ~~~~~~~~ \lb{TauOne}
\\
 {\cal T}_2 & \equiv &  { \pi^2 \over 2 k_I }  ~,~~~~~~~~~~~~k_t = k_{12} + k_{34}
 \la{TauTwo}
      \eea
      where we have done the integral. We have only kept the leading $\eta_0$ dependence for small $\eta_0$ which gives us
      the correlators of the dimension one part of $\varphi$.
       The term ${\cal T}_2$,  \nref{TauTwo},  is a bit subtle. Naively all the contributions from the various pieces of the contour cancel out.
       However, as it happened for the three point function \nref{ThreePole}, we still get a finite piece when we take into account
       the precise $i\epsilon$ prescription for each piece of the contour.

      We see that the result is perfectly regular at $k_{12} = k_I$ or $k_{34} = k_I$.
      But we do  have    singularities
      at $k_{12} + k_I \to 0$, $ k_{34} + k_I \to 0$.
      We  also have a singularity at $k_t \to 0$. In this case ${\cal T}_1$, \nref{TauOne}, goes as
      \be
    {\cal T}_1 \sim k_t ( \log k_t) { 1 \over [ -(k_{12})^2 + k_i^2  ] } ~,~~~{\rm as}~~~ k_t \to 0
     \ee
     We see that the singularity is very mild because the interaction is relevant and becomes small at high energies.
     Note that the function is still non analytic at $k_t=0$ because its first derivative is not finite at $k_t=0$.
     In general the singularity is produced by
     \be
     \int d\eta e^{ i k_t \eta }  \eta^{ n -4} {\cal A}'_{n, {\rm flat}}( -\eta \vec k_i)
     \ee
     where $n$ is the number of particles, which is four in this case, but three in \nref{ThreePole}.
     $ {\cal A}'_{n, {\rm flat}} $ is the stripped   (no energy conservation delta function) flat space amplitude.
     Since the singularity comes from  $\eta \to -\infty $ we are
     taking the high energy limit of the amplitude.

     \begin{figure}[h]
\begin{center}
\includegraphics[scale=.5]{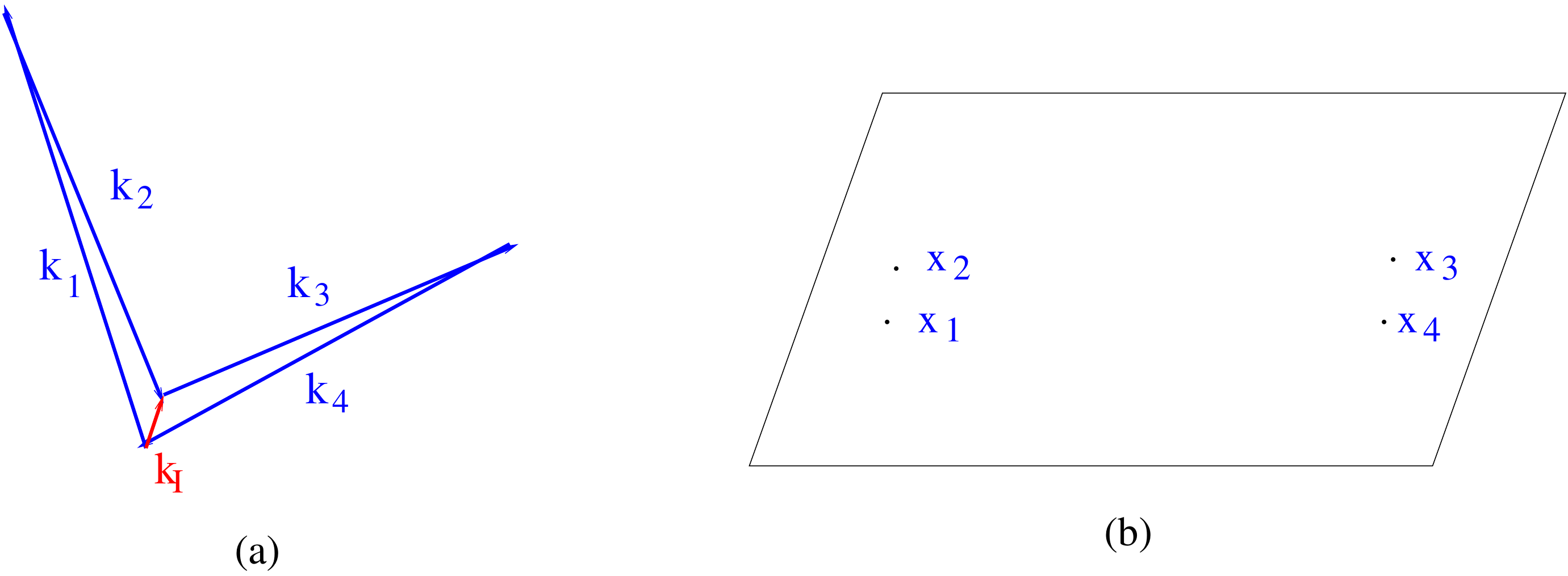}
\caption{Kinematic configuration in the Operator Product Expansion limit or OPE limit. (a) In momentum space the momentum
$\vec k_I = \vec k_1 + \vec k_2$ is much smaller than the rest. (b) In position space, points form two pairs with  the two pairs far
from each other.  Similar limits for higher point functions when a partial sum of the momenta is zero.}
\label{OPELimit}
\end{center}
\end{figure}

     \begin{figure}[h]
\begin{center}
\includegraphics[scale=.5]{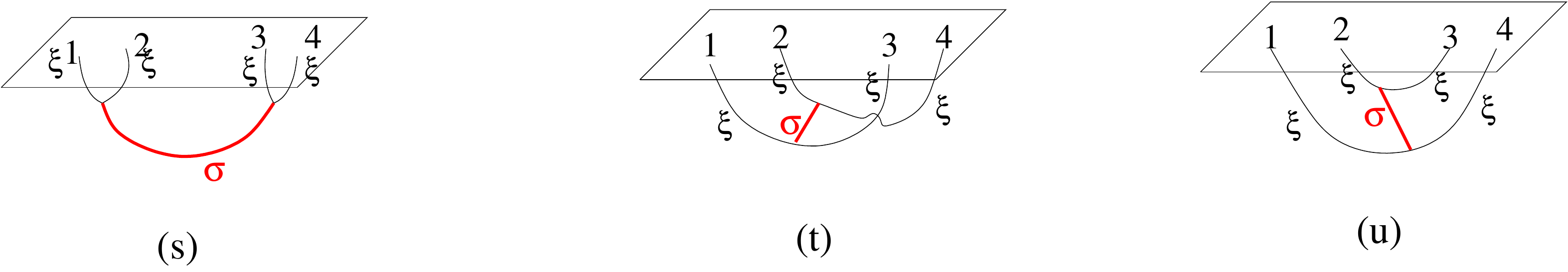}
\caption{ Diagrams containing the exchange of a massive field. We consider
external lines which are massless,  $\xi$, or conformally coupled scalars $\varphi$. The $s$ diagram is the only one
that can contribute non-analytic terms in the small $\vec k_I = \vec k_1 + \vec k_2$ expansion. }
\label{stuDiagrams}
\end{center}
\end{figure}

     A particularly interesting limit is the $k_I \to 0$ case. This is interesting because we can access it within the space of physical momenta.
     The non-analytic parts of the correlator in this limit determine the long distance correlators in a operator product expansion
      limit where two points
     are much closer to each other than they are to the other two points\footnote{ Note that this limit looks like an operator product expansion on the three dimensional boundary, but it is not
      an operator product expansion in the four dimensional sense, since the proper distance between the operators is
      always larger than the Hubble radius, even for those that are ``near'' each other in the OPE limit.}.
      See figure \ref{OPELimit}, see \cite{Suyama:2007bg,Assassi:2012zq}.

     When $k_I \to 0$ we get
     \be \la{OPEfour}
      \langle 4pt \rangle \sim  { \eta_0^4 \lambda^2
      \over  8  k_1 k_2 k_3 k_4 }   \left[  { \pi^2  \over k_I }  - 2 { k_I \over k_{12} k_{34} }  + \cdots \right]
      \ee
      The first term is what we expect from the operator product expansion with an intermediate
       operator of dimension $\Delta =1$. This term comes only from ${\cal T}_2$ in \nref{FourPoly}.
       The second term looks like the contribution from an intermediate operator of dimension $\Delta =2$.
      The dots denote terms that are analytic in $\vec k_I$ at $\vec k_I =0$ or higher order contributions that can be interpreted
      as the contributions of the descendents of the previous operators.
      Note that a term going like $k_I$ is not analytic as a function of $\vec k_I$, since it involves a square root. However, in the
      dots in \nref{OPEfour} there are terms which are independent of $\vec k_I$, which are formally bigger than the second term but
      will not give rise to a contribution in position space in the OPE kinematic region of figure \ref{OPELimit}(b).
        In this particular example, the second term in
      \nref{OPEfour} could, in principle, also be a descendent of the first but the powers of $\pi$ suggest otherwise.
      Note that ${\cal T}_1$ does not contribute to the $1/k_I$ term, despite the explicit factor in \nref{TauOne}.  Note that the
      non analytic terms in the small $\vec k_I$ expansion can only arise from the diagrams in figure \ref{stuDiagrams}(s), since
      there is nothing special happening at small $\vec k_I$ in the diagrams in figure \ref{stuDiagrams}(t),(u).

      The analog computation in Anti-de-Sitter is very similar. But there are two differences. First we only have one of the integrals,
      similar to  $I_{++}$. Also, the propagator of the intermediate particle is different in such a way that the four point function is
      proportional to only ${\cal T}_{1}$. In the OPE limit this produces only a factor of $k_I$ which reflects a dimension two intermediate
      state. Indeed a conformally coupled scalar in de-Sitter gives rise to a dimension two operator on the boundary if we impose the
      Dirichlet-like boundary conditions\footnote{ For this  mass  there is also an alternative choice of
      boundary condition \cite{Klebanov:1999tb}. In that
      case we get only the contribution from the dimension one operator. In other words, only the $1/k_I$ term but not the $k_I$ term
      in the analog of \nref{OPEfour}. } \cite{Witten:1998qj,Gubser:1998bc}. The OPE limit in Anti-de-Sitter was analyzed extensively,
      starting with \cite{Freedman:1998bj}. In conclusion, in AdS we only see a single operator in this OPE limit, but in de-Sitter we
      see two operators. This will become more clear when we look at the contribution from a scalar field with a generic mass.

     We have focused on the singularities in the small $\vec k_I$ limit in momentum space.
      We have argued that such terms give rise to   interesting power
      law behavior in position space. We are interested in these because they signal the existence of new particles.
      Let us now comment on other contributions to the position space OPE which are less interesting from our point of view.
      In position space there are further contributions to the OPE limit which
      arise as logarithmic terms, $\log (x_{13}^2/x_{12}^2 )$, see figure \ref{OPELimit}(b) Such terms arise when we do   the Fourier transform.
      If we focus on the regime $\vec k_1 \sim - \vec k_2$, then if the large $k_1$ limit of the four point function goes as
      $ 1/k_1^3$, then we will end up with an integral of the form $\int d^3 k_1{  e^{ i \vec k_1 .(\vec x_1 - \vec x_2) }
      \over k_1^3 } $, which   gives rise to a $\log |x_{12}|$ from the large $k_1$ integration  region.
      Terms that are analytic in $\vec k_I$ can indeed give rise to such terms. For example, look at the four point function
      \nref{FourDSff} which for large $\vec k_1 \sim - \vec k_2$ contains a factor behaving as  ${ 1 \over k_1 k_2   k_t } \sim
       { 1 \over k_1^3   }$.
      Indeed we know that the position space version of the four point function \nref{FourDSff} has the form \cite{Usyukina:1993ch}
      \bea
      \langle \varphi(\vec x_1) \cdots \varphi(\vec x_4) \rangle & \propto &  { 1 \over x_{13}^2 x_{24}^2 }
       { 1 \over( z - \bar z )} \left[ 2 {\rm Li}_2(z) - 2 {\rm Li}_2(\bar z ) + \log (z \bar z) \log { 1 - z \over 1 - \bar z }
       \right]
       \cr  {\rm with}~~~
&&       z \bar z = { x_{12}^2 x_{34}^2 \over x_{13}^2 x^2_{24} } ~,~~~~~~~(1-z)(1-\bar z) =  { x_{14}^2 x_{23}^2 \over x_{13}^2 x^2_{24} }
       \eea
       where the variables $z$ and $\bar z$ are defined implicitly by solving the last equations. We see that as $x_{12} \to 0$ we
       get a logarithmic singularity.
       This can be interpreted as the leading correction due to the anomalous dimension of the operator $\varphi(x_1)^2$ in
       the $\varphi^4 $ theory.
       This is just a simple explicit example, but the fact that such logarithmic terms arise is a generic feature of the
       perturbative AdS or dS diagrams, see \cite{Freedman:1998bj}. We have seen that these arise from the fourier transform from momentum
       to position space. These terms are less interesting for our purposes because they do not signal the existence of new
       particles. They simply reflect interactions between the particles that already exist. Such terms can arise both from a contact interaction
       as in figure \ref{FourDiagrams}(a) or from analytic terms in $\vec k_I$ in the exchange diagrams in figure
        \ref{stuDiagrams}(t),(u).

       \subsubsection{Four point function in de Sitter with an intermediate massive scalar field}

       We can now consider conformally coupled scalars $\varphi$ coupled to a general massive scalar field $\sigma $ with
       an  interaction vertex  $\int \lambda \varphi^2 \sigma $.
       The late time expectation value is given by the following expression
       \bea
       \langle \varphi_{\vec k_1}(\eta_0) \cdots \varphi_{\vec k_4}(\eta_0) \rangle' & = &{ \eta_0^4 \, 2^2 \, \lambda^2 \over 16 k_1 k_2 k_3 k_4 }
         I_E(k_{12},k_{34},k_I) + {\rm other ~diagrams}
        \la{JscalDef}
       \\
       I_E(k_{12},k_{34},k_I)  &\equiv &   I_{++} + I_{+-} +I_{-+} + I_{--}
     \cr \la{FourInteg}
    I_{\pm \pm }  &=& (\pm i) (\pm i) \int_{-\infty}^0 { d\eta \over \eta^2 } e^{ \pm i k_{12} \eta} \int_{-\infty}^0 { d\eta' \over \eta'^2 }
     e^{\pm  i k_{34} \eta'}  \langle \sigma_{\vec k_I} (\eta) \sigma_{-\vec k_I}(\eta') \rangle'_{\pm \pm } ~~~~~~~~
    \eea
    where the $\pm$ indicate the  type of branch in the integration contour. The factor of $2^2$ in the numerator is a symmetry
    factor. Here $
   \langle \sigma_{\vec k_I} (\eta) \sigma_{-\vec k_I}(\eta') \rangle'_{\pm \pm }  $ denotes the propagator along the appropriate branch, given in
    \nref{KeldyshC}. There are two other diagrams and they result from the replacements $(1,2,3,4) \to (4,1,2,3), ~(2,3,4,1)$ respectively.
    The functions $I_{\pm\pm }(k_{12},k_{34}, k_I)$ obey a differential equation that is a consequence of the
    wave equation for the field $\sigma$. Namely, from the integral expressions \nref{FourInteg} one can show that the
    equation $(\nabla_x^2 -m^2) \langle \sigma(x) \sigma(x') \rangle_{++} ={ i \over \sqrt{-g}} \delta^4(x-x') $, or
    \be
    \left[ \partial_\eta^2   - { 2 \over \eta} \partial_\eta   + ( k_I^2 + { m^2 \over \eta^2 } )  \right] \langle \sigma_{k_I}(\eta) \sigma_{k_I}(\eta') \rangle'_{++}  =- i  \eta^2 \delta(\eta - \eta')
    \ee
     implies the equation\footnote{ In deriving this equation we drop some boundary terms at $\eta =0$. This is correct for generic masses
     of the intermediate state but might be incorrect for special masses, such as $m^2=2$, the conformally coupled scalar, or $m=0$.
     Indeed the expressions we will obtain are singular for $\mu = i/2$ or $\mu =  3 i/2$, see \nref{GPM}. For this special masses
     it is simpler to do the explicit computation.  }
    \bea \la{DiffEqn}
  && \left[  ( k_{12}^2 - k_I^2 ) \partial_{k_{12}}^2 + 2 k_{12} \partial_{k_{12} }  + (m^2 -2) \right] I_{++} = {  1
   \over k_{12} + k_{34} }
   \cr
   && \left[ ( \WM^2 -1) \partial_\WM^2 + 2 \WM \partial_\WM + (m^2 -2)  \right] G_{++}(\WM,\ZM) = {1\over \WM + \ZM }  \la{DiffEquMaSc}
   \\
    && {\rm with}~~~~
    \WM \equiv { k_{12} \over k_I} ~,~~~~~~ \ZM \equiv { k_{34} \over k_I } ~,~~~~~~ I_{\pm \pm } = { 1 \over k_I  } G_{\pm\pm}(\WM,\ZM)
    \eea
    We also have the same equation in terms of the $\ZM$ variable.
    We want solutions that are symmetric under the exchange $\WM \leftrightarrow \ZM$.
    The function $I_{--}$ obey an identical  equation to \nref{DiffEquMaSc}.
    The functions
    $I_{+-}$ and $I_{-+}$ obey the same  equation but with zero in the right hand side of \nref{DiffEquMaSc},
    since  $(\nabla_x^2 -m^2) \langle \sigma(x) \sigma(x') \rangle_{+-} =0$.

    Using the explicit expressions for the propagator we can either do the integrals \nref{FourInteg} or solve the equations
    \nref{DiffEquMaSc}. The equations contain a
    bit less information than the explicit integral expression. We recover the missing information by imposing the correct boundary
    conditions at the possible singularities of the equation. The possible singularities are $\WM = \pm 1$ and $\WM= \infty$.
    We demand that the solution is regular at $\WM=1$, and we impose that the leading singular behavior at $\WM = -1$ is properly
    normalized, as we will discuss in more detail below, in \nref{NorCon}.

    For the three point function we had derived the equation \nref{CoCouSc} from the conformal invariance condition. This is
    identical to the homogeneous version of \nref{DiffEquMaSc}.  For the
    four point function it is also possible to study the constraints coming from conformal invariance. If we make the
    further assumption that the four point function has a structure as in \nref{JscalDef}, with a function $I_E$ that depends only
   $k_{12}, k_{34}$ and $k_I$, then conformal invariance gives
    us the condition
    \be  \la{FourConfInv}
   ( D_\WM - D_\ZM ) G_{\pm\pm } = 0 ~,~~~~~D_\WM =
  ( \WM^2 -1) \partial_\WM^2 + 2 \WM \partial_\WM    -2
    \ee
 Of course, this implied by \nref{DiffEquMaSc}. Is the converse true? It is certainly not, since \nref{FourConfInv} does not
    know about the mass of the intermediate field. We can make a conformal invariant function by imposing
    \be
    ( D_\WM + m^2 ) G_{++}(\WM,\ZM) = r(\WM,\ZM) ~,~~~~~~~~ ( D_\ZM + m^2  ) G_{++}(\WM,\ZM) = r(\WM,\ZM)
    \ee
    Here we have included the mass as an arbitrary constant in the left hand side and we consider a more general function in the
    right hand side. For consistency, this function should also obey
    \be \la{rEqn}
    ( D_\WM - D_\ZM ) r =0
    \ee
     For large $m^2$ we can solve for $G$ in terms
    of a power series expansion of the form
    \be
    G_{++}(\WM, \ZM) = { 1 \over( D_\WM + m^2)  }  r =  { 1 \over m^2  } \left( r - { D_\WM r \over m^2  } + \cdots \right)
    \ee
    This gives a function which is symmetric in $\WM,~\ZM$, to all orders , thanks to   \nref{rEqn}. We can make $r$ a rational
    function. The simplest example is to take $r = 1/(\WM + \ZM)$ which is what we had above, in \nref{DiffEquMaSc}.
     We could also replace it by
    $r = D_\WM [1/(\WM + \ZM)] $. This corresponds to a bulk theory with an additional  contact $\varphi^4$ interaction tuned
    so as to cancel the term coming from $\sigma $ exchange.
     Proceeding in this way we can add an arbitrary set of contact interactions.

     The function   $G_{+-}$ is particularly simple because it obeys the homogeneous version of equation \nref{DiffEquMaSc}.
    Then the solution is simply a product of the form
    \be
    \la{ProdForm}
    G_{-+}(\WM,\ZM) =
    F(\WM) F(\ZM)
    \ee
     It is this simple product because there is a unique (up to scaling) solution
     of the homogeneous version of the equation \nref{DiffEquMaSc}
     which is regular at $\WM=1$ and $\ZM=1$.
    Here the function $F$ is the same as the one that appeared in the expression for the three point function \nref{HypThPt}.
    The normalization condition can be set by looking at the region near $\WM \sim \ZM \sim -1$. The singular
    contribution comes from the large $\eta, ~\eta'$ region of the integral in \nref{FourInteg}. In this region, the
    $\sigma$ propagator simplifies and we can write
   \be \la{NorCon}
   I_{+-} |_{\WM \sim \ZM\sim \, -1}  \sim  \int { d\eta \over \eta^2 } e^{ i k_{12} \eta}\int  { d\eta' \over \eta'^2 } e^{ - i k_{34} \eta'}
    { \eta \eta'  \over 2  k_I } e^{ i k_I ( \eta - \eta') }   \sim { 1 \over 2 k_I } \log(1 + \WM) \log(1 + \ZM)
    \ee
    This fixes the function
    \be \la{GPM}
    G_{+-}(\WM,\ZM) ={ \pi^2 \over 2 \cosh^2(\pi \mu)}
  F(\half + i \mu , \half - i \mu, 1; { 1 - \WM\over 2} )
    F(\half + i \mu , \half - i \mu, 1; { 1 - \ZM \over 2} )
    \ee
 We have the same expression for $G_{-+}$.

    We were not able to find an analytic solution of the equations for the $G_{++}$ or $G_{--}$ functions for generic masses\footnote{
    For $m^2 =2$ which is the conformally coupled scalar,  we can   check that
    $ k_I  {\cal T}_1$ in \nref{TauOne}  obeys the equation \nref{DiffEquMaSc}. This is an example of an analytic solution for
    a special mass.
     Note that ${\cal T}_{2}$ is a homogeneous solution of the equation for
    this particular mass.   $m^2 =2$ is a bit special since the derivation of the
    equations \nref{DiffEquMaSc} is not valid due to some boundary terms that come when we integrate by parts to derive
    the equation. For example the function $G_{+-} \propto \log(\WM +1) \log( \ZM +1) $ does not obey the homogeneous
    equation. This issue does not arise for generic masses.  }. However, it is
    easy to derive an approximate expression in the large $\WM$ and/or $\ZM$ limit.
    For this purpose, let   us focus on the one of the integrals, say the $\eta$ integral, and let us compare
    the integral for the $+$ contour versus the integral with the $-$ contour.
  \be
   I_{+  \pm} \propto \int { d\eta \over \eta^2 } e^{ i k_{12} \eta } \langle \sigma(\eta) \sigma(\eta') \rangle'_{+\pm} ~,~~~~~~
   I_{-  \pm} \propto  -\int { d\eta \over \eta^2 } e^{ - i k_{12} \eta } \langle \sigma(\eta) \sigma(\eta') \rangle'_{-\pm}
   \la{LateIntr}
   \ee
  The difference between the two integrals is  that we replace $k_{12} \to - k_{12}$ in the first factor and that we
  consider a different propagator in the second factor.
  There are  situations where the propagator will actually be the same.
  For example, if we consider a regime where the two points are spacelike separated in position space, then we expect that the two propagators
  are the same. This means that the non-analytic piece in $k_I$ of the propagators is the same
   $    \langle \sigma_{k_I}(\eta) \sigma_{k_I}(\eta') \rangle'_{+\pm}  |_{\rm non-analytic}  =  \langle \sigma_{k_I}(\eta) \sigma_{k_I}(\eta') \rangle'_{-\pm} |_{\rm non-analytic}  $.  See \nref{LongMomTwo} for the explicit expression.
  This is precisely the region we are interested in in the OPE limit.
  Then we can say
  \be \la{AnCoSq}
   I_{+ \pm }(k_{12} , k_{34},k_I) = - I_{-\pm}( e^{ i \pi } k_{12} , k_{34},k_{I} )  ~,~~~~~~k_{I} \ll k_{12}
   \ee
   This equation is true for the non-analytic pieces in $\vec k_I$.
   Here the precise analytic continuation has been chosen to be  consistent with the proper $i\epsilon$
   prescription for each contour. This enables us to compute a part of the $I_{++}$ or $I_{--}$ integrals
   from the knowledge of the $I_{+-}$ or $I_{-+}$ integrals \nref{GPM}.

   \subsubsection{ OPE  limit of the four point function with a massive intermediate scalar }
   \la{OPEIntScalar}

   Here we consider the OPE limit, see figure \ref{OPELimit}.
   This involves taking the
     $ \vec k_I \to 0$ limit of the four point function, focusing on the
   non-analytic terms as a function of $ \vec k_I$.
   In this case, in \nref{FourInteg}
    we can replace the correlators $\langle \sigma_{\vec k_I}(\eta) \sigma_{-\vec k_I}(\eta') \rangle'_{\pm \pm }$ by their
   long distance expression given in \nref{LongMomTwo}. This long distance expression is independent of the $\pm$ subindices.
   Inserting this into the expressions for the correlator \nref{FourInteg} we find that  we can integrate separately
   each of the two terms in \nref{LongMomTwo}. For each of these two terms,
   the integrals $I_{\pm\pm}$ factorize
   as $I_{\pm \pm} (k_{12},k_{34},k_I) \propto J_\pm (k_{12})  J_\pm (k_{34})  $ with\footnote{ In \nref{LongMomTwo} we had
   factors $ ( \eta \eta' )^\Delta $ which we should really view as $ ( -\eta)^\Delta ( -\eta')^\Delta $.}
   \be \la{JpmInt}
   J_\pm(k_{12})  =\pm i  \int_{-\infty}^0 { d \eta \over \eta^2 } e^{ \pm i k_{12} \eta } ( -\eta)^{\Delta }  =
   - (\mp i )^{ \Delta } (k_{12})^{ 1-\Delta } \Gamma( \Delta -1 )
   \ee
   \bea
    \left[
    J_+
     (k_{12} ) + J_- (k_{12} )
      \right]
   \left[ J_+ (k_{34}) + J_- (k_{34}) \right]
  &=&  2 ( 1 +\cos(\pi \Delta)  ) (k_{12} k_{34})^{ 1-\Delta } \Gamma( \Delta -1)^2 \la{Expecv}
 \\
 &=&  2 ( 1 + i \sinh \pi \mu ) (k_{12} k_{34})^{ -\half - i \mu } \Gamma( \half + i \mu)^2  \notag
   \eea
  Of course, the expressions in \nref{JpmInt} are consistent with the analytic continuation \nref{AnCoSq}.
    For the second term in \nref{LongMomTwo} we get the same as above but with $\mu \to - \mu$.
   Putting all of this together, including the numerical factors,  we get
   \bea \notag
   \langle \varphi(1) \cdots \varphi(4) \rangle'_{k_I \to 0 } & \sim &
   { \eta_0^4  \lambda^2    \over 4 k_1 k_2 k_3 k_4 } I_E
   \eea
   \bea
   I_E & \sim & { 1 \over  2 \pi\sqrt{ k_{12} k_{34} }     }
   \left[  \left( { k_I^2 \over 4  k_{12} k_{34} } \right)^{ i \mu} (1 + i \sinh \pi \mu) \Gamma( - i \mu)^2
   \Gamma(\half + i \mu)^2 +
   \right.
   \cr &+&
   \left.
   \left( { k_I^2 \over 4  k_{12} k_{34} } \right)^{- i \mu} (1 - i \sinh \pi \mu) \Gamma(  i \mu)^2
   \Gamma(\half - i \mu)^2
   \right]  \la{CoScsmallki}
   \eea

   Note that the region of integration contributing to \nref{JpmInt} is of order $-\eta \sim 1/k_{12}, ~1/k_{34}$, which is
   much smaller than the characteristic values of $\eta$ where the massive propagator behaves non-trivially, which is
   $-\eta \sim 1/k_I $. We see that in this regime we are very directly measuring the long distance propagator for the
   massive field. Note that $k_{ 12} \sim 2 k_1 \sim 2 k_2$ as  $\vec k_I \to 0$. Of course, in expression
   \nref{CoScsmallki} we can replace $k_2 \sim k_1$ and $k_3 \sim k_4$, $k_{12} \sim 2 k_1$ and $k_{34} \sim 2 k_3 $.

  Notice that \nref{CoScsmallki}  contains oscillations in the logarithm of the ratio $k_I^2/ ( k_{12} k_{34} ) $.
   The logarithm of this ratio is basically the total number of efolds for which the pair has existed, adding the efolds for
   each member of the pair.
   Notice that the phase of the oscillations is also fixed. It is determined by the mass.
   Notice that conformal symmetry constrains the  OPE limit so that we
   can predict the existence of the two terms in \nref{CoScsmallki}, but
   the computation of the phase required more information. It required solving the equation for the massive field in order to
   connect the simple boundary condition at early times with the late time behavior of the propagator \nref{LongMomTwo}.
     The amplitude of the effect goes as $e^{ - \pi \mu }$ for large $\mu$.
   We can view this leading piece, going as $e^{- \pi \mu }$, as arising from the interference of two processes. One is when
   we have the usual gaussian evolution for the $\varphi$ fields. The other is when we create a pair of massive $\sigma $ particles
   which then decay into $\varphi$ excitations.
   Notice that the probability to create a pair of massive particles goes as $e^{ - 2 \pi \mu }$. We are getting something bigger, of
   order $e^{ -\pi \mu}$, because we have an interference effect that is sentitive to the amplitude and not the square of the
   amplitude for creating the pair.
   Therefore, if we look at the sky we cannot say for certain where a pair has actually been created, we can only detect the
   deviation in the probability distribution.
   Notice that indeed the leading term in the $e^{ - \pi \mu}$ expansion, comes from the $I_{++}$ or $I_{--}$ integrals, from
   the $ i \sinh \pi \mu $ factors in \nref{CoScsmallki}. The $I_{+-}$ or $I_{-+}$ integrals go like $e^{ - 2 \pi \mu }$. These
   represent physical processes where one produces an actual $\sigma$ particle and it involves the square of the amplitude to
   produce them.

   This effect is very  quantum mechanical since we see the oscillations of the wavefunction of a quantum particle.
   Here we are seeing the
   oscillations because we have the interference between producing the pair and not producing the pair of particles.

   We will now repeat the computation in a more indirect way.
    It is conceptually the same as what we did above, but it is easier to generalize to   the higher
   spin case.

  We now start from the explicit expression for
   $G_{+-}$ written in \nref{GPM}. The $k_I \to 0 $ limit implies that $\WM , ~\ZM \to \infty$.
   Therefore we simply expand \nref{GPM} for large values of the arguments. Each of the hypergeometric functions in
   \nref{GPM} contains two different powers of its large argument
   \be \la{LargHy}
  F(\WM) _{p \gg 1 } \sim { 1 \over 2 \sqrt{\pi } \WM^{\half } } \left[ (2 \WM )^{ - i \mu} \Gamma(- i \mu) \Gamma( \half + i \mu )
  + (2 \WM )^{  i \mu} \Gamma( i \mu) \Gamma( \half - i \mu ) \right] + \cdots
   \ee
    When we multipy the powers coming from each of the
   hypergeometric function we obtain four terms, only two of such terms give rise to non-analytic dependence in $k_I$.
   Focusing on these terms we find
   \bea
  I_{+- \, p,\, p' \gg 1}  &\sim &
    { 1 \over 4 \pi   (k_{12} k_{34})^{\half} }
  \left[ ( 4 \WM \ZM )^{i \mu} { \Gamma(i \mu)^2  \Gamma( \half - i \mu)^2 } +
    ( 4 \WM \ZM )^{-i \mu} { \Gamma(-i \mu)^2   \Gamma( \half + i \mu)^2 } \right] ~~~~~~~~
      \eea

  In the small $k_I$ limit, we can  obtain the results for the other integrals by performing the analytic
   continuation in \nref{AnCoSq}.
   Adding them all up we get again \nref{CoScsmallki}.

     \subsection{ Four point function of massless scalars and an intermediate massive scalar field }

     Here we consider the diagram of the form in figure \ref{FourDiagrams}(b) but with massless external fields. We denote
     the massive external fields by $\xi $.  We write the coupling as $ \int \lambda (\nabla \xi)^2 \sigma$.
     As we have seen in the section on the three point function, we could get the result
     for this case by applying a differential operator on the result for the four point function of conformally coupled scalars.
     \bea
     \langle \xi_{\vec k_1} \cdots \xi_{\vec k_4} \rangle' &=&
     { \lambda^2 \,    \over 4 k_1^3 k_2^3 k_3^3 k_4^3 } \left[ J_E(k_1,k_2,k_3,k_4,k_I) + {\rm other~ diagrams} \right] \la{FMsca}
       \\
       J_E&\equiv & O_{12} O_{34} I_E(k_{12} ,k_{34}, k_I)  \la{O12O34}
       \eea
       where $I_E(k_{12},k_{34},k_I)$ is exactly the same function we had for the conformally coupled scalar in \nref{JscalDef}.
      The operator $O_{12}$ is given in \nref{OpZetak}.
      Notice that the dependence on $(k_1 - k_2 )$ or $(k_3 -k_4)$ is rather simple and
      it comes  exclusively
      from the operators $O_{12}$ and $O_{34}$.

     In the small $k_I$ limit,  the operator $O_{12} $ simplifies.
      $k_1 - k_2 \ll k_{12}= k_1+k_2$ and the operator acts on a   power of $k_{12}$ to give
      \be \la{LargeO}
       O_{12}  k_{12}^{a} \to { 1 \over 8 } (a -1)(a -2) k_{12}^{ a +2 }
       \ee
       Acting in this way on \nref{CoScsmallki} we obtain
           \bea \la{FourZeroSca}
  && J_E(k_1,k_2,k_3,k_4,k_I)_{k_I \to 0} \sim  { 1\over 128 \pi }
  {  (k_{12} k_{34})^{ 3 \over 2}   }
  \times
  \cr
  &&~~~\times
   \left[\left({ k_I^2 \over 4 k_{12} k_{34} } \right)^{i \mu} (1 + i \sinh \pi \mu){ ( { 3 \over 2 }  + i \mu)^2 ( { 5 \over 2 } + i \mu)^2\Gamma(-i \mu)^2   \Gamma( \half + i \mu)^2 } +\right.
  \cr
  && ~~~\left. +\left({ k_I^2 \over 4 k_{12} k_{34} } \right)^{- i \mu} (1 - i \sinh \pi \mu)  { ( { 3 \over 2 } - i \mu)^2 ( { 5 \over 2 } - i \mu)^2\Gamma(i \mu)^2   \Gamma( \half - i \mu)^2 }
    \right]
   \eea

   \subsubsection{   Four point function with a spinning particle intermediate state}

   In this subsection we consider the situation with a spinning particle as the intermediate state.
   We   show how the spin information appears in the expression of the four point function in the
   OPE limit.  We also derive the precise mass and spin dependence of the phase.

   We start from the computation of the $I_{+-}$ integrals which have the structure of a product of two
   three point functions. More precisely, it is a product of two point functions contracted with the  inverse two point
   function of the intermediate massive particle.
    It includes a sum over all the polarization states of the intermediate massive spining particle. This has the
    schematic form
    \be \la{FourHelSum}
     I_{+-} = \sum_m \langle O_2 O_2 O_{\Delta , m} \rangle' { 1 \over k_I^{2\Delta -3}  I_2(\Delta, m) }  \langle  O_{\Delta ,- m}
      O_2 O_2 \rangle'
     \ee
     where the sum runs over all the helicity components of the operator with spin. Here $I_2(\Delta, m )$ is the two point function
     coefficient for each helicity, see \nref{Itwodef}. The expressions for $\langle O_2 O_2 O_{\Delta m}\rangle $ can be obtained
     by fourier transforming \nref{ThreePos} and then projecting onto the various helicity components using \nref{EpEptil}.

    Here we are going to focus only on the small $\vec k_I$ limit of this expression. More precisely we will be interested in the
    non-analytic terms in the small $\vec k_I$ expansion. These are the terms that give rise to long distance correlators in position
    space. For this purpose we need the small $k_I$ expansion of the three point functions.
    This  expansion   involves two powers of $k_I$ \nref{Sqannonan}.
     For \nref{Sqnonan} we get the same factors of $I_2(\Delta, m)$ that we got in the two point function \nref{Itwodef}.
      The $\hat k_I$ dependence of $I_2( \vec \epsilon , \vec \tilde \epsilon , \hat k_I) $ comes only through such factors.
      In other words, if were were to set $I_2(\Delta , m) \to 1$, then we would have no dependence on $\hat k_I$, all the dependence
      would be on the angle between $\vec \epsilon$ and $\vec k_1$.
      For this reason
      \nref{Sqan} is independent of $\hat k_I$. For small $\vec k_I$ we can write the three point function as
           \be \la{ThreeHelExp}
     \langle O_2 O_2 O_{\Delta , m}\rangle'_{k_I \to 0}  \sim   A_m k_{12}^{ 1 - \Delta} k_I^{ 2 \Delta -3} I_2(m) + B_m k_{12}^{ \Delta-2}
    \ee
   where
    $A_m$ and $B_m$ depend on the angles only through  the angular momentum $m$ projection of $(\hat k_1)^s $ on to the
    $\hat k_I$ axis. In other words, they involve the spherical harmonics,
     $A_m , ~  B_m \propto Y_{s \, m}(\theta,\phi)  $ with $ \cos \theta = \hat k_1 . \hat k_I $.
      At first sight
    one is tempted to keep only the first term in \nref{ThreeHelExp} since that is the only non analytic term in $k_I$. But we should
    be careful, we are interested in the non-analytic term of the final answer in \nref{FourHelSum}, and we do see that we have
    a non-trivial power of $k_I$ from the inverse two point function in \nref{FourHelSum}.
     Therefore the non-analytic terms in $k_I$ in \nref{FourHelSum} are given by
    \be \la{HeIpm}
    I_{+-} \sim \sum_m A_m A_{-m}    (k_{12} k_{34} )^{ 1 - \Delta} k_I^{   ( 2 \Delta -3) } I_2(m) + B_m B_{-m}  (k_{12} k_{34} )^{   \Delta-2}
    k_I^{  - ( 2 \Delta -3) }  { 1 \over I_2(m)}
     \ee
      We now see that both terms  in \nref{ThreeHelExp} contribute to the expansion.

      \begin{figure}[h]
\begin{center}
\includegraphics[scale=.5]{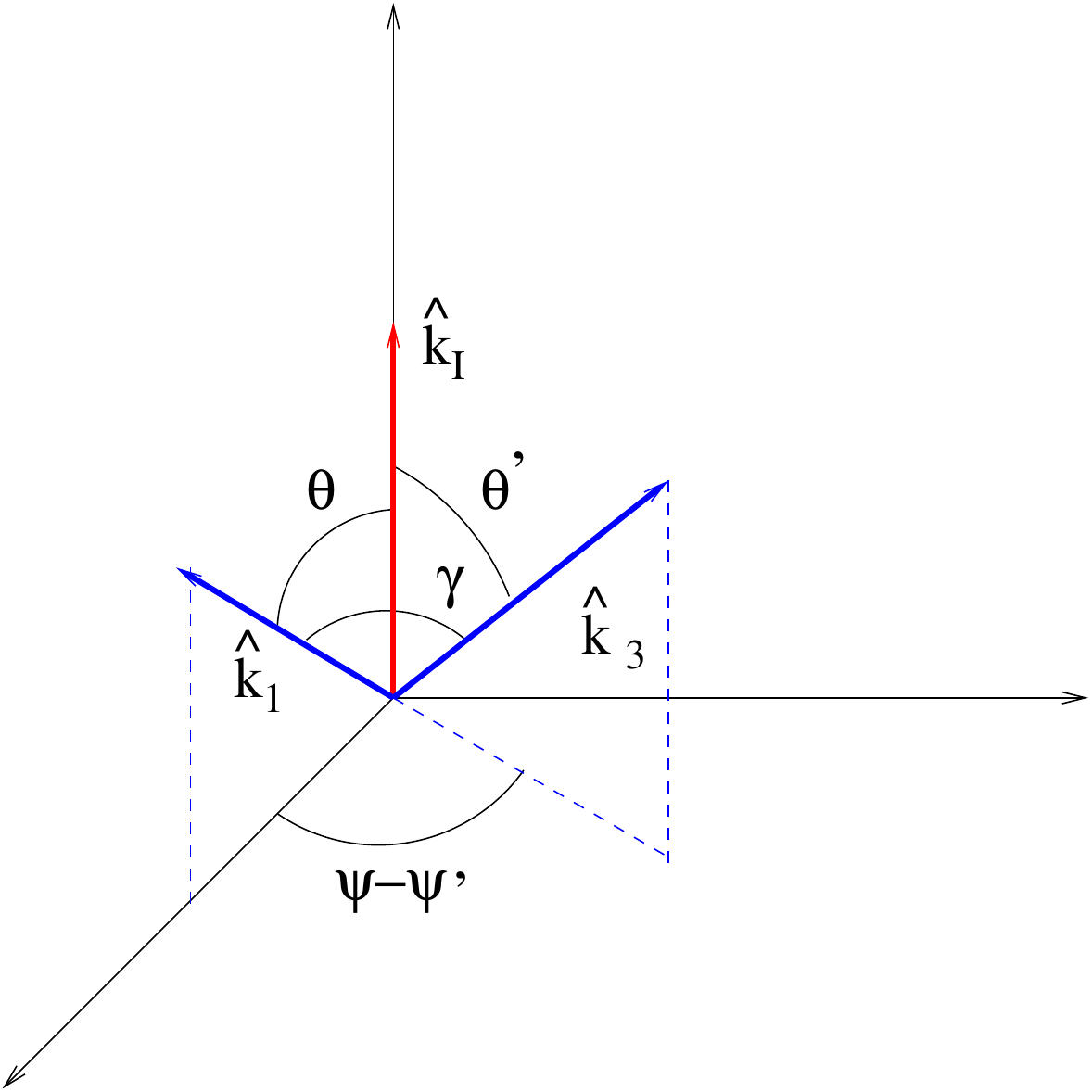}
\caption{ Definition of the angles that enter into the answer. $\gamma$ is the angle between $\hat k_1$ and $\hat k_3$.
$\psi - \psi'$ is the angle between the projections of $\hat k_1 $ and $\hat k_3$ onto the plane perpendicular to $\hat k_I$.
Finally, $\theta$ and $\theta'$ are the angles between $\hat k_I$ and $\hat k_1$, $ \hat k_3$ respectively. }
\label{Angles}
\end{center}
\end{figure}

     Now the sum
      $\sum_m A_m A_{-m}$ should be a function of only $\vec k_1 $ and $\vec k_3$  which is projected onto the spin $s$ contribution.
      Therefore it is the Legendre polynomial of the relative angle
      $P_s(\cos \gamma)$. This is not quite what we have in \nref{HeIpm} since we also have the $I_2(m)$ contributions.
      These contributions can be taken into account as follows. Let us first define the angles
     \bea
   &&  
   \cos \gamma = \cos \theta \cos \theta'  + \sin \theta \sin \theta' \cos(\psi - \psi')  \la{CosGamma}
     \\
     && \cos \theta = \hat k_1 . \hat k_I ~,~~~~\cos \theta' = \hat k_3 . \hat k_I ~,~~~~~~\cos \gamma = \hat k_1 . \hat k_3
     \eea
     and $\psi$ and $\psi'$ are the angles specifying the directions of $\hat k_1$ and $\hat k_3$ in the plane transverse to $\vec k_I$.
       We then  expand
     $P_s(\cos \gamma) $ in definite helicity components
     \be \la{LegUsual}
     P_s( \cos \gamma) = \sum_{m=-s}^s e^{ i m (\psi - \psi') }   \hat P^m_s(\theta) \hat P^m_s(\theta')
     \ee
     where the $\hat P^m_s$ are a version of the associated Legendre polynomials but with a
     different overall normalization. Here we just define them as the coefficients of the Fourier expansion of the left hand side
     of \nref{LegUsual}.
        In other words, we write the left hand side using \nref{CosGamma} and we expand in powers of   $e^{ i (\psi - \psi') }$.

     We now define  the functions $\Theta_\pm$
     \bea
     \Theta_\pm =\Theta_\pm(\theta,\theta', \psi-\psi')&=& \sum_{m=-s}^s e^{ i m (\psi - \psi') } \hat P_s^m(\theta) \hat P_s^m(\theta') \left[ I_2(\Delta, m)\right]^{\pm 1} \la{HelPh}
     \\
     I_2(m)  &=&    { \Gamma( \half + m + i \mu ) \Gamma( \half + s - i \mu) \over \Gamma(\half + m - i \mu)
      \Gamma( \half + s + i \mu ) } \la{ItwoPh}
      \eea
      In other words, we define these functions by expanding the left hand side of \nref{LegUsual} in powers of $e^{ i (\psi - \psi')}$
      and then we multiply each power by the appropriate $I_2(\Delta , m)$ factor.
      For example, for $s=2$ we get
      \be
      \Theta_+ = P_2(\cos \gamma) -    { i  3 \mu \over { 3   } + i 2 \mu} \sin 2 \theta \sin 2 \theta' \cos (\psi - \psi' )
      + { \mu ( 1 + 3 \cos 2\theta)(1 + 3 \cos 2 \theta' ) \over ({ 3   } + i 2 \mu)( 1   + i  2 \mu ) }
      \ee
      We see that for $\mu =0$ we get the usual Legendre polynomial. This is simply due to the fact that \nref{ItwoPh} becomes one
      for $\mu=0$. This is true for all spins.
      It is also interesting to analyze the case of $\mu \to \infty$. In this case we see that $\nref{ItwoPh}$ becomes $(-1)^{s-m}$.
      This is equivalent to replacing $\psi' \to \psi'' + \pi$, and then changing the overall sign of $\cos \theta'$ and $\sin \theta'$ we
      get
      \be
     \Theta_\pm|_{\mu \to \infty} \sim
      P_s(\cos \tilde \gamma) ~,~~~~~~{\rm with} ~~~~ \cos \tilde \gamma = - \cos \theta \cos \theta' + \sin \theta \sin \theta'
      \cos (\psi - \psi' )
      \ee
      We see that $\cos \tilde \gamma$ is the inner product of $ \hat k_1$ with a reflected version of $\hat k_3$ along the $\hat k_I$
      direction, that flips the sign of only $\cos \theta' \to -\cos \theta'$. This is what we expect from the discussion around
      figure \ref{SpinCorrelation}.

       We also have some helicity  independent factors which
      can be computed by looking at the expansion of the highest helicity component.
      We can write  the three point function as (see appendix \ref{ThreeMomSpin})
      \be \la{Highhe}
      \langle O_{2}(\vec k_1 ) O_{2  } (\vec k_2 ) O_{\Delta, s  } (\vec k_3 ) \rangle' = [ \vec \epsilon . (\vec k_1 - \vec k_2)]^s k_I^{\Delta -2 - s } G_s(\YM) + \cdots
      \ee
      where the dots are terms of the form $  [ \vec \epsilon . (\vec k_1 - \vec k_2)]^{s-i} (\vec k_I . \vec \epsilon)^i$ with $i>0$. These
      neglected terms do not contribute to the piece with largest angular momentum around the $\hat k_I$ axis.
      The function $G_s$ obeys
     the equation  \nref{CoCouSpin} . Its solution is
    \be  \la{ThreeSpin}
    G_s(\WM )  \propto  { \Gamma( \half + s + i \mu ) \Gamma( \half + s - i \mu)
    } ~_2F_1( \half + s + i \mu, \half + s - i \mu ,
    1 + s ; { 1 - \WM \over 2 } )
    \ee
    where $\WM = ( k_1 + k_2 )/k_I$. Here and below the symbol $\propto $ means that we have neglected spin dependent numerical
    factors (such as $2, ~ s!, ~2^s$),
    but we have included the dependence on the mass.
     This is analytic at $\WM=1$ as expected. As $\WM \to -1$, it is normalized so that it matches the answer expected from
      the three point vertex
     \be
     \int  \varphi \nabla_{i_1} \cdots \nabla_{i_s} \varphi \sigma_{i_1 \cdots i_s}  \to \int { d\eta \over \eta}
     \eta^s e^{ i \eta ( \WM + 1) } \propto { 1 \over (\WM  + 1)^s }  \la{NorCond}
     \ee
     The large $\WM $ behavior of \nref{ThreeSpin}, together with the definition in \nref{Highhe} implies that we have an expansion as in
     \nref{ThreeHelExp} with
     \bea
     A_s &\propto &   { \Gamma(\half +s+ i \mu) }    { \Gamma(-i \mu)   } 2^{- i \mu}
     \cr
     B_s &\propto &     { \Gamma(\half+s - i \mu ) }  { \Gamma(i \mu)    }2^{ i \mu}
     \la{AsBs}
   \eea
        where the ratio $A_s/B_s$ is correctly given by the ratio of the right hand side, with no extra factor.

      Putting this together
      with the angular dependent factors we get
  \bea
   I_{+-  }   &\propto &  { 1 \over     (k_{12} k_{34})^{\half} }   \left[ \left( { k_I^2 \over 4 k_{12} k_{34} } \right)^{i \mu} { \Gamma(-i \mu)^2   \Gamma( \half + s+i \mu)^2 } \Theta_+ +
   \right.
   \cr
   && + \left.
     \left( { k_I^2 \over 4 k_{12} k_{34} } \right)^{ - i \mu}  { \Gamma(i \mu)^2   \Gamma( \half +s - i \mu)^2 } \Theta_- \right]
   \eea

    So far we obtained just one of the integrals. We can obtain the rest of them by using the analytic continuation
    \nref{AnCoSq}. We had derived that formula for the case of the scalar intermediate state, but we expect the same answer for
    a spinning particle since it only depends on the exponential factor in \nref{LateIntr}.
    This leads to an expression similar to \nref{CoScsmallki}
    \bea \la{FourSqSpin}
   \left. I_4^E \right|_{k_I \ll k_{12},~k_{34} }  &\propto &     { 1 \over     (k_{12} k_{34})^{\half} }   \left[ \left( { k_I^2 \over 4 k_{12} k_{34} } \right)^{i \mu}
   ( 1 + i \sinh \pi \mu) { \Gamma(-i \mu)^2   \Gamma( \half + s+i \mu)^2 } \Theta_+ +
   \right.
   \cr
   && + \left.
     \left( { k_I^2 \over 4 k_{12} k_{34} } \right)^{ - i \mu} ( 1 - i \sinh \pi \mu) { \Gamma(i \mu)^2   \Gamma( \half +s - i \mu)^2 } \Theta_- \right]
   \eea
     The angular dependence is all contained within the factors $\Theta_\pm$ given in \nref{HelPh}.
   This $I_4^E$ should be inserted into \nref{JscalDef} in order to get the actual correlator.

     \subsection{Four point function of massless scalars and an intermediate massive field with spin}

   Here we consider external particles which are massless. In this case we have not
   worked out the analog of the operator $O_{12}$ in \nref{O12O34}.
   However, it is still possible to obtain the expression when $k_I\to 0$. The first observation is that the
   angular dependence will be the same as in the previous case. The reason is the following. When we consider
   a three point function of two scalar operators and a general operator with spin, then the ratio of the sizes of different
   helicity components is independent of the dimensions of the two scalars, in the limit $k_I \to 0$. This follows easily from the
   expressions for the fourier transform of three point functions that we derived in \nref{Sqnonan} and \nref{Sqan}. In both of these
   expressions the $\epsilon $ dependence does not mix with the dependence on $\Delta_1$ and $\Delta_2$.
   Therefore the only possible difference relative to our discussion for the conformally coupled scalar can only come from the overall
   phase. This can arise due to the different behavior for the maximal helicity component.

   We have seen that this maximal helicity component had the form given in \nref{MaxHel3}, \nref{as3}.
   For large $\YM$ the function $G_s(\YM)$ has the power law behavior
   \be \la{largepGs}
   G_s \sim  A_s \YM^{ 1 - \Delta -s }  + B_s \YM^{ \Delta - 2 - s}
   \ee
   with $A_s$ and $B_s$ given in \nref{AsBs}. Here we used that the normalization condition at $\YM =-1$ is basically the
   same as for the conformally coupled scalar \nref{NorCond}, see \nref{NorCondxi}.
  Then the power law behavior of the maximal helicity component is $a_s$ given in \nref{as3} is, for large $\YM$,
  \be
  a_s \sim { A_s\over 4 } { \Delta + 1 + s \over \Delta - 3 + s} \YM^{ 3 -\Delta -s}   +
   { B_s \over 4} { \Delta -s - 4 \over \Delta -s }  \YM^{ \Delta   - s}  \la{Expfuc}
  \ee
   This produces simple extra $\Delta $ dependent
   factors   relative to the answer for the conformally coupled scalar.

   Keeping track of these extra factors we can immediatly write the answer for the small $k_I$ limit
   \bea
   \langle \xi_{\vec k_1} \cdots \xi_{\vec k_4} \rangle' &=& { \lambda^2 \over 4 k_1^3 k_2^3 k_3^3 k_4^3 } { k_I^3 J_E}  \la{FourZeroSp}
   \\
   \left. k_I^3 J_E \right|_{k_I \ll k_{12} ~k_{34} } &\propto&
   ( k_{12} k_{34} )^{3\over 2 } \times
\cr
  &\times & \left[
   \left({ k_I^2 \over 4 k_{12} k_{34} } \right)^{i \mu} (1 + i \sinh \pi \mu)  { ( { 5 \over 2 } +s+ i \mu)^2\over ( { 3 \over 2 } +s- i \mu)^2 }{\Gamma(-i \mu)^2   \Gamma( \half + s  + i \mu)^2 } \Theta_+ +
   \right.
   \cr
   && +
   \left.
     \left({ k_I^2 \over 4 k_{12} k_{34} } \right)^{- i \mu} (1 - i \sinh \pi \mu)
                  { ( { 5 \over 2 } +s- i \mu)^2\over ( { 3 \over 2 } +s+ i \mu)^2}
                  {\Gamma(i \mu)^2   \Gamma( \half +s- i \mu)^2 } \Theta_- \right] \notag
   \eea
   where the angular dependent factors are the same as before \nref{HelPh}.
   For $s=0$ this formula does {\it not} reduce to \nref{FourZeroSca}. This is simply because here, in \nref{FourZeroSp}, we have normalized
   the answer to a vertex of the form $\int \xi \nabla_{i_1} \cdots \nabla_{i_s} \xi \sigma_{i_1 \cdots i_s}  $, which reduces to
 $\int \xi^2 \sigma$ for spin zero. On the other hand  in \nref{FourZeroSca} we used the vertex $\int (\nabla \xi)^2 \sigma$.
 The diference is indeed a factor of $[ ({ 3 \over 2 } + i \mu)({ 3 \over 2} - i \mu) ]^2   = m^4$ as expected from \nref{TwoVertexF} (one factor of $m^2$ per vertex).

 \section{Three point function in inflation}

 In the previous sections we have considered correlation functions in de-Sitter space. We have argued that
 they are almost completely fixed by the de-Sitter isometries.
 In this section we consider correlators in an inflationary situation. We will assume a standard slow roll scenario
 with an inflaton field $\phi$ which gives rise to curvature fluctuations $\zeta$.
 In this situation, the de-Sitter isometries are slightly broken by the presence of a background time derivative
 $\dot \phi_0$. This breaks the scale and special conformal invariance of the correlators.
 Since $\dot \phi_0$ is small, this breaking is small and the amount of violation of the symmetries can be controlled.
 In fact, we will see that the symmetries continue to determine the quantities that we are interested in.

In inflation, the fluctuations of the scalar field are well approximated by a massless field $\xi$, $ \phi(t,x) =    \phi_0(t) + \xi(t,x)$.
Then the curvature fluctuation $\zeta $ is given by  (see \cite{Maldacena:2002vr} for the precise definition of $\zeta$)
\be
\la{ZetaXi}
\zeta = - { H \over \dot \phi_0 } \xi
\ee
 When a correlator for $\xi$ does not vanish in
de-Sitter, then the corresponding correlator for $\zeta$ is given by the relation in \nref{ZetaXi}.

\begin{figure}[h]
\begin{center}
\includegraphics[scale=.7]{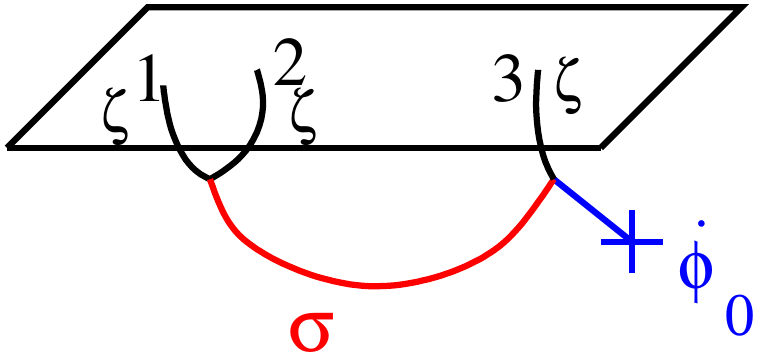}
\caption{ Contribution to an inflationary three point function from an intermediate massive particle. The blue line with a cross is
an insertion of $\dot \phi_0$ which is the homogeneous time derivative of the inflaton.     }
\label{ThreeInflation}
\end{center}
\end{figure}

 The presence of an extra  massive scalar field  gives rise to a four point function for $\xi$ in the de-Sitter case.
 In the inflationary situation it also gives rise to a three point function.
  We can view this three point function as
 coming from the same diagram that was giving rise to the four point function in de-Sitter, but with one external particles set to
 $\dot \phi_0$.  See figure \ref{ThreeInflation}.
 One important difference is that  now we only have three non-zero momenta to play with. In this case
 $\vec k_I = \vec k_3$ and $\vec k_4 =0$.

  \subsection{Intermediate massive scalar field}

  Before discussing the general case, we concentrate on the case of a massive intermediate field.
 We can easily analyze the coupling on the right vertex of figure \ref{ThreeInflation}
    which comes from $\lambda \int  (\nabla \phi)^2 \sigma $  upon replacing one of the fields $\phi\to  \phi_0(\eta)$ and the
    other $\phi \to \xi$.  We get
    \be \la{VertInfl}
    \lambda \int { d \eta \over \eta^2 } [ - \partial_\eta \xi \partial_\eta \phi_0 \sigma ] \longrightarrow
     \lambda \int { d \eta \over \eta^2} [ - k_3^2  \eta e^{ i k_3 \eta } ]\partial_\eta \phi_0 \sigma \longrightarrow
    \lambda k_3^2 \dot \phi_0 \int { d \eta \over \eta^2 } e^{ i k_3 \eta } \sigma
    \ee
    where $\dot \phi_0 = - \eta \partial_\eta \phi_0 $ is the proper time derivative.   We have assumed that $\dot \phi_0$ varies slowly enough
    that we can take it out of the integral. The remaining integral has precisely the same form as the integrals we had when we were
    dealing with the coupling to conformally coupled fields, except that we set $k_4=0$. The vertex on the left side of   figure
    \ref{ThreeInflation} is the same as the one we discussed in the de-Sitter case. It can be represented by acting with an operator
    on the results for the conformally coupled field.
    In summary, the correction to the inflationary three point function due to a massive field has the form
    \be \la{3Inf}
    \langle \xi_{\vec k_1} \xi_{\vec k_2} \xi_{\vec k_3} \rangle = \dot \phi_0(\eta_3^*) { \lambda^2 2^2 k_3^2 \over 8  k_1^3 k_2^3 k_3^3}  O_{12} I_E( k_{12},k_3, k_3)  + {\rm
    other ~diagrams}
     \ee
     where  $I_E(k_{12},k_{34},k_I)$ is the four point function in the de-Sitter case for  external
     conformally coupled scalars, see \nref{JscalDef} and
     $O_{12}$ is given in \nref{OpZeta}.  Here $\eta_3^*$ is the value of $\eta$ such that $\eta_3^* k_3 \sim 1 $, corresponding to the time
     that the mode $k_3$ crosses the horizon.  The other diagrams correspond to two other permutations of the 1,2,3 labels.

     We can now obtain the  small $k_3$ limit of  this expression. This is not identical to the small $k_I$ limit we took in the previous
     section  because
    now, $k_{34} =k_I =k_3$, while before we had assumed $k_I \ll k_{34} $. Only the diagram explicitly described in \nref{3Inf} contributes
    non-analytic terms in $k_I$, the others contribute with analytic terms in the small $k_I$ limit.

    As before, it is easiest to write an  expression for the $I_{+-}$ contribution. This is a special case of  \nref{ProdForm},
    except that now we have $\ZM=1$. In other words, we have
    \be
    I_{+-}(\WM,1) = { 1 \over k_3 } F(\WM) F(1)
    \ee
      We can also write a similar (the same) expression for $I_{-+}$. We can compute the large $\WM$ limit using \nref{LargHy}.
      We can also use
      \be
      F(1) = { 1 \over \sqrt{2} }  { \pi \over \cosh \pi \mu }
      \ee
      For large $\WM$,
      we can find   $I_{++}$ and $I_{--}$ by doing the analytic continuation in
     \nref{AnCoSq} for the $\WM$ variable. For example, we can get $I_{--}(\WM , 1) = -  I_{+-}( e^{ - i \pi} y, 1) $.
     In this way we can find
     \bea
     I_E(k_{12},k_3,k_3)|_{k_3 \to 0}  &= & { 1 \over \sqrt{ k_3 k_{12} } } { 1 \over  \sqrt{2 \pi } } { \pi \over \cosh \pi \mu}
     \left[ \left( {  k_3 \over 2 k_{12} } \right)^{ i \mu} (1 + i \sinh \pi \mu) \Gamma(- i \mu) \Gamma( \half + i \mu) +
     \right.
     \cr
 &+&     \left.
      \left( {  k_3 \over 2 k_{12} } \right)^{- i \mu} (1 - i \sinh \pi \mu)\Gamma( i \mu) \Gamma( \half - i \mu)
      \right]
     \eea
   We can now act with the operator $O_{12}$ using \nref{LargeO} to obtain
    \bea
    &&\langle \xi_{\vec k_1} \xi_{\vec k_2} \xi_{\vec k_3} \rangle'_{k_3 \to 0} ={ \dot \phi_0(\eta_3^*) \lambda^2 \over 2 k_1^3 k_3^3 } J(\mu, {k_3 \over k_{1} } )
    \\
    J(\mu, {k_3 \over k_{1} }) &\equiv &{ 1 \over 4\sqrt{   \pi }}
    { \pi^2 \over \cosh^2 \pi \mu }   \left( {k_3 \over k_{1 } } \right)^{3\over 2 }
    \times
    \cr
    &&~~~~ \times \left[ \left( { k_3 \over 4 k_{1} } \right)^{- i \mu}{(1 - i \sinh \pi \mu )  ( { 5 \over 2 } - i \mu)( { 3 \over 2 } - i \mu) \Gamma( i \mu) \over \Gamma( \half + i \mu)} +
    \right.
    \cr
     && +  \left.
    \left( { k_3 \over 4 k_{1} } \right)^{ i \mu}{(1 + i \sinh \pi \mu )  ( { 5 \over 2 } + i \mu)( { 3 \over 2 } + i \mu) \Gamma( - i \mu) \over \Gamma( \half - i \mu)}
    \right] \la{JDefi}
    \eea
    These are the correlators of the scalars, we can obtain the correlators of $\zeta$ by using \nref{ZetaXi} to obtain
    \bea
     \left.  { \langle \zeta_{\vec k_1}  \zeta_{\vec k_2}  \zeta_{\vec k_3}  \rangle'
    \over 4 \langle \zeta_{\vec k_1} \zeta_{ -\vec k_1} \rangle' \langle \zeta_{\vec k_3}\zeta_{-\vec k_3} \rangle' }
    \right|_{k_3 \to 0 }   &= &- C J(\mu ,{k_3 \over k_{1}} )   \la{FinResScaZ}
    \\
    C &\equiv &  { \dot \phi^2_0(\eta^*_3) \lambda^2  \over 2 H^2  } = {   \epsilon }   M_{pl}^2 \lambda^2 \la{CDefi}
    \eea
    with $J$ defined in \nref{JDefi}. We have also restored the factors of the Hubble constant. $\epsilon$ is the standard slow roll
    parameter, $\epsilon = { \dot \phi^2 \over 2 H^2  M_{pl}^2} $.
     We have  used the standard leading order form for the two point function
    \be
    \langle \zeta_{\vec k} \, \zeta_{-\vec k} \rangle' = { H^4 \over \dot \phi_0^2 } { 1 \over 2 k^3 }
    \ee
  %
    In \nref{FinResScaZ} we have expressed the ratio between the three point function and a product of two point functions.
    It is interesting to compare this to the ratio of two point functions, since that ratio appears in the local non-gaussianity which is
    sometimes used a reference comparison (in that case the ratio in \nref{FinResScaZ} is defined as  $ { 3 \over 5 }  f_{NL}$).

\begin{figure}[h]
\begin{center}
\includegraphics[scale=.5]{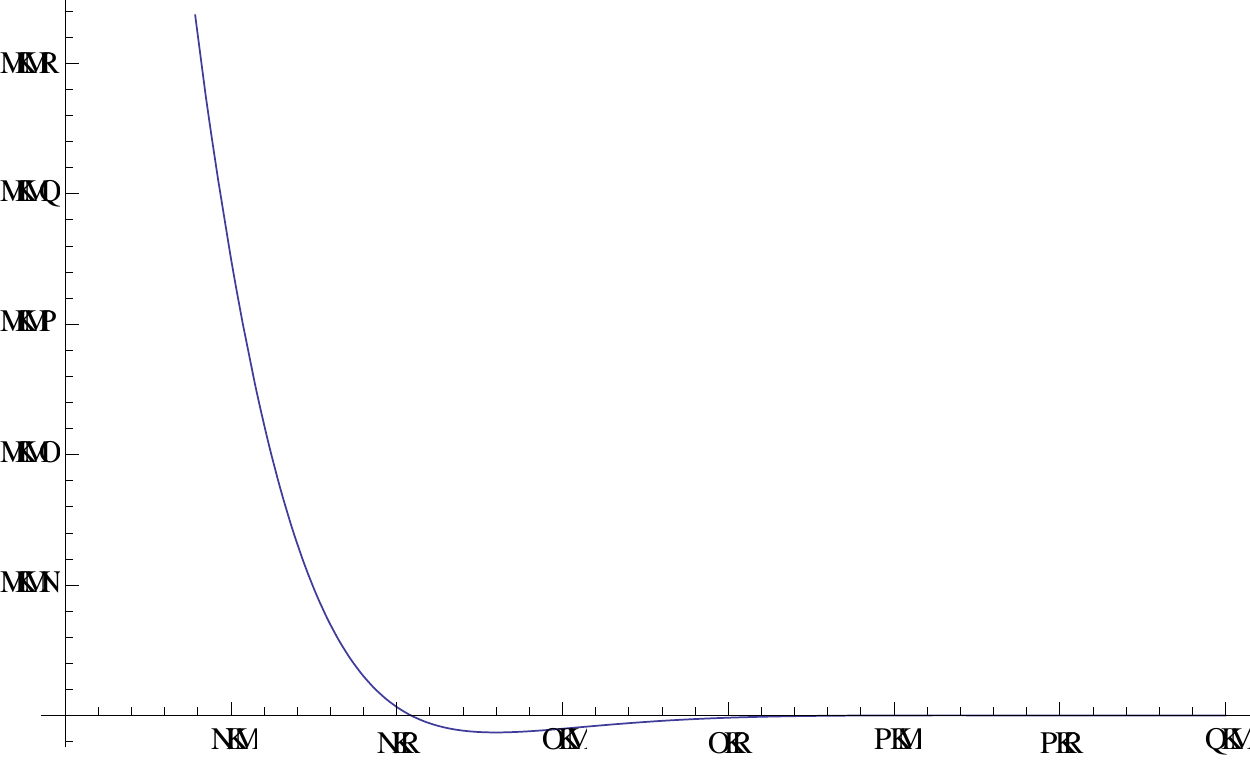} ~~~~~~\includegraphics[scale=.5]{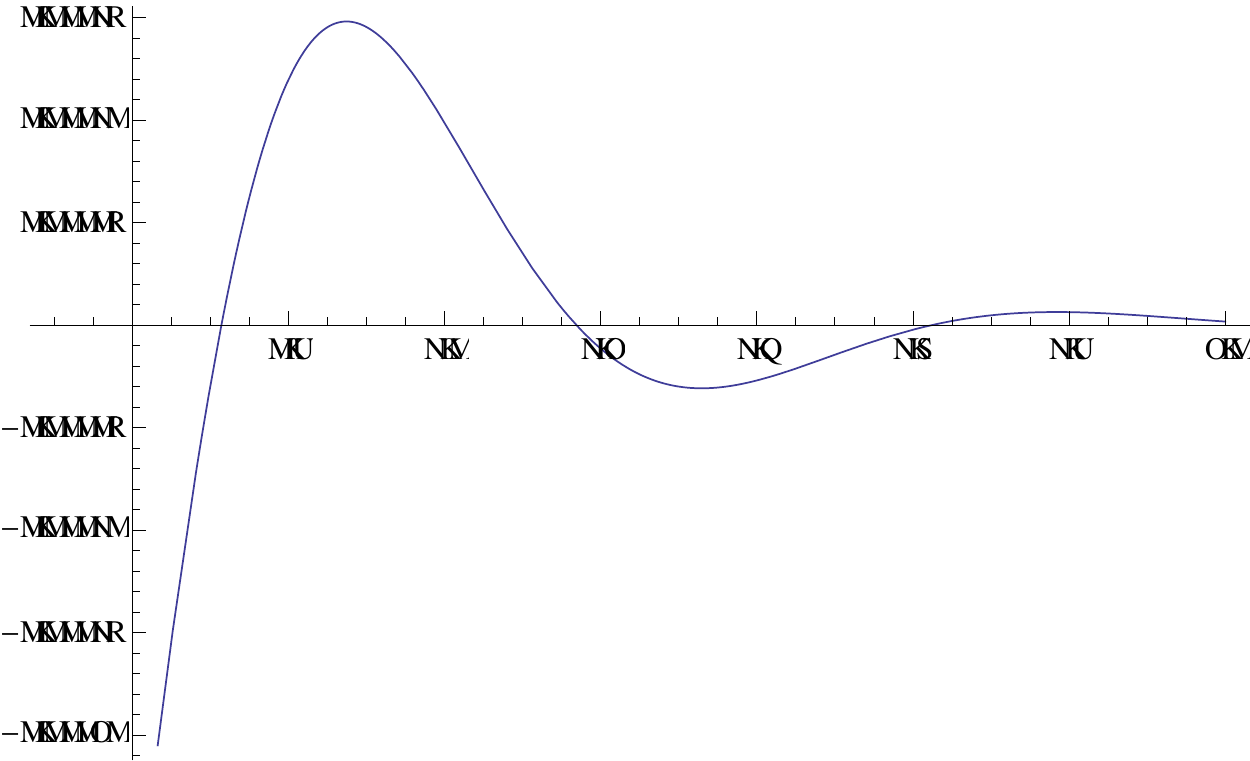}
\caption{ These are plots of the function $J(\mu, {k_3 \over k_1} )$. The left side corresponds to $\mu =1$, the right one to
$\mu =3$. The vertical axis displays  the numerical value of $J$.
 The horizontal axis shows $\log_{10} ({k_1\over k_3} )$ which is the number of decades separating $k_1$ and $k_3$.    }
\label{JPlots}
\end{center}
\end{figure}

    We plot the  function $J$ in \nref{JDefi} for a couple of values of $\mu$  in figure \ref{JPlots}.

    It should be noted that \nref{FinResScaZ} represents an additional contribution to the three point function beyond the
    one present in ordinary inflation \cite{Maldacena:2002vr}. For ordinary single field
    inflation the three point function in this limit is given by the consistency condition and
    gives a constant value for the ratio in the left hand side of \nref{FinResScaZ}.
    However, as explained in \cite{Creminelli:2004yq,Pajer:2013ana},
     this constant value is simply saying that the effects of the long mode are not observable for
    a short mode observer, any physical effect in the single field theory would be down by $k_3^2/k_{1}^2$ \cite{Creminelli:2011rh}.
     Notice that the
    effect we are talking about is bigger, going like $ ( k_3/k_1)^{3\over 2 }$. It is bigger because we now have a second field that
    leads to physical effects. The function in \nref{FinResScaZ} is challenging to measure because of several small factors.
    First we have the slow roll factor   $\epsilon$.
    The function $J$ is also very small. First it is small because of the
    dilution factor $ \left( { k_3 \over k_{1} } \right)^{3 \over 2 } $. Second, as $\mu$ becomes large, there is the exponential
    suppression factor $e^{ - \pi \mu }$. The oscillations are more rapid for larger $\mu$, but the overall amplitude is correspondingly
    smaller.
    All these suppression factors suggest that one is not going to be able to measure this from CMB measurements alone, unless
    $\lambda M_{pl}$ is particularly large. Generically we expect that $\lambda M_{pl} \sim 1 $ or greater.
     With the 12cm tomography \cite{Loeb:2003ya} the prospects  seem to be better, at least in theory.

    \subsection{ Intermediate massive field with spin}

    Here we consider the case of an intermediate massive field with spin. As we have done before, we concentrate first on
    the expression for $I_{+-}$. This expression can be written in terms of products of a three and two point function of a form similar to \nref{FourHelSum}
 \be \la{ThreeHelSum}
     I_{+-} =  \langle O_3 O_3 O_{\Delta , m=0} \rangle' { 1 \over k_I^{2\Delta -3}  I_2( 0) }  \langle  O_{\Delta ,m=0}
       O_3 \rangle' _{\rm Inf}
     \ee
where only the helicity zero part contributes. This the case because in the two point function on the right
hand side there is just a single  momentum.
    The three point function on the left is the same as the de-Sitter one. The scalar operators have dimension three,
as appropriate for the massless scalars.  The two point function is a new object.
 It is the two point function of a
    massless scalar $\xi$ and an operator with spin.
 It would be zero in de-Sitter, but it can be non-zero in an inflationary background.
 This two point function is constrained by lorentz symmetry and scaling to be
    of the form
    \be  \la{tptInf}
     \langle O_3(-\vec k) \epsilon^s . O(\vec k) \rangle'_{\rm Inf} \propto   \dot \phi_0  k^{\Delta -s }  ( \vec \epsilon . \vec k)^s
    \ee
    Note that only the longitudinal piece of the spinning operator contributes. We have put in a factor of $\dot \phi_0$ since
    we expect it to be proportional to this factor. Note that we are imposing scaling symmetry since a constant $\dot \phi_0$
    does preserves the time translation symmetry. Of course, at higher orders we will also have a violation of scaling.
    The equation for  violation of special conformal symmetry was derived in
 \cite{Kundu:2014gxa}
    \be \la{WardId}
    ( \vec b . \vec K_1 + \vec b . \vec K_3) \langle O_3(\vec k_1) \, \epsilon^s . O(\vec k_3) \rangle'_{\rm Inf} =
    \left[ \vec b . \vec \partial_{k_2} \langle O_3(\vec k_1) O_3(\vec k_2) \, \epsilon^s . O(\vec k_3) \rangle'_{\rm dS} \, \right]_{\vec k_2 =0}
    \ee
We can understand this equations as follows. The special conformal symmetry is roughly  a scaling
transformation with a parameter that is linear in $\vec x$, or the form $\vec b . \vec x$. A scaling
transformation is related to a change in $\zeta $. The position dependence of the generator
 translates into a derivative
with respect to momentum which we see in \nref{WardId}. For more details see \cite{Kundu:2014gxa}.

  So we see that  \nref{tptInf} is then related
 to the three point function in a pure de-Sitter background.
    The conditions for  conformal symmetry imply that the highest helicity term   in the three point function obeys a simple equation.
    In addition, the various angular momentum components are related to each other.
In the end the three point function is evaluated at a point where
$ \vec \epsilon. (\vec k_1 - \vec k_2) = \vec \epsilon . \vec k_3 $ (since $\vec k_2 =0$),
  which picks out the zero helicity component,
     and $\XM = \YM =1$, see equation \nref{correla}. Since the highest helicity obeys the simplest equation and
the boundary condition is set at $\YM =-1$, we first solve the equation to get the highest helicity
component at $\YM =1$ and then we use the constraints of conformal symmetry to go down to the
zero helicity component.
    We are
 interested in looking at \nref{WardId} for the particular case that $\vec b . \vec k_3 =0$. This implies that
    only terms of the form $\vec b . \vec \epsilon$ are relevant in both the left and right hand sides of the equation.
    This equation is analyzed and solved in detail in   appendix \ref{TwoPtInflat}.
 The final result is that the two point function can be written
    in terms of the function $G_s(\YM)$ that describes the maximal helicity part of the
three point function \nref{MaxHel3} as
    \be \la{TwoInf}
    \langle O_3 \epsilon^s. O \rangle' =  \dot \phi_0 {  \Gamma(\Delta-1) \over  \Gamma(\Delta -1 + s) } { G_s(1) \over
   [ \Delta(\Delta -3) - s (s-3) ] } k^{\Delta } (\vec \epsilon. \hat   k)^s
    \ee
    and from \nref{ThreeSpin} we get
\be
G_s(1) \propto  \Gamma( \half + s + i \mu ) \Gamma(\half + s - i \mu )
\ee

  We now go back to \nref{ThreeHelSum} and look at the answer for small $\vec k_I$.
We only need to expand the first factor of \nref{ThreeHelSum}  for
large $\YM$.
 The expansion of the first factor for large $\YM$ can be obtained by taking the expansion of the maximal angular momentum
    component given in \nref{Expfuc} and then converting to the zero angular momentum part by using \nref{Sqnonan} and
    \nref{Sqan} to obtain
    \be
     \langle O_3 O_3 O_{\Delta ,m=0} \rangle' \propto P_s(\cos \theta) \left[ A_s{ \Delta + 1 + s \over \Delta -3 +s}
      k_1^{3-\Delta} k_3^{2\Delta-3} I_2(0) + B_s{ \Delta -s - 4 \over \Delta -s}
       k_1^{\Delta} \right]
       \ee
       with $A_s$ and $B_s$ given in \nref{AsBs}. Here $\cos \theta = \hat k_1 . \hat k_3$.
   Together with \nref{TwoInf} we can write the small $k_3$ expansion as
   \bea
   I_{+-} &\propto & P_s(\cos \theta) (k_1 k_3)^{3\over 2} {
   \Gamma( \half + s + i \mu)^2 \Gamma( \half + s - i \mu )^2 \over \Delta(\Delta -3) -s (s-3)  } \times
   \cr
   &&\times \left[ \left( { k_3 \over 4 k_1} \right)^{ i \mu }
   { { 5 \over 2 } + i \mu +s \over  { 3\over 2 } - i \mu -s }
   { \Gamma( {1 \over 2} + i \mu ) \over   \Gamma( \half + i \mu + s) }
   {\Gamma(- i \mu) \over \Gamma( \half - i \mu+ s  ) } +
   \right.
   \cr
   &
   +& \left.   \left( { k_3 \over 4 k_1} \right)^{- i \mu }{  { 5 \over 2 } - i \mu +s \over { 3\over 2 } + i \mu -s }
    { \Gamma( {1 \over 2} - i \mu ) \over   \Gamma( \half - i \mu + s) }
     {\Gamma( i \mu) \over \Gamma( \half   + i \mu+ s  ) }
   \right]
   \eea

  The result for $I_{-+}$ is the same. And the $I_{++}$ and $I_{--}$ contributions can be found in the small $k_3 $ limit by
  using \nref{AnCoSq}. Adding these contributions and converting  to correlators of $\zeta$ we get
  \bea
     \left.  { \langle \zeta(\vec k_1)  \zeta(\vec k_2)  \zeta(\vec k_3)  \rangle'
    \over 4 \langle \zeta(\vec k_1) \zeta( -\vec k_1) \rangle' \langle \zeta (\vec k_3) \zeta(-\vec k_3) \rangle' }
    \right|_{k_3 \to 0 }   & \propto  &- C P_s(\cos \theta)   J_s(\mu ,{k_3 \over k_{1}} )   \la{FinResSpiZ}
    \\
    C &\equiv &  { \dot \phi^2_0(\eta^*_3) \lambda^2  \over 2 H^2  } = {   \epsilon }   M_{pl}^2 \lambda^2
   \eea
   \bea
   J_s(\mu , { k_3 \over k_1} )   &=&
   {\Gamma( \half + s + i \mu)  \Gamma( \half + s - i \mu )  \over  ( s-{ 3 \over 2})^2 + \mu^2  } { \pi \over \cosh \pi \mu } \times
   \cr
   &&\times \left[ \left( { k_3 \over 4 k_1} \right)^{ \32 +i \mu } (1 + i \sinh \pi \mu)
   { { 5 \over 2 } + i \mu +s \over  { 3\over 2 } - i \mu -s }
   {\Gamma(- i \mu) \over \Gamma( \half - i \mu   ) } +
   \right.
   \cr
   &
   +& \left.   \left( { k_3 \over 4 k_1} \right)^{\32 - i \mu } (1 - i \sinh \pi \mu) {  { 5 \over 2 } - i \mu +s \over { 3\over 2 } + i \mu -s }
     {\Gamma( i \mu) \over \Gamma( \half   + i \mu   ) }
   \right]
   \eea
    We have not computed the overall numerical coefficient in \nref{FinResSpiZ}. It is simply an $s$ dependent numerical constant.
    It can be computed easily using the techniques in this paper once the normalization for the higher spin three point interaction
    $ \lambda \int \xi \nabla_{i_1} \cdots \nabla_{i_s} \xi \sigma_{i_1 \cdots i_s } $ is chosen.

\begin{figure}[h]
\begin{center}
\includegraphics[scale=.5]{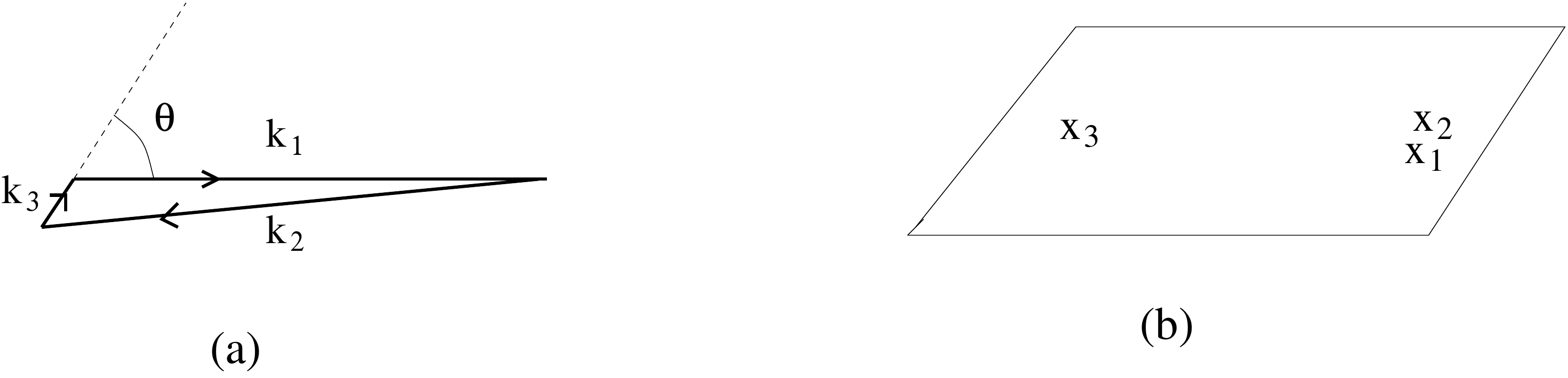}
\caption{ (a) Squeezed momentum contribution where $k_3 \ll k_1 , ~k_2$. Massive fields with spin give rise to a
dependence on the angle $\theta$ between the vectors $\vec k_3$ and $\vec k_1 \sim - k_2$.
(b) In position space, the non analytic terms in $k_3$ give rise to correlators at long distances in position space $|x_{3,2}|, ~|x_{31}| \gg |x_{12}| $.    }
\label{ThreeSqueezed}
\end{center}
\end{figure}

    Note that for $s=0$ \nref{FinResSpiZ} reduces to \nref{FinResScaZ} up to a factor of $(m^2)^2 $ due to the different definition of
    the three point vertex.
   An important point about \nref{FinResSpiZ} is the presence of the angular dependence through the Legendre polynomial
   $P_s(\cos \theta)$, where $\theta$ is the angle between $\hat k_3$ and $\hat k_1$. See figure \ref{ThreeSqueezed}.

 \section{ Comments on loop corrections}

  A theme in this paper has been the idea that the squeezed limit of three point functions or
  four point functions contains information about the quasinormal mode spectrum of the theory in de-Sitter space.
  Our discussion so far has been limited to tree level effects. In this subsection we discuss briefly
  some aspects of loop corrections.

\begin{figure}[h]
\begin{center}
\includegraphics[scale=.5]{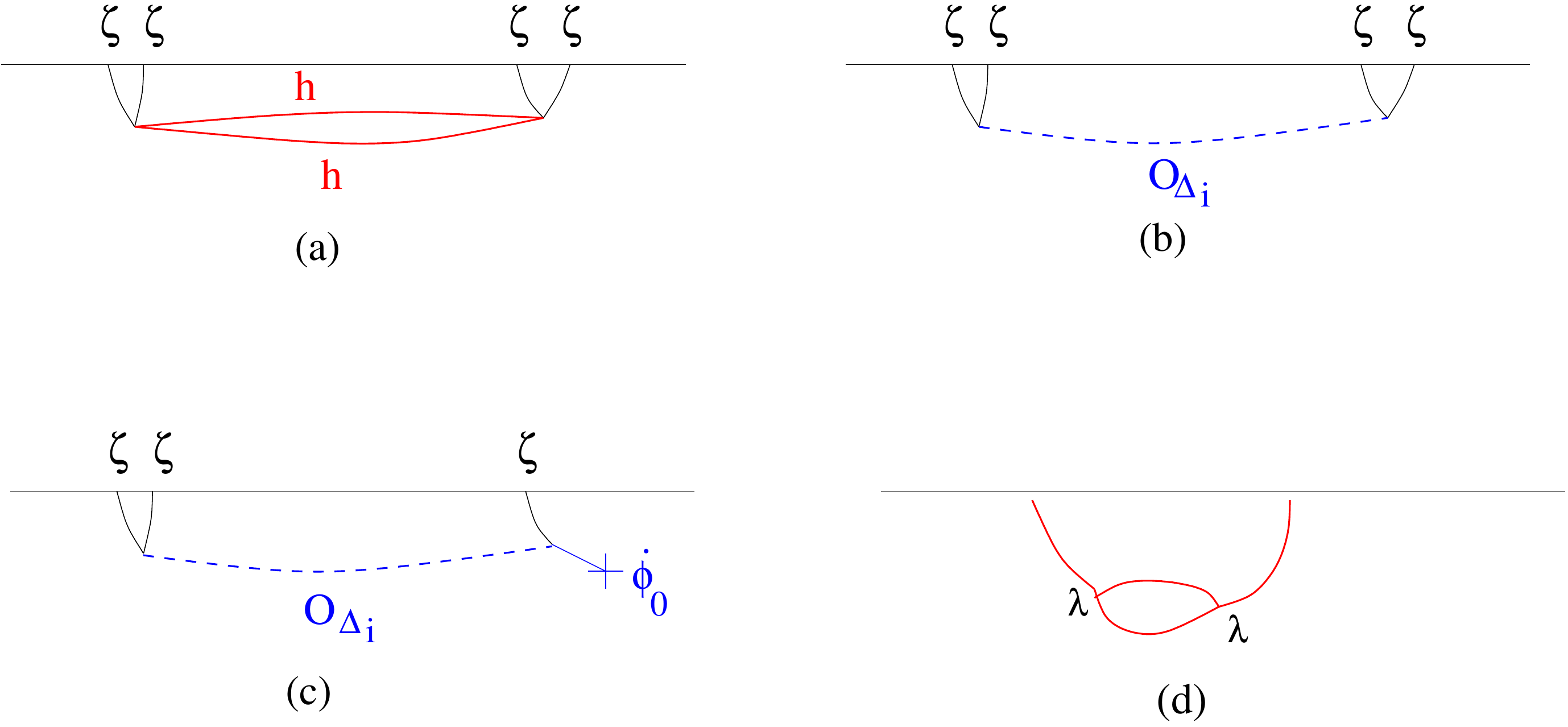}
\caption{ (a)   Example of a one loop diagram arising from the coupling between the inflaton and the Higgs field
$\int (\nabla \phi)^2 h^\dagger h $. (b) Example of a diagram where we have a coupling to a general four
dimensional operator $\int (\nabla \phi)^2 {\cal O}^{(4)}$ and we decompose the four dimensional operator in
terms of three dimensional operators with definite three dimensional conformal dimensions. The correlator
in the OPE regime is governed by these dimensions. (c) When we go to an inflationary background we can
also compute contributions to the three point function. (d) Loop correction to the two point function of a massive field
due to a cubic self interaction.      }
\label{LoopDiagram}
\end{center}
\end{figure}

One effect of loop corrections is the  following shift to the late time behavior of
  two point functions \cite{Marolf:2010zp,Krotov:2010ma} (see figure \ref{LoopDiagram}(d))
    \be
  \alpha \eta^{ { 3 \over 2} + i \mu }  + \beta \eta^{ {3 \over 2} - i \mu }  ~~\longrightarrow \tilde \alpha \eta^{ { 3\over 2} + \delta \gamma+ i (\mu + \delta \mu) } + \tilde \beta \eta^{ { 3\over 2} + \delta \gamma - i (\mu + \delta \mu) }
  \ee
  The term $\delta \mu$ can be interpreted as a one loop shift in the mass of the field. The qualitatively new feature is
  the presence of a nonzero $\delta \gamma >0$, which leads to a faster decay at late times.
   This
  was interpreted in \cite{Marolf:2010zp,Jatkar:2011ju}  as due to the decay of the particle \cite{Bros:2008sq}.
    In flat space the decay might  be forbidden
 by energy conservation, but in de-Sitter it is allowed. The particle can
  decay into two particles of the same mass, and these two particles can fall into the cosmological horizon. When
  all particles have the same mass we get
   $\delta \gamma \sim \lambda^2  e^{ - \pi \mu }$ \cite{Bros:2008sq,Jatkar:2011ju},
  showing  the exponential suppression.
  Note that the presence of this extra term is no problem for the interpretation of each coefficient as an operator
  in the conformal field theory. At three level we had that the two operators were related by $
  \Delta \to 3 -\Delta$. This was a consequence of the fact that the operator obeyed the free field equation in de Sitter.
  Once we include interactions this is no longer the case.
   We can think of the exponents as the quantum mechanical quasinormal mode frequencies,
  which are complex. They were complex at tree level and they continue to be complex now. The only new thing is that
  the correlators are decaying a bit
  faster than what is expected solely from the dilution due to the expansion of the universe.

  \subsection{Coupling to a general operator in de-Sitter}

  Another interesting loop effect arises when we have an intermediate loop. For example, if the inflaton couples to the
  Higgs field during inflation, then we can have a coupling of the form
  $\int ( \nabla \xi)^2 hh^\dagger $. This will then give rise to a loop diagram contribution, see figure \ref{LoopDiagram}(a).
 More generally, we can consider a coupling to a general four dimensional operator
   $ \lambda  \int \varphi^2 {\cal O}^{(4)} $. This operator could be the pair of Higgs fields $h^\dagger h$, for example.
   Then the long distance correlator is determined in terms of the three dimensional late time operators that
   the four dimensional operator ${\cal O}^{(4)}$ gives rise to
   \be \la{SumOp}
   {\cal O}^{(4)} \sim  \sum_j  (- \eta)^{\Delta_j} O_{\Delta_j } (\vec x)
   \ee
   The operators $O_{\Delta_j } (\vec x)$ have definite scaling dimensions under
   three dimensional conformal transformations.
   Note that the operator ${\cal O}^{(4)}$ is an operator in a massive non-conformal theory, so it does not need to have
   a well defined four dimensional scaling dimension. If the operator ${\cal O}^{(4)}$ indeed has
   a  definite four dimensional scaling dimension, then
   we can say that the smallest $\Delta_j$ is equal to the scaling dimension of the four dimensional operator.

  We then focus on the two point functions of the three dimensional operators
   \be \la{Ope3}
   \langle O_{\Delta_j} (\vec x ) O_{\Delta_j}(\vec 0) \rangle = { c_j \over |\vec x|^{ 2 \Delta_j } }
   \ee
   Using  \nref{JpmInt} \nref{Expecv} and \nref{FTPower}  we obtain
   \bea
   I_{j\,\pm \pm} &=& (\pm i )(\pm i) \int^0_{-\infty} { d \eta\over \eta^2} e^{ i k_{12} \eta}  \int^0_{-\infty} { d \eta'\over {\eta'^2}} e^{ i k_{34} \eta'}
   \int d^3 x e^{ i \vec k_I . \vec x } { (-\eta)^{\Delta_j} ( - \eta')^{\Delta_j } \over |\vec x |^{ 2 \Delta_j } } c_j
   \cr
   I_{j\,E} &=& \sum I_{j \, \pm \pm }
   \cr
 I_{j \, E}  &=& \left( { k_I^2 \over k_{12} k_{34} }\right)^{\Delta} { k_{12} k_{34} \over k_I^3 }
     8 \pi \sin (\pi \Delta_j) ( 1 + \cos( \pi \Delta_j)) \Gamma( \Delta_j-1)^2 \Gamma( 2 - 2 \Delta_j) c_j
      ~~~~~~~~ \la{Fourca}
   \eea
   Inserting this expression for $I_E$ into   \nref{JscalDef} we get the contribution to the OPE limit  from the operator $O_{\Delta_i}$ in
   \nref{SumOp}. The full contribution is obtained by summing over all $j$ appearing in \nref{SumOp}.
   This is the contribution in the case that the external particles are conformaly coupled scalars.

    The contribution in the case that we have  external massless scalars coupled as $ \lambda
   \int (\nabla \xi )^2 { \cal O}^{(4)}$ is simply
   given by inserting \nref{Fourca} into \nref{O12O34} and \nref{FMsca}. We get a final answer of the form
   \bea
   \langle \xi_{\vec k_1} \cdots \xi_{\vec k_4} \rangle_{k_I \to 0} &\sim& { \lambda^2 \over 4 k_1^3 k_3^3 k_I^3}
    \sum_j \left( { k_I^2 \over k_{12} k_{34} }\right)^{\Delta_j}
   \times
   \\
&\times &    8 \pi \sin (\pi \Delta_j) ( 1 + \cos( \pi \Delta_j))
   \Delta_j^2(\Delta_j+1)^2 \Gamma( \Delta_j-1)^2 \Gamma( 2 - 2 \Delta_j) c_j ~~~~~~~~~~~
   \eea

   As a more explicit example consider the case that the four dimensional operator is given by the square of two
   massive fields, ${\cal O}^{(4)}  =\sigma^2$, as in the Higgs case.
  We can  read off  the three dimensional operators that appear by taking the expectation value
   $\langle \sigma^2(\eta , \vec x) \sigma^2( \eta',0) \rangle$ which is twice the square of the single field expectation
   value, given in \nref{LongPosTwo}
   \bea
   2 \langle \sigma(\eta,\vec x) \sigma(\eta',0)\rangle^2
    & \sim &  { 1 \over 8 \pi^5 }
   \left[ \left( { \eta \eta' \over |\vec x |^2 }\right)^{ 3 + i 2\mu} \Gamma(-i \mu)^2 \Gamma({ 3 \over 2 } + i \mu)^2 +
   \right. \label{SqPropa}
   \\
   &+& \left.
   \left( { \eta \eta' \over |\vec x |^2 }\right)^{ 3 - i 2\mu} \Gamma(i \mu)^2 \Gamma({ 3 \over 2 } - i \mu)^2 +
   2 \left( { \eta \eta' \over |\vec x |^2 }\right)^{ 3  }
    { ( { 1 \over 4 } + \mu^2) \pi^4 \over \mu^2 \sinh^2 ( 2 \pi \mu) }
   \right] \notag
   \eea
     From here we can see that there are three leading operators appearing, with dimensions
     \be \la{ThreeDim}
     \Delta_1 =3 + 2 i \mu, ~ ~~~~ \Delta_2 = 3 - 2 i \mu,~~~~~~~ ~\Delta_3 =3
     \ee
     The coefficients $c_j$ are simply the factors of
   each power of $  \left( { \eta \eta' \over |\vec x |^2 }\right)^{ \Delta_j}$. Here it is clear that the coefficient $c_j$
   depends on more information than $\Delta_j$ \footnote{For example the third term in \nref{SqPropa} has a
$\mu$ dependent coefficient while it correspond to a $\Delta =3$ case.} .
  The dimensions in \nref{ThreeDim} arise from the addition of two  pairs of particles, the subtraction of two pairs of particles
  or from adding a pair and removing a pair of particles.
  The coefficients are all suppressed by factors of $e^{-2 \pi \mu }$ for
  large $\mu$.

   \subsection{Coupling to a general operator in an inflationary background }

  We can also consider the case of an inflationary background where the OPE contribution arises from a diagram
  as in figure \ref{LoopDiagram}(c). In this diagram only the operator on the left is going to late times and can be expanded as
  in \nref{SumOp}.
  Nevertheless it is still possible to find a simple expression for the two point correlators of the four dimensional operators
  in this limit
  \be
  \langle {\cal O}^{(4)}(\eta , \vec x ) {\cal O}^{(4)}(\eta', \vec 0 ) \rangle|_{\eta \sim 0 } \sim \sum_j  (-\eta)^{\Delta_j}
   {  (-\eta' )^{\Delta_j }\over (  \vec x^2 - \eta'^2 + i \epsilon )^{\Delta_j} }
   \ee
  where we have considered a small $\eta$ but a finite $\eta'$. This is determined by  de-Sitter
  symmetry plus the
  assumed late time behavior \nref{SumOp}.   One easy
  way to understand this is by putting first the operator at some fixed value of $\vec x$ and $\eta'$ and then use the
  de-Sitter isometries to move it to a general location\footnote{More explicitly, set $\vec x =0$, then \nref{SumOp} plus
  the dilatation symmetry implies that the answer is $ ( \eta/\eta')^{\Delta}$. We can then use a special
  conformal transformation
  which moves the bulk point but leaves the position of the boundary point fixed to get the general answer.}.
  The rest of the computation is the same as what we did in the case that we were exchanging an intermediate
  free field. In fact, we do not need to do it again, we can simply  express the answer we got for a free field in terms
  of $\Delta = \32 + i \mu$, and then substitute $\Delta \to \Delta_j$.
  The final result is
  \bea
   { \langle \zeta_{\vec k_1} \zeta_{\vec k_2} \zeta_{\vec k_3 } \rangle'
   \over 4 \langle \zeta_{\vec k_1} \zeta_{-\vec k_1} \rangle'  \langle \zeta_{\vec k_3} \zeta_{-\vec k_3} \rangle' }
&=& - C \sum_{j} \tilde J[ - i(\Delta_j- \32), { k_3 \over k_1} ] \,  { c_j \over
c_{ {\rm free}} (\Delta_j) }  \la{InflGOp}
\\
c_{  {\rm free}}(\Delta) &\equiv & { 1 \over 4 \pi^{5/2} } \Gamma({3 \over 2} - \Delta) \Gamma(\Delta)
\eea
with the following definitions.  $C$  is the same as in \nref{CDefi}.  The  $ c_j$ were defined in \nref{Ope3} and
$c_{  {\rm free}} (\Delta) $ is the coefficient we get when we had the operator ${\cal O}^{(4)}$ being a single massive free field,
expressed in terms
of the dimension.
Finally, the functions $\tilde J$ are obtained from \nref{JDefi} writing
\be
J(\mu , {k_3 \over k_1} ) = \tilde J( \mu , {k_3 \over k_1} ) + \tilde J (-\mu , {k_3 \over k_1 } )
\ee
where $J$ is given in \nref{JDefi} and $\tilde J(\mu,{k_3 \over k_1})$ corresponds to the part of  $J$ containing
the  power   $\left({k_3 \over k_1}\right)^{ \32 + i \mu}$.

It was noted in \cite{Green:2013rd} that for $\Delta_i < 2$, this is the dominant correction to the three point function
and that for $2 < \Delta_i  $ there is a bigger contribution of the form $k_3^2/k_1^2$. However, this term
has the form of a field redefinition in position space, since it is local in $k_3$. Therefore, in principle,
 we can still unambiguously
identify the subleading contribution with a $ 2< \Delta_i $. More physically, we can say that the corrections
of the form $k_2^{2 n}/k_1^{2 n}$ are saying that the statistics of the short modes depend on the
features of the long mode at the location we are measuring the short mode. On the other hand,
a correction which is not analytic in $\vec k_3$ implies that the statistics of the short modes
depend on the shape of the long mode {\it somewhere else}, it is not determined uniquely by the
local value of the short mode.

  Of course, this whole discussion is also valid when the operator ${\cal O}^{(4)}$
  is a conformal operator. In that case, the leading
  three dimensional operator also has dimension equal to the scaling dimension of the four dimensional operator.
  Several aspects of this situation were discussed in \cite{Green:2013rd}. Other simple examples include
the Yang Mills field strength, with $\Delta =4$ or pairs of
  Standard Model fermions   with $\Delta =3$ if they are massless.

\section{Conclusions}

Inflationary cosmology provides us with a natural high energy accelerator.
The late universe represents the detector output of this accelerator.
The presence of new particles, beyond the
inflaton, lead to subtle imprints on the cosmological primordial fluctuations. Of course, the main features in
these
 fluctuations are set by the inflaton.
 The fact that we can approximate inflaton fluctuations by a nearly free field
implies that these fluctuations are very close to gaussian. Self interactions of the inflaton, or interactions between the
inflaton and other particles give rise to non-gaussianities. In fourier
space, these non-gaussianities are a function of the momenta of the fluctuations.
When we look at the simplest three point function, we have a function of a triangle. When this triangle has a very small
side it is interesting to look at the behavior of the correlator as a function of the ratio between the small side and the
other two large sizes $k_3/k_1$. As a function of this ratio the non-gaussianity displays a characteristic power law
behavior which signals the presence of new particles
\cite{Chen:2009zp,Baumann:2011nk,Assassi:2012zq,Noumi:2012vr}.
 This power law behavior is a generic feature
of correlation functions for quantum field theory in de-Sitter space. The particular powers that appear
seem to be connected to the quasinormal modes for the quantum field theory in the static coordinate
patch. We have not given a proof of this fact, but it is true in the examples we considered and it seems
to be a reasonable expectation.

Massive particles with masses of order $H$ give rise to interesting signatures in the
non-gaussianities. These signatures arise for very reasonable couplings which we generically expect
to be present.   They contain information about the mass and the spin of
the particle.
 When $m > { 3 H \over 2}$  there are oscillations as a function of the ratio between the
 scales. These oscillations  arise due
to the quantum mechanical nature of the problem and the fact that we are doing
an interference experiment. It is the ``cosmological double slit experiment''.
The oscillation frequency tells us about the mass. The phase of the oscillation is
also fixed in terms of the mass \cite{Noumi:2012vr}.  Other oscillatory effects were discussed in
\cite{McAllister:2008hb,Behbahani:2012be}  and
they differ because the oscillations are due to oscillatory features in the potential
which are in phase with the background evolution. In other words, in those models the oscillations depend
on the value of $k$ while here they depend only on the ratio between two momenta.
In principle, with enough data, one can   distinguish between these two types of oscillatory behavior.

In terms of measurability, if the couplings are Planck suppressed, then it seems impossible to measure this
through the CMB or large scale structure. (See e.g. \cite{Dalal:2007cu,Matarrese:2008nc,Slosar:2008hx,Baumann:2012bc}
 for a discussion of measuring these effects via large scale structure.)
 But it might be possible using the 21cm tomography \cite{Loeb:2003ya}.
It would be nice to analyze   what range of masses and
couplings are detectable.

We should stress that the effects we have discussed in this paper
are not the {\it largest} non-gaussianities.
They are the effects that most clearly reveal the presence of new particles.
It is likely that the non-gaussianity will be discovered first, and only then, a more precise study
will reveal the new particles.

In this paper we have worked in the approximation of a quantum field theory in de-Sitter space. In this
case the conformal symmetry is exact and its action is very clear. If we consider a theory with dynamical
gravity, the situation is slightly less clear because local operators are not proper gauge invariant observables.
Also the correlators are not de-Sitter invariant in the sense we discussed here \cite{Ghosh:2014kba}.
 This is due to the
fact that the graviton propagator cannot be chosen to be de-Sitter invariant. This lack of invariance should
cancel out in proper gauge invariant observables. But the standard four point function we have been
considering is not a proper gauge invariant observable. It was shown in \cite{Ghosh:2014kba}
 how to modify the action
for the conformal generator to make the four point function conformal invariant. It would be nice to
see if a statement like \nref{3ptExin} can make sense to higher orders in the gravitational perturbation theory.

It would also be nice to see how new particles affect correlation functions with external gravitons. In
other words, we could also use non gaussianities that involve tensor modes in order to detect new
particles.

{\bf Acknowledgements }

We thank M. Porrati and M. Zaldarriaga for discussions.
We are supported in part by U.S. Department of Energy grant
de-sc0009988.

\appendix

\section{ Derivation of the two point function in momentum space}
 \lb{TwoPtApp}

 Let us write down the special conformal generator in momentum space.
 Its action on a scalar operator was written in \nref{SpecConfP}. To describe its
 action on an operator with spin, it is conventient to introduce a null vector
 $\vec \epsilon$, $\vec \epsilon . \vec \epsilon =0$ and to contract the operator
 with this vector: $ \epsilon^s . O_s \equiv \epsilon_{i_1} \epsilon_{i_2} \cdots \epsilon_{i_s} O_{i_1 i_2\cdots i_s} $.
 Here $O_{i_1 \cdots i_s}$ is a traceless symmetric tensor. Our choice of $\epsilon$ produces also a traceless symmetric tensor
 when we take $s$ factors of $\vec \epsilon $.
 Then the action of the special conformal generator is given by
  \be \la{SpecConfSpin}
  \vec b . \vec K = - ( \Delta -3) 2 \vec b . \vec \partial_k -  [  ( \vec b . \vec k)( \vec \partial^2_k) - 2( \vec k . \vec \partial_k )(\vec b . \vec \partial_k )]
  + 2 [ (\vec \epsilon . \vec \partial_k)(\vec b. \vec \partial_{\epsilon} ) - ( \vec b . \vec \epsilon) ( \vec \partial_k . \vec \partial_\epsilon ) ]
  \ee
 When we take the derivatives with respect to $\epsilon$ we can treat its three components as independent. The error we make by not taking into
 acount the $\epsilon^2 =0$ condition cancels out at the end. This is easily seen since the last bracket vanishes when acting on a term of the form $\epsilon^2$.

 We now consider a two point function and we make an ansatz of the form
 \bea \la{TwoPtAn}
\langle \epsilon^s . O_s \, \tilde \epsilon^s O_s \rangle'  &= &   k^{ 2 \Delta - 3 } h  ~,~~~~~~~~~ h = \sum_{i=0}^s c_i \rho^{ i }  (\alpha \beta)^{s-i}  ~,~~~~~~
\\
 &&\alpha \equiv { \vec \epsilon . \vec k \over k }  ~,~~~~~~~~ \beta
 \equiv  { \vec { \tilde \epsilon } . \vec k \over k } ~,~~~~~~ \rho \equiv \vec \epsilon . \vec { \tilde \epsilon }
\eea
Note that $h$ is dimensionless. When we act with the conformal generator we think of the momentum
in \nref{TwoPtAn} as that of the first operator.
When we act with the conformal generator \nref{SpecConfSpin}  we can move
through the factor of $k^{ 2 \Delta -3 } $ so that now the operator only acts on $h$. Note that
 the part of the operator acting purely on $k^{ 2 \Delta -3}$
vanishes due to the usual conformal invariance condition for scalar operators.
We only have to worry about the case where only one of the derivatives of the operator \nref{SpecConfSpin} acts on
$k^{ 2 \Delta -3}$ and the other acts on $h$, as well as the case where both act on $h$.
Note that the term $ \vec k . \vec \partial_k $ measures the total power of $k$ and it can be
commuted past the $\vec b . \vec \partial_k$, producing a $-(\vec b . \vec \partial_k)$ term in the process.
  We then use that $\vec k . \vec \partial_k$ produces a simple factor of $(2 \Delta -3)$.

We will also assume that we choose $\vec b$ in such a way that
\be \la{Condb}
\vec b . \vec k =0 ~,~~~~~~ \vec b . \vec \epsilon =0 ~,~~~~~~ \vec b . \vec {\tilde \epsilon} \not =0
\ee
This condition implies that the term with $( \vec b . \vec k )$ in  \nref{SpecConfSpin} does not
contribute. We then obtain the
  following expression for the conformal
invariance equation
\be
\left[ (\Delta -1) ( \vec b . \vec \partial_k) + ( 2 \Delta -3) { (\vec \epsilon .\vec  k )\over k^2}  ( \vec b . \vec \partial_\epsilon ) +
 ( \vec \epsilon . \vec \partial_k )( \vec b . \vec \partial_\epsilon)   \right] h =0
\ee
Here we have used \nref{Condb}  in order to drop various terms, includding the last term in the spin factor
in  \nref{SpecConfSpin}.
Now we use the following properties of $\alpha, \beta, \rho$.
\bea
& & \partial_{\vec k} \alpha = { 1 \over k } ( \vec \epsilon - \hat k \alpha ) ~,~~~~~~
\partial_{\vec k} \beta = { 1 \over k } ( \vec {\tilde \epsilon}  - \hat k \beta) ~, ~~~~
\partial_{\vec k} \rho =0
\\
&& \partial_{\vec \epsilon } \alpha = \vec k ~,~~~~~~~ \partial_{\vec \epsilon } \beta =0 ~,~~~~~~
\partial_{\vec \epsilon } \rho = \vec {\tilde \epsilon }
\eea
Using these properties it is possible to show that \nref{SpecConfSpin}   becomes
\bea
 0&=&\left[ ( \Delta -1) \partial_\beta + ( 2 \Delta -3) \alpha \partial_\rho + [ - \alpha^2 \partial_\alpha + ( -
\alpha \beta + \rho ) \partial_\beta ] \partial_\rho    \right] h
\eea
Writing $h = (\alpha \beta)^s F( { \rho \over \alpha \beta} )$ we find that the above equation becomes
\bea
 & & (y-2)  y F''  +  [(\Delta -s) ( 2 -y) -1 ] F' + s ( \Delta -1) F =0 \longrightarrow
\\ && \longrightarrow  F = ~_2F_1(\Delta -1, -s, \Delta -s-{ 1 \over 2} ; { y \over 2})  ~,~~~~~ y \equiv { \rho \over \alpha \beta }
\la{HyperTwoPt} \eea
This equation is of the hypergometric form and we are interested in the solution that is a polynomial
for integer $s$\footnote{
  It is the Jacobi polynomial  $P_s^{ (\Delta -s -3/2 ,-1/2)}(1-y) $.}.
Writing $\vec \epsilon $ and $\vec {\tilde \epsilon}$ as in \nref{EpEptil}, with $\vec k = (0,0, k)$,
 we find that
$y = 1 + \cos( \psi -\psi')$. Inserting this in \nref{HyperTwoPt} we can see that it is proportional to
 \bea
 h &=& \sum_{m=-s}^s e^{ i m \chi} { (-1)^{s-m} (2 s)! \over (s-m)! (s+m)!}   { \Gamma(\Delta   - 1 -m ) \over \Gamma(\Delta -s -1) } { \Gamma(\Delta     -1 +m) \over \Gamma(\Delta +s -1)}
\\
&=& e^{ - i s \chi}  ~_2F_1( \Delta -s-1 , - 2 s , 2-s -\Delta ; - e^{ i \chi} )  \la{TwoHyApp}
\eea
with $e^{ i \chi} \equiv e^{ i { (\psi - \psi')  }  } $.

In position space the two point function has a very simple expression
\bea
\langle 2pt \rangle &\sim & { [ X_1 . X_2 (V_1 . V_2) - ( X_1 . V_2)(X_2.V_1) ]^{ s} \over (X_1.X_2)^{\Delta+s} }
\\
&& \sim { \left[ x_{12}^2 ( \epsilon_1 . \epsilon_2) -
2  {( \epsilon_1 . x_{12} ) (\epsilon_2 .
x_{12} )  } \right]^s \over |x_{12}|^{2 \Delta + 2 s} } \la{PosTwoPt}
\\
X& =& (X^+,X^-,\vec X) = ( 1 , -\vec x^2 , \vec x ) ~,~~0= X^+ X^- + \vec X^2 ~,~~~~V = ( 0 , - 2 \vec \epsilon . \vec x , \vec \epsilon ) \notag
\eea
where the first expresion is written in embedding coordinates, summarized in the last line, with $V. X = V^2 =0$.
We have checked in some cases that the Fourier transform of \nref{PosTwoPt} does indeed match \nref{HyperTwoPt}.

Let us record here the momentum space expressions for $s=1$
\be
\langle 2 pt \rangle_{s=1} =  k^{ 2 \Delta - 3 } \left[   e^{ - i \chi} + 2 { (2- \Delta ) \over \Delta -1 } + e^{ i \chi} \right]
\ee
The expression for $s=2$ was given in \nref{SpinTwoEx}.

When $\Delta ={ 3 \over 2 }$ we see that the phases simplify and we get a result that the angular dependence goes as $h \propto ( \vec \epsilon . \vec{ \tilde
\epsilon } )^s $. We think that this is due to the fact that in this case, we are getting that the prefactor $k^{ 2 \Delta -3 } \to 1$ so that the Fourier transform is
completely local. In this case we expect that the non-local parts in $x$ should arise from a term that has a $\log k$ in momentum space. The simple angular dependence
arises because the local term should have a simple angular dependence.

Another interesting limit is $\Delta \to \infty$. This corresponds to the massive case.  Note that since \nref{TwoHyApp} is a rational function of $\Delta$,
 it does not matter in which direction in the complex plane we take $\Delta$ to infinity. Then the result becomes
 \be \la{LargeDelTwo}
 h \propto ( \vec  \epsilon . \vec {\tilde \epsilon }  - 2 \, \vec \epsilon .   \hat k \, {\vec {\tilde \epsilon} }.\hat k )^s \propto ( \cos(\psi -\psi') -1)^s
 \ee
The combination that appears here is similar to the one appearing in the numerator in position space,
\be \la{invPos}
 ( \vec  \epsilon . \vec {\tilde \epsilon }  - 2 \, \vec \epsilon .   \hat x \, {\vec {\tilde \epsilon} }.\hat x )^s
\ee
see \nref{TwoSpinPos}, where $\hat x = { \vec x_{12} \over |\vec x_{12} | } $. This combination
is related to the following fact. If we have two points on the boundary. Then they are left invariant
by an $SO(3)$ subgroup of the conformal group. That $SO(3)$ subgroup acts like ordinary rotations around
each of the two points. However, one of the rotations is conjugated by a reflection along the
line that joins the two points.  Notice that
 \be
 \vec  \epsilon . \vec {\tilde \epsilon }  - 2 \, \vec \epsilon .   \hat x \, {\vec {\tilde \epsilon} }.\hat x = \vec  \epsilon . R_{\hat x} \vec {\tilde \epsilon }
 \ee
 where $R_{\hat x}$ is a reflection along the $\hat x$ direction.

\section{ Three point function with a pair of $\Delta =3$ operators and a general scalar operator}
\la{ConfInvDel3}

In this appendix we obtain  the expression for the correlator of two scalar operators with dimension
$\Delta_1 = \Delta_2 =3$  and one
scalar operator with general dimension $\Delta$ by using the special
conformal invariance conditions in momentum space.
Using \nref{SpeConfSca} we can write
\be \la{ConfIDel3}
\left\{0=
 \sum_{i=1}^2 (b.\vec k_i) [ -2 { 1\over k_i } \partial_{k_i} + \partial_{k_i}^2 ]  + (b. \vec k_3) [ -2(\Delta -2) { 1\over k_3 } \partial_{k_3} + \partial_{k_3}^2 ]
 \right\} k_3^{\Delta } F( { k_1\over k_3},{k_2 \over k_3} )
 \ee
 Picking first $\vec b$ so that $\vec b . \vec k_3 =0$ and using the momentum conservation condition, we end up with the equation
\bea
&& [ -2 { 1\over k_1 } \partial_{k_1}   + 2  { 1\over k_2 } \partial_{k_2}+ (  \partial_{k_1}^2 - \partial_{k_2}^2 )] F =0
\\
&& 0=
[  2( \XM \partial_\YM - \YM \partial_\XM) + (\YM^2 -\XM^2) \partial_\XM \partial_\YM ] F =0 ~,~~~~~~
\XM = { k_1 - k_2 \over k_3} ~,~~\YM= { k_1 + k_2 \over k_3 } ~~~~~~~~
\la{12perpEqn}
\eea
 Defining  $Q = \XM^2$ and $P = \YM^2$, then we find that \nref{12perpEqn} becomes
\be \la{nefeqg}
0=  [ ( \partial_P - \partial_Q) + (P-Q) \partial_P \partial_Q ] F =0
\ee
We take then $\partial_Q$ of the above and get
\be
( -1 + (P-Q) \partial_P ) \partial_Q^2 F =0
\ee

Then this implies that   $F = u_0(P) + Q u_1(P)  + P w_1(Q) + w_0(Q)$.
In other words, $F$ is a polynomial of up to degree 1 in $Q$ or $P$.\footnote{ If we had assumed that the operators at 1 and 2 both have dimensions
$ \Delta_e$, we would have obtained that it is a polynomial of degree $  \Delta_e -2$ in $Q$, when this quantity is an integer.  Otherwise,
we   get a more complicated function. }
Inserting this now into \nref{nefeqg} we get that the variables separate
and we get
\be
\partial_P u_0 + ( P\partial_P -1) u_1 = \lambda = \partial_Q w_0 + ( Q \partial_Q -1) w_1
\ee
We will be interested in solutions with a simple $Q$ dependence. Thus we set $w_1 = w_0=0 = \lambda$.
Then we see that we can determine $u_0$ from $u_1$.
We can also write
\be
\la{condig}
F = U(\YM) + { 1 \over 4 } (\YM^2 -\XM^2) G(\YM) ~,~~~~~~~~~~~~ U'(\YM) =- \YM G(\YM)
\ee

We can now take $\vec b $ along $\vec k_3$ in \nref{ConfIDel3}  to obtain
\bea
  0 &=& -(\Delta -3)\Delta (\YM^2-\XM^2) F + 2 \XM ( -2-\YM^2 +\XM^2) \partial_\XM F + (\YM^2-\XM^2 )( \XM^2-1)  \partial_\XM^2 F  +
  \cr
  &&+  2 \YM( 2 + \XM^2-\YM^2  ) \partial_\YM F
  + 2 \YM \XM (\YM^2 -\XM^2) \partial_\XM \partial_\YM F + ( \YM^2 -\XM^2)(  \YM^2-1)  \partial_\YM^2 F
  \eea

We can now insert the expansion \nref{condig}.
 We then find a polynomial in $\XM$ whose highest coefficient is $\XM^2$, once we use the equation for the derivative of $U$ in \nref{condig}.
  This
highest coefficient gives the equation
\be \la{equfih}
( \YM^2-1)G''(\YM) + 2 \YM G'(\YM) - ( \Delta^2 - 3 \Delta + 2 ) G(\YM)=0
\ee
which is the same as the equation for the $\Delta_1=\Delta_2 =2$ case \nref{CoCouSc}.
The  subleading coefficient gives
\be \la{solvt}
U(\YM) ={ ( \YM^2 -1)( G(\YM) - \YM G'(\YM) ) \over \Delta(\Delta -3) }
\ee
This equation is compatible with \nref{condig} \ and \nref{equfih}.
Thus this the final solution, we solve \nref{equfih}  and determine $U$ from \nref{solvt}.

 In the particular case of the $ (\nabla \zeta)^2 \sigma $ interaction, we have seen that we can get the answer
 by acting with the operator in   \nref{OpZeta} on the result for a conformally coupled scalar with an $\int \varphi^2 \sigma$ interaction.
 This then produces the expression
\be \la{Del3ExprConf}
 G(\YM) =- 2f  + 2\YM f' + (\YM^2-1) f''   =  \Delta(\Delta-3) f  ~,~~~~~~ U(\YM) = (\YM^2-1) ( f - \YM f')
 \ee
 where $f = G_{\rm conf}/2$ where $G_{\rm conf}$ is the Fourier transform when the two scalar operators have $\Delta =2$
 (see \nref{CoCouSc}).
 This is compatible with \nref{solvt} and it gives the properly normalized interaction with the vertex in \nref{TwoVertexF}.

\section{ Thee point functions of two scalars and a general spin operator}
  \la{ThreeMomSpin}

\def\Malpha{\gamma}
\def\Mbeta{\delta}

In this section we present the derivation of the three point functions with two scalar operators with special dimensions and a general
operator with spin with general dimension.
Of course, the expressions of the three point functions are simple  in position space, see \nref{ThreePos}.
We present the derivation in momentum space, which, in particular is giving us the expressions for the Fourier transform of \nref{ThreePos}.

We consider the correlator
\bea \la{correla}
 \langle O_{\Delta_e} O_{\Delta_e} \epsilon^s O_\Delta \rangle &= &|k_1 + k_2|^{ \Delta + 2 (\Delta_e -3) -s }   \sum_{i=0}^s \Malpha^i \Mbeta^{s-i} a_i(\XM, \YM)
 \\
 &&\XM \equiv { |k_1| - |k_2| \over |k_1 + k_2| } ~,~~\YM \equiv { |k_1| + |k_2| \over |k_1 + k_2| } ~,~~ \Malpha =    \vec \epsilon . (\vec k_1 - \vec k_2)
 ~,~~~\Mbeta =    \vec \epsilon . (\vec k_1 + \vec k_2 )  \notag
\eea
Since we have written all momenta in terms of $\vec k_1 $and $\vec k_2$, we can forget about the action of the special conformal generator on the third operator.
We now consider a vector $\vec b$ that obeys
\be\la{propofbc}
\vec b . \vec k_1 =0 ~,~~~~~ \vec b. \vec k_2 =0 ~,~~~~~ \vec b . \vec \epsilon \not =0
\ee
Then the action of the conformal generator can be written as
\bea
\vec b. \vec K_1 + \vec b . \vec K_2 &= &   ( \vec b . \vec \partial_{k_1} +  \vec b. \vec \partial_{k_2} ) ( -2(\Delta_e -2) +  \vec k_1 . \vec \partial_{k_1} +
 \vec k_2 . \vec \partial_{k_2} ) +
( \vec b . \vec \partial_{k_1} -\vec b. \vec \partial_{k_2} ) (   \vec k_1 . \vec \partial_{k_1} - \vec k_2 . \vec \partial_{k_2} )
\cr
&= &  (\Delta -2)  ( \vec b . \vec \partial_{k_1} + \vec b.\vec \partial_{k_2} )   +
( \vec b . \vec \partial_{k_1} - \vec b.\vec \partial_{k_2} ) (   \vec k_1 . \vec \partial_{k_1} - \vec k_2 . \vec \partial_{k_2} )
\eea
where we used the overall scaling dimension of \nref{correla}.
We will need the following equations
\bea
& &( \vec b . \vec \partial_{k_1} - \vec b.\vec \partial_{k_2} ) \Malpha = 2 (\vec \epsilon . \vec  b) ~,~~~~
( \vec b . \vec \partial_{k_1} -\vec b. \vec \partial_{k_2} ) \Mbeta =0
\cr
& & ( \vec b . \vec \partial_{k_1} + \vec b.\vec \partial_{k_2} ) \Malpha = 0  ~,~~~~( \vec b . \vec \partial_{k_1} +\vec b. \vec \partial_{k_2} ) \Mbeta =
2 (\vec \epsilon . \vec b)
\cr
&&( \vec k_1 . \vec \partial_{k_1} - \vec k_2 . \vec \partial_{k_2} )\Malpha = \Mbeta ~,~~~~
( \vec k_1 . \vec \partial_{k_1} - \vec k_2 . \vec \partial_{k_2} )\Mbeta = \Malpha
\cr
&&( \vec k_1 . \vec \partial_{k_1} - \vec k_2 . \vec \partial_{k_2} ) |k_1 + k_2 | = |k_1 + k_2 | \XM \YM
\cr
&&( \vec k_1 . \vec \partial_{k_1} - \vec k_2 . \vec \partial_{k_2} ) \XM=  \YM(1-\XM^2) ~,~~~~
( \vec k_1 . \vec \partial_{k_1} - \vec k_2 . \vec \partial_{k_2} )\YM = \XM(1-\YM^2)
\eea
We then obtain
\bea \la{conengegd}
 0&=&\left\{
\partial_\Malpha \left[ \YM(1-\XM^2) \partial_\XM + \XM(1-\YM^2) \partial_\YM + ( \Delta +2(\Delta_e-3)-s) \XM \YM + \Mbeta \partial_\Malpha + \Malpha \partial_\Mbeta \right]
+ (\Delta -2) \partial_\Mbeta \right\} \times
\cr
& &\times \sum_{i=0}^s \Malpha^i \Mbeta^{s-i}
   a_i (\XM,\YM)
\eea
which implies the following recursion  relation for the $a_i$ coefficients
 %
\bea
a_{i-1} &= & - { i  \over  (s-i+1) (\Delta - 2 + i )} \left[ \left\{\XM   (1-\YM^2) \partial_\YM  + \YM(1- \XM^2) \partial_\XM + (\Delta +2(\Delta_e-3) -s) \XM \YM \right\}  a_i +
\right.
\cr
&& \left. +  
(i+1) a_{i+1}
\right]
 \la{recreserlgd}
\eea
This recursion relation determines all the $a_i$ coefficients once we know $a_{s}(\XM, \YM)$. (We should set $a_{s+1} =0$ in \nref{recreserlgd}).
Notice that $a_s$ corresponds to the highest helicity component around the $\hat k_3$ axis. The crucial point is that conformal symmetry relates the
various helicity components.

Note that we have not yet imposed all the special conformal invariance conditions, only the subset in \nref{propofbc}.
The other conditions imply equations that determine $a_s$.
These equations can be most easily obtained by picking $\vec b$ so that it obeys
\be
\la{borthep}
\vec b . \vec \epsilon =0
\ee
 This still leaves two independent
components for $\vec b$. In order to write the remaining equations is it convenient to go back to the ansatz \nref{correla} and replace
$\vec k_1 + \vec k_2 = - \vec k_3$. In that way, the coefficients depend only the $|k_i|$, $i=1,2,3$ and the only dependence on the vectors is
in $\Malpha$ and $\Mbeta$. One then finds that the $\epsilon$-independent part of the conformal generators \nref{SpecConfSpin} does not act on
$\Malpha$ or $\Mbeta$ (there is a cancellation between different terms in \nref{SpecConfSpin}). Finally, the $\epsilon$ dependent piece in
\nref{SpecConfSpin} has a simple action.
The result is that we get two equations, one per $b$ component that obeys \nref{borthep},  of the form
\bea
0&=& \left[ \half ( \tilde K_1 - \tilde K_2 ) - 2 \Mbeta \partial_\Malpha { 1 \over k_3 } \partial_{k_3} \right]  |k_3|^{ \Delta + 2 (\Delta_e -3) -s }
 \sum_{i=0}^s \Malpha^i \Mbeta^{s-i} a_i(\XM, \YM)\la{Qe1}
 \\
0&=& \left[\tilde K_3 - \half ( \tilde K_1 + \tilde K_2) + 2 \Mbeta \partial_\Mbeta { 1 \over k_3 } \partial_{k_3}
 \right]  |k_3|^{ \Delta + 2 (\Delta_e -3) -s }
 \sum_{i=0}^s \Malpha^i \Mbeta^{s-i} a_i(\XM, \YM) \la{Qe2}
 \\
 && \tilde K_i \equiv  - 2 (\Delta_i -2) { 1 \over k_i} \partial_{k_i} + \partial_{k_i}^2  ~,~~~~~~~\XM = { k_1 - k_2 \over k_3} ~,~~~~~\YM= {k_1 + k_2 \over k_3 }
 \eea
 Here $\tilde K_i$ is the same operator we encountered when we considered the action of special conformal generators on functions that depended
 only on $k_i= |\vec k_i|$. The $\partial_{k_i}$ derivatives in \nref{Qe1}, \nref{Qe2}, by definition,   do not act on $\Malpha,~\Mbeta$.
 An  important property of the equations \nref{Qe1},  \nref{Qe2} is that they do not mix the highest helicity component $a_s$ with the rest of the
 functions. Therefore we get two closed equations for the highest helicity component which are
 similar to the equations we would get for a scalar correlator, but different in detail because of the extra power of $k_3^{-s}$ in the above ansatz.
 The two equations are
 \bea \la{FEqus}
0&=&\left[  2(\Delta_e-2) ( \XM \partial_\YM - \YM \partial_\XM ) + (\YM^2 - \XM^2) \partial_\XM \partial_\YM \right] a_s(\XM ,\YM)
 \\
 0 &=& \left\{ -[ (\Delta -3)\Delta - 4 \Delta_e^2 +18 \Delta_e + 4 \Delta_e s -s(s+9) -18] (\Delta_e-3)(\YM^2-\XM^2)   - \right.
 \cr
 &&  2 \XM[ 2\Delta_e -4  -(1+s) \YM^2 + (5 - 2 \Delta_e + s) \XM^2  ] \partial_\XM   +
  \cr
  &&+  2 \YM[ 2\Delta_e -4  -(1+s) \XM^2 + (5 - 2 \Delta_e + s) \YM^2  ] \partial_\YM
 \cr
 &&  \left. +(\YM^2-\XM^2 )( \XM^2-1)  \partial_\XM^2  +  ( \YM^2 -\XM^2)(  \YM^2-1)  \partial_\YM^2  \right\} a_s(\XM,\YM) \la{SEqus}
  \eea
We will now discuss these equations in turn for two special cases, $\Delta_e=2,~3$,  which corresponds to the conformally coupled   and
  massless cases respectively.

 It will be interesting for us to solve the recursion relations \nref{recreserlgd} at $\XM=\YM=1$. This is a point where the $a_i(\XM,\YM)$ functions are
 required to be analytic. Then we see that the terms with derivatives drop out since they have a factor of $(1-\XM^2)$ or $(1-\YM^2)$. We can simply solve the
 recursion relations by going back to \nref{conengegd}, setting $\XM= \YM=1$ and then making an ansatz of the form
 $  \sum_{i=0}^s \Malpha^i \Mbeta^{s-i}
   a_i (1,1) = \delta^s h(\gamma/\delta ) $. Replacing this in the equation \nref{conengegd} we find
   \be
    \sum_{i=0}^s \Malpha^i \Mbeta^{s-i}
   a_i (1,1) =  \Mbeta^s (1-{ \Malpha \over \Mbeta}  )^{\Delta_e -2} ~_2F_1(\Delta + \Delta_e -3,-s + \Delta_e -2,\Delta_e -1 ;{ 1 - {\Malpha \over \Mbeta } \over 2} )
 \la{RecSolOne} \ee

\subsection{Two $\Delta_e=2$ operators and a general operator with spin}

In this subsection we set $\Delta_e =2$ in  the above discussion.
  The equation \nref{FEqus} becomes very simple, and it says that $a_s$ depends only on $\YM$ or only on $\XM$, as
  we will see below we want the solution that depends only on $\YM$,  $a_s(\YM)$.
 Inserting this into the second equation \nref{SEqus}, we find
 \be \la{aseqn}
 0 = \left[ ( \YM ^2 -1) \partial_\YM^2 + 2 (1+s) \YM \partial_\YM  - ( \Delta -1 + s)(\Delta - 2 -s) \right] a_s(\YM)
 \ee
 which reduces to \nref{CoCouSc} when $s=0$.
 The solution that depends only on $\XM$ is set to zero by demanding regularity at $\XM= \pm 1$.
For large $\YM$ this solutions behave as
\be \la{LargeYbe}
a_s(\YM) \sim \YM^{ 1- \Delta -s} ~,~~~~~\YM^{\Delta -2 -s }
\ee

   Inserting this solution into the recursion relations \nref{recreserlgd} we get all the other helicity components.
For example, for massive spin two we get
\bea
a_1( \XM,\YM) & =& { 2 \XM \over \Delta} \left[ (\YM^2 -1) \partial_\YM a_2(\YM) - \Delta \YM a_2(\YM) \right]
\\
a_0(\XM,\YM) &=& { 1 \over \Delta(\Delta-1) }\left[ (1 + ( 3 + 2 \Delta)\XM^2) \YM(\YM^2-1) \partial_\YM a_2 +
\right.
\cr
&& \left.  +\{ \Delta(\YM^2-1) + \XM^2
( 4 + 4 \Delta - \Delta^2 - 4 \YM^2 - 5 \Delta \YM^2 + 2 \Delta^2 \YM^2) \}a_2 \right]
\eea
Here we have used the equation for $a_2$, \nref{aseqn}, to simplify the expressions. In general, we find that $a_i$ is a polynomial
of degree $s-i$ in $\XM$. Note that he singularity at $\Delta =1$ is the same as the one we discussed around \nref{SpinTwoEx}.

For the   $s=1$ case we get
\be
a_0 = - { \XM [ ( 1 - \YM^2) \partial_\YM  + (\Delta -3) \YM] a_1(\YM) \over \Delta -1}
\ee
Note that for odd spin, the full answer is odd under the $1 \leftrightarrow 2 $ exchange.
So that if the operators $O(\vec k_1)$ and $O(\vec k_2)$ are identical the
three point function vanishes identically.

Let us go  back to general $s$. For $\XM =\YM=1$ we can solve the recursion relations as in \nref{RecSolOne}. Specializing also to the case that
$\Malpha = \Mbeta $ we get that
\be \la{azone}
 \sum_{i=0}^s a_i(1,1) = { s ! 2^s \Gamma(\Delta -1) \over \Gamma(\Delta -1 +s)  } a_s(1,1)
 \ee
 This relation is important if we want to to set $\vec k_2 =0$ and $\vec k_1=\vec k_3$. This sets $\Malpha = \Mbeta$ and in addition. Then the
 relation \nref{azone} is specifying the value of the zero angular momentum component in terms of the maximal solution $a_s(\YM)$ evaluated at one. This
 relation is useful because $a_s(\YM)$ obeys the simple equation \nref{aseqn}.

It is possible to solve the recursion relations explicitly for large $\YM$. For large $\YM$,  $a_s$ has two possible power law behaviors, coming
from the large $\YM$ solution of \nref{aseqn}, see \nref{largepGs}.
Inserting each of these two possible powers  in the recursion relations \nref{recreserlgd} we obtain results consistent with \nref{Sqannonan}, where
each term comes from each of the two possible powers at large $\YM$.

\subsection{Two $\Delta_e = 3 $ scalars and a general operator with spin}
 \la{DelThreeSpAp}

Here we set $\Delta_e =3$ in the  discussion of appendix \ref{ThreeMomSpin}. The first equation \nref{FEqus} becomes identical to \nref{12perpEqn}
whose solution is given in \nref{condig}
\be \la{as3}
a_s(\XM,\YM) = U_s(\YM) + { 1 \over 4 } ( \YM^2 -\XM^2) G_s(\YM) ~,~~~~~~~~\partial_\YM U_s(\YM) = - \YM G_s(\YM)
\ee
Inserting \nref{as3} into \nref{SEqus} we get a couple of equations.
The first is that $G_s(\YM)$ obeys the same equation as in \nref{aseqn}
The second equation gives $U_s$ in terms $G_s$
\be \la{u3}
U_s(\YM) = { - 2 s \YM^2 G_s + (\YM^2 -1) ( G_s - \YM \partial_\YM G_s )  \over \Delta(\Delta-3) - s (s-3) }
\ee
which for $s=0$ reduces to \nref{solvt}. Again, inserting these solutions in the recursion relations we get the rest of the $a_i(\XM,\YM)$.

 We now consider the solution of these recursion relations for $\XM=\YM=1$.
This is given by \nref{RecSolOne}. Setting $\Delta_e=3$ and $\Malpha = \Mbeta $ in \nref{RecSolOne} we find that it vanishes
$ \sum_{i=0}^s a_i(1,1) =0$.
However, for our application, we need
 to take the derivative of the expression \nref{RecSolOne} with respect to $\partial_\Malpha - \partial_\Mbeta$ before setting them equal.
This gives a nonzero answer.
We finally get
\bea
( \partial_\Malpha - \partial_\Mbeta) \left. \sum_{i=0}^s \Malpha^i \Mbeta^{s-i}
   a_i (1,1) \right|_{\Malpha = \Mbeta =1} & = &  { s ! 2^s \Gamma(\Delta ) \over \Gamma(\Delta -1 +s)  } a_s(1,1) \la{Der3One}
   \\ \notag
   &=&  { s ! 2^s (\Delta-1) \Gamma(\Delta-1 ) \over \Gamma(\Delta -1 +s)  } { (-2 s) G_s(1) \over [\Delta(\Delta -3) - s(s-3)] }
\eea
where we have also used \nref{as3} and \nref{u3} at $\XM=\YM=1$ (using that $G_s(\YM)$ is analytic at $\YM=1$).

 It is also interesting to consider the case $\YM =-1$ which corresponds to the singularity at $k_{12} = - k_I$. In this case we expect the singularity to be
 of the form
 \be \la{NorCondxi}
 \int \xi \nabla_{i_1} \cdots \nabla_{i_s} \xi  \sigma_{i_1 \cdots i_s} \to  k_1 k_2 \int { d \eta \over \eta} e^{ i k_{12} \eta + i k_I \eta } \propto { \YM^2 - \XM^2 \over
 ( \YM+1)^{s} }
 \ee
 where we focused on the most singular piece. We see from \nref{as3} that this will come from the second term in $a_s$, since the $U$ term involves an integral
 that will reduce the order of the singularity at $\YM=-1$. This imposes the same condition on $G_s$ as \nref{NorCond}, namely $G_s \sim { 1 \over (\YM+1)^s } $ at
 $\YM \sim -1$.

\subsection{Calculating the two point function in inflation}
\la{TwoPtInflat}

In this subsection we calculate the two point function of a dimension three scalar operator (associated to a massless field) and
a general dimension $\Delta$ operator with spin.
We begin with the identity
\be \la{AnWaAp}
  ( \vec b . \vec K) \langle O_3(-\vec k) \epsilon^s . O(\vec k) \rangle'_{\rm Infl} \propto
  \vec b . \vec \partial_{k_2} \langle O_3(\vec k_1) O_3(\vec k_2) \epsilon^s O(\vec k) \rangle'  |_{\vec k_2 =0 }
  \ee
  where $\vec k = - (\vec k_1 + \vec k_2) $. We consider $\vec b$ so that $\vec b . \vec k =0$. This will be enough to determine the two point function.
  The ansatz for the two point function was given in \nref{tptInf}. The left hand side of \nref{AnWaAp} is
  \be \la{actConfAn}
   ( \vec b . \vec K) \langle O_3(-\vec k) \epsilon^s . O(\vec k) \rangle'_{\rm Infl} = - 2 (\Delta -1) k^{\Delta -s} (\vec \epsilon . \vec k)^{s-1} (\vec \epsilon . \vec b)
   \ee
   We have only one generator acting on the left hand side since we can write all the explicit momenta of the two point function in terms of the momentum of the
   second operator, by using the momentum conservation equation.
   We can now compute the right hand side of \nref{AnWaAp}. The first observation is that if the derivative acts on a scalar, such as $k_i$, then by the time
   we set $\vec k_2\to 0$ the direction will become proportional to $\vec k$, which contracted with $\vec b$ will give zero due to our condition on $\vec b$.
   Then the only terms will come from the derivatives of terms that contain $\epsilon . \vec k_2 $. This will give rise to a derivative of the form
   $\partial_\Malpha -\partial_\Mbeta $. This gives us a result proportional to   \nref{Der3One}. Equating this to \nref{actConfAn} gives us
   \be
   \langle O_3(-\vec k) \epsilon^s . O(\vec k) \rangle'_{\rm Infl} \propto \dot \phi_0 { \Gamma(\Delta -1) \over  \Gamma(\Delta -1 + s) } { G_s(1) \over
   [ \Delta(\Delta -3) - s (s-3) ] } k^{\Delta } (\vec \epsilon. \hat  k)^s
   \ee

\end{document}